\documentclass{mn2e}

\usepackage[dvips]{graphics}
\usepackage{lscape}

\newif\ifAMStwofonts

\title{Present-day scaling relations of submm galaxies: origin of 
spheroidal sytems}

\author[T. Takagi, H. Hanami and N. Arimoto]
 {T. Takagi$^{1, 2, 3, 4}$\thanks{E-mail: t.takagi@kent.ac.uk}, 
 H. Hanami$^{5}$ and N. Arimoto$^{6, 7}$ \\ 
$^1$ Centre for Astrophysics and Planetary Science, University of Kent,
Canterbury, Kent, CT2 7NR, UK \\     
$^2$ Blackett Laboratory, Imperial College, 
Prince Consort Road, London, SW7 2BZ, UK \\
$^3$The Institute of Space and Astronautical Science, 3-1-1 Yoshinodai, 
Sagamihara, Kanagawa
229-8510, Japan \\
$^4$Department of Physics, Rikkyo University, 3-34-1 Nishi-Ikebukuro,
Toshima-ku, Tokyo 171-8501, Japan\\ 
$^5$Physics Section, Faculty of Humanities and Social Sciences, Iwate
University, Morioka, 020-8550, Japan \\
$^6$Institute of Astronomy, School of Science, University of Tokyo,
2-21-1 Osawa, Mitaka, Tokyo 181-0015, Japan \\
$^7$National Astronomical Observatory,
2-21-1 Osawa, Mitaka, Tokyo 181-8588, Japan 
}

\date{}

\begin{document}

\maketitle

\begin{abstract}
We analyse the spectral energy distributions (SEDs) of 23 submm 
galaxies and 3 {\it ISO}-detected EROs, all of which have 
the spectroscopic redshifts, 
by using an evolutionary SED model of starbursts. 
This SED model allows us to investigate intrinsic properties 
of starbursts, such as the starburst age and 
the mean stellar metallicity, as it takes into account 
the chemical evolution. Also, the intrinsic size of 
the starburst region is estimated from observed SEDs.
Using this SED model, we predict colour, magnitude, and size 
of present-day descendants of submm galaxies, and derive scaling 
relations, such as the present-day colour-magnitude and size-magnitude 
relations. 
We argue that submm galaxies are the progenitors of present-day 
elliptical galaxies, provided that 
the initial mass function (IMF) of submm 
galaxies is slightly flatter than the Salpeter IMF. 
In this case, we find that 
1) the mean present-day magnitude of submm galaxies is similar to 
that of $L^*$ elliptical galaxies, 
2) the present-day colour-magnitude relation
is consistent with that of elliptical galaxies, 
3) the present-day size-magnitude relation of elliptical galaxies
can be reproduced if massive submm galaxies consist of 
multiple starburst regions.  
We estimate the effect of feedback in submm galaxies
as a function of starburst age. 
It is found that starburst regions in submm galaxies are likely to 
be self-regulated; i.e. the effect of feedback 
is nearly balanced with the self-gravity of starburst regions. 
\end{abstract}

\section{Introduction}
Submillimetre (submm) galaxies are the 
key to understand the process of galaxy formation 
(e.g. Smail, Ivison \& Blain 1997; 
Hughes et al. 1998; Eales et al. 1999; Scott et al. 2002).
It has been found that 
submm galaxies are massive starburst galaxies at high redshifts; 
i.e. gas mass is estimated to be as large as
$\sim 10^{11}$ M$_\odot$ from the 
CO line emission (Frayer et al. 1998, 1999;
Ivison et al. 2001; Genzel et al. 2003; Neri et al. 2003), 
and the median redshift of $z=2.4$ is recently found 
spectroscopically (Chapman et al. 2003a). 
If observed submm fluxes originate mainly from star formation, 
the star formation rates (SFRs) of 
bright submm galaxies are estimated to be 
over $10^3$ M$_\odot$ yr$^{-1}$ (e.g. 
Smail et al. 2002; Chapman et al. 2003a), large enough to 
produce a massive elliptical galaxy ($L>3$ $L^*$) within 
$\sim 1$ Gyr (comparable to the Hubble time at $z=2.5$). 
This estimate seems to be confirmed 
with the follow-up observations at X-ray, 
which suggest the negligible contribution of the AGN to 
submm fluxes (Alexander et al. 2003).  
These observations indicate that submm galaxies are the most 
plausible candidates of the progenitors of present-day 
elliptical galaxies. 

The evolutionary link between submm galaxies and 
elliptical galaxies is also suggested by the statistical 
properties of submm galaxies. 
The comoving number density of submm galaxies with the 
850 $\mu$m flux of $> 8$ mJy is comparable to that of 
present-day ellipticals with $L \sim$ 3 -- 4 $L^*$ 
(Chapman et al. 2003a). 
Another clue to the evolutionary 
link could be given by the clustering properties of submm galaxies, 
which should be strong if they are the progenitors of massive 
elliptical galaxies. This would be tested 
by large submm surveys in near future. 


Theoretically, 
the semi-analytic methods (SAMs) are extensively applied to 
explain the statistical properties of elliptical galaxies (e.g. 
Baugh et al. 1996;
Kauffmann et al. 1996; 
Cole et al. 2000; 
Benson et al. 2001;
Diaferio et al. 2001;
Springel et al. 2001).
The SAMs show that most of massive galaxies 
are formed at $z \la 1$ as a result of successive merging of 
smaller objects. Therefore, it is difficult to explain 
the statistical properties of submm galaxies with the standard SAMs. 
This problem comes from the uncertainty in the relation between 
the dynamical evolution of the dark matter and 
the star formation processes in the early 
stage of galaxy formation (e.g. Binney 2004). 
In order to avoid this uncertainty, 
we may need an alternative method to 
investigate the evolutionary link between submm 
galaxies and elliptical galaxies. 

Elliptical galaxies are known as a fairly homogeneous family; 
e.g. they exhibit very small scatter in the colour-magnitude
relation (e.g. Bower et al. 1992) 
and the relation between the $\mbox{Mg}_2$ index and the 
velocity dispersion $\sigma$ (Dressler et al. 1987), 
i.e., their properties are well characterized as a sequence of mass. 
It is also remarkable that they occupy a two-dimensional sheet (fundamental 
plane) in the three dimensional space defined by the velocity 
dispersion $\sigma$, 
the effective radius $r_e$, and the 
mean surface brightness within $r_e$ (Dressler et al. 1987; 
Djorgovski \& Davis 1987). 
If submm galaxies evolve into elliptical galaxies, 
their present-day properties must be consistent 
with these scaling relations. This kind of test 
can provide the other piece of evidence for the evolutionary 
link between submm galaxies and elliptical galaxies. 

In this paper, we analyse submm galaxies with an 
evolutionary model of spectral energy distribution 
(SED) for starburst galaxies, which is presented by 
Takagi, Arimoto \& Hanami (2003a; hereafter TAH03). 
In this model, the chemical evolution of starburst 
region is considered; 
i.e. if we specify the evolutionary phase of submm galaxies 
as starbursts, gas mass and stellar mass of submm galaxies 
can be estimated, together with their chemical abundances. 
Also, the intrinsic size of 
starburst region can be estimated from the observed SEDs, since 
the geometry of starburst region determines its
optical depth for a given dust mass. 
By using this model, we predict the present-day 
colour, magnitude and size of submm galaxies from 
observed SEDs, and produce the present-day scaling 
relations for submm galaxies, i.e. the colour-magnitude 
relation and the size-magnitude relation. 
Then, these scaling relations are compared with those of 
elliptical galaxies, in order to prove the evolutionary link 
between submm galaxies and elliptical galaxies. 
This approach is new and complementary to the previous studies. 

The structure of this paper is as follows. We describe 
practical methods of SED diagnostics of submm 
galaxies in Section 2. We summarise the results of 
SED fitting in Section 3. 
We confront the predicted
present-day characteristics of submm galaxies with those 
of present-day elliptical galaxies in Section 4. 
Then, we discuss 
a possible evolutionary scenario of submm 
galaxies in Section 5.
Our conclusions are given in Section 6.
Throughout this paper, we adopt the cosmology of 
$\Omega_m =0.3$, $\Omega_\Lambda =0.7$ and $H_0 =75$ km sec$^{-1}$ Mpc$^{-1}$.

\section{Prescription of SED diagnostics for submm galaxies}



\subsection{Evolutionary SED model of submm galaxies}
We consider the evolution of starburst regions in submm galaxies. 
Each starburst region is assumed to be 
a dynamically isolated system in which the chemical enrichment 
proceeds with effective mixing of gas by supernova 
feedback. 
The SED of such starburst region is determined by the 
starburst age for a given star formation history and 
the spatial distribution of stars and dust, i.e. geometry. 
The adopted SED model is the same as the model 
applied to nearby starbursts in TAH03, except for the 
initial metallicity and the initial mass function (IMF). 
Here we briefly describe this SED model. 


\subsubsection{Evolutionary model of starburst regions}
We approximate the star formation history in a starburst region 
with an infall model of Arimoto, Yoshii \& Takahara (1992). 
In this model, the overall star formation history is described 
with two time-scales, i.e. the gas infall time-scale $t_i$ and 
the star formation time-scale $t_*$. 
We adopt the simplest case, in which a
starburst is characterized by only one evolutionary time-scale $t_0$, 
i.e., we assume $ t_i=t_* \equiv t_0$.
Thus, we specify the star formation history with $t_0$, which 
is important only for the absolute time-scale of starburst events.
Note that the chemical evolution as a function of $t/t_0$ 
is almost independent of the practical choice of $t_0$ (TAH03). 
Since this is also true for the properties of the SED, specifically 
when $t_0 \ga 50$ Myr, we can derive the chemical properties of 
starbursts from the observed SED, irrespective of the adopted
value of $t_0$ (i.e. the star formation history). 
Practically, all models are calculated with $t_0=100$ Myr. 
Once the starburst age ($t/t_0$) is specified with the total 
mass of initial gas $M_T$, we can derive the gas mass $M_g$ 
and stellar mass $M_*$ in the starburst region, together
with the metallicities of gas and stars. 

We assume that the metallicity of pre-galactic gas clouds 
$Z_i$ is negligibly small (i.e. $Z_i$=0).
It is possible that pre-galactic gas clouds are chemically 
enriched with the former star formation activity. 
However, our conclusions are independent of practical choice 
of $Z_i$ when $Z_i \la 0.1 Z_\odot$. 

A simple model of dust evolution is adopted; 
i.e. we assume that the dust-to-metal ratio 
$\delta_0$ is constant. The value of $\delta_0$ depends 
on the dust model. We adopt three dust models, i.e. 
the model for dust in the Milky Way (MW), LMC and SMC. 
For the MW, LMC and SMC,  
the values of $\delta_0$ derived from the extinction curve 
and the spectra of cirrus emission are 0.40, 0.55, and 0.75, 
respectively. See TAH03 and  Takagi, Vansevi\v cius, \& Arimoto 
(2003b) in detail. 

\subsubsection{Geometry of starburst regions}
We assume that stars in submm galaxies are centrally concentrated 
as those in elliptical galaxies. 
The stellar density distribution is given by the 
King profile with log($R_t/R_c) = 2.2$ where 
$R_t$ and $R_c$ are the cut-off and core radius, respectively, and 
the adopted value is typical for elliptical galaxies (Combes et al. 1995). 
As in TAH03, dust is assumed to distribute 
homogeneously within $R_t$. 

Note that TAH03 show that the effective radii of 
ultraluminous infrared galaxies (ULIRGs) 
derived from SEDs 
are consistent with the observed ones at 
$J$, $H$, and $K$-band. In this geometry, 
the longer the observed wavelength, the smaller 
the observed effective radius, so far as 
the observed flux is dominated by stellar emissions. 
This is because stars in 
the central region contribute to the observed flux 
more and more with decreasing the optical depth. 
TAH03 also show that this trend is consistent with 
the observations of ULIRGs. Therefore, the adopted 
geometry is found to be suitable for nearby ULIRGs. 

Since the amount of dust is given 
by the chemical evolution, the geometry of starburst 
region determines the optical depth by dust. 
Therefore, 
one of the important parameters of this SED model is a compactness 
factor $\Theta$ of starburst regions, defined by
\begin{equation} \label{radius}
\frac{R_t}{1 \mathrm{kpc}} =
\Theta \left( \frac{M_*}{10^9 \mathrm{M}_{\odot}} \right)^\gamma\;,
\end{equation}
where $\gamma =1/2$ is adopted, resulting in the constant surface 
brightness for constant $\Theta$. 
Note that the SED feature is preserved for different values 
of $M_*$ when $\gamma =1/2$, 
since the source function within the starburst is conserved. 

The optical depth of starburst regions, here defined with the 
column density of dust measured from the centre to outer edge, 
is a function of starburst age and $\Theta$. 
By using $M_*$ and $\Theta$ derived from 
the SED fitting, we calculate the intrinsic effective radius $R_e$ 
(i.e. effective radius with no dust)\footnote{
Hereafter, we simply use the term `effective radius' 
for this meaning in this paper. 
}, which is given by 10.75$\times R_c$ in the adopted geometry. 

In Section 4.2, we point out that massive 
submm galaxies may consist of multiple starburst regions. 
So far, photometric data are available only for a whole 
galaxy especially at longer wavelengths. Therefore, 
it is difficult to analyse the SED of each starburst region. 
If the SED of starburst regions in submm galaxies is similar 
to each other, the derived starburst age and $\Theta$ for a 
whole galaxy can be interpreted as those of each starburst region. 
This condition is 
such that each starburst is triggered almost simultaneously 
and the intrinsic bolometric surface brightness 
of each starburst region is similar to each other, like nearby 
self-regulated starbursts (TAH03). 



\subsubsection{Evolution of stellar populations}
A population synthesis code developed by Kodama \& Arimoto (1997) is
used to calculate the spectral evolution of stellar populations. 
In this model, the effects of stellar metallicity are explicitly 
taken into account in spectra. 
This is important to consistently 
calculate spectra of passively evolving 
galaxies, which mainly depend on the metallicity and 
the age of stellar population (Worthey 1994). 

We adopt the two IMFs, 
i.e. the Salpeter IMF with the power raw index of 
$x=1.35$ (IMF$_{1.35}$), 
and a top-heavy IMF with the flatter slope $x=1.10$ (IMF$_{1.10}$). 
Such a flat IMF has been suggested by Kodama \& Arimoto (1997)
for elliptical galaxies, and found for the OB associations 
in the Milky Way (Massey et al. 1995). 
For both the IMFs, we adopt
the lower and upper mass limit of 0.1 M$_\odot$ and 60 M$_\odot$, 
respectively. 

The choice of IMF is important for the present-day magnitudes 
of submm galaxies, since the mass-to-light ratio is 
smaller for the flatter IMF. Also, the present-day colours
of submm galaxies depend on the IMF, since the yield of 
chemical enrichment is higher for the flatter IMF.

\subsubsection{Radiative transfer with dust}
The SED from UV to submm of starburst regions is calculated 
for each starburst age, $\Theta$ and the extinction curve (MW, LMC and SMC), 
by using the code for the radiative transfer used by TAH03. 
In this code, 
isotropic multiple scattering is assumed and the self-absorption of
re-emitted energy from dust is fully taken into account. 
The temperature of dust grains is calculated for each dust size and 
constituent at each radial grid. 
For very small grains, the temperature fluctuation is calculated 
consistently with the radiative transfer. See Takagi et al. (2003b) 
in detail.

\subsection{Sample galaxies} 
The sample includes extremely red objects (EROs) detected at the 
MIR--submm wavelengths, submm galaxies found with SCUBA 
blank field surveys, submm galaxies amplified by the gravitational 
lens, submm galaxies in the proto-cluster region, and  host galaxies 
of gamma-ray bursts (GRBs) detected at the submm wavelengths. 
Although 3 of {\it ISO}-detected EROs 
do not have observed flux at submm wavelengths, 
we call the sample galaxies as submm galaxies for simplicity. 

We consider submm galaxies which have spectroscopic redshifts only. 
Therefore, the secure optical identification is inevitable. 
The spectroscopic redshift is necessary 
to significantly reduce the uncertainty of the SED fitting. 

Also, the sample galaxies need to have extensive photometric data enough 
to perform the SED fitting. We consider galaxies which have more than 
3 photometric data in the optical-NIR bands\footnote{
Although 2 GRB hosts have only 2 optical-NIR 
photometric data, we find that the constraint on the model parameters
is acceptable. This is because the observed SEDs are rather 
peculiar, which is reproduced only with very young and optically 
thick SED models. Therefore, we include these galaxies in the sample.}.
In total, we collect 23 galaxies detected at submm wavelengths 
and 3 {\it ISO}-detected 
EROs from the literature (including private communication). 
The observed properties of the sample galaxies are summarised in Table 1. 
In Appendix A, we summarise the observations of the sample galaxies. 
For SMMJ02399-0136, we perform the SED fitting with two possible 
optical counterparts L1 and L2 (Ivison et al. 1998).

\subsection{SED-fitting method} 
The best-fitting SED model is searched by the $\chi^2$ minimization 
technique from a prepared set of SED models.
We calculate the SED models for 10 different 
starburst ages ($t/t_0 = 0.1$ -- 6.0) 
and 16 different compactness factors ($\Theta$=0.3 -- 3.0) 
for each type of extinction curve; i.e. the best-fitting 
SED model is selected from a total of 480 SED models. 
This model set is made for both IMF$_{1.35}$ and IMF$_{1.10}$. 
The upper limits of flux are taken into account in the 
fitting process, i.e. models exceeding the 3 $\sigma$ upper 
limits are simply rejected. 


\section{Results of SED fitting}
The fitting results with IMF$_{1.35}$ and IMF$_{1.10}$ are 
summarised in Table 2a and 2b, respectively. 
The results of SED fitting are shown in 
Figure \ref{sed} for IMF$_{1.10}$ which 
is more suitable for submm galaxies than IMF$_{1.35}$ (see section 4). 
Note that the fitting 
parameters are $t/t_0$ and $\Theta$ with the choice of the extinction curve. 
The normalization of the SED model is adjusted by $M_T$. 
The rest quantities are derived from the best-fitting SED model. 
In Appendix B, 
we show the contour maps of $\Delta \chi^2$ for the 
sample used in the main analysis given in Section 4 and 5. 
Also, the typical values of estimated errors 
are shown in Figure \ref{cm_uv}, \ref{kormendy}, and \ref{lbol}.

Table 3 summarises the results of SED fitting, giving the 
number of galaxies for each bin of age and optical depth, and 
the resulting extinction curve. 
We find that $45$ \% of submm galaxies have 
$t/t_0 \le 1$, in a marked contrast to 
nearby starbursts, showing only $\sim $23 \% 
have $t/t_0 \le 1$.
Thus, a significant fraction of submm galaxies are 
found to be young. 
The optical depth $\tau_V$ 
of submm galaxies has a peak at 10 -- 20, 
which is similar to that of nearby ULIRGs 
(cf. TAH03). 

The large fraction of apparently young starbursts in submm galaxies 
may be caused by an extra AGN contribution 
to the rest-frame UV light. Since we have no 
further information on the presence of AGN for most of 
submm galaxies, we hereafter assume that they are intrinsically 
young. This assumption does not change our main conclusions which 
are derived from old ($t/t_0 \ge 2.0)$ submm galaxies. 

EROs detected in the MIR -- submm wavelengths 
are found to be the oldest starburst galaxies 
($t/t_0 \ga 5$) in the sample. 
This means that colours of stellar populations 
should be intrinsically red, in order to 
reproduce extremely red colours. 
Note that the derived stellar masses for EROs are systematically 
higher than those of the other sample, due to 
higher mass-to-light ratio.

GRB host galaxies (2 out of 3) have unique SEDs 
characterized by young stellar populations ($t/t_0<1$) 
and large optical depth ($\tau_V \sim 60$). 
Although their optical depths are the largest among our sample, 
the optical/NIR SEDs are not very red. 
This is because only stars near the surface of the starburst 
region actually contribute 
to the observed UV-NIR SED in such a large $\tau_V$. 

Spectroscopic observations and X-ray detections suggest that 
6 sample galaxies (EROJ164023, ISOJ1324-2016, 
PDFJ011423, CUDSS14.13, SMMJ02399-0136 and SMMJ02399-0134) 
harbour AGN.
As a result of SED fitting, we find a clear 
MIR excess over the SED model for 4 sources 
(EROJ164023, PDFJ011423, CUDSS14.13, and SMMJ02399-0134)
out of 6, 
which can be attributed to hot dust components 
around AGNs. On the other hand, the observed MIR fluxes 
of the other sample (HR10, CUDSS14F, and CUDSS14A) 
are reasonably explained solely by the starburst SED model. 
When the observed SEDs, except for the MIR excess, 
are well explained by the SED model, we assume that AGN 
dominates only at MIR wavelengths, and therefore the 
derived parameters are still usable for the following analysis. 

For 4 sample galaxies 
(EROJ164023, N2 850.1, SMMJ123629.13+621045.8, 
SMMJ131225.7+424350), 
we find no reasonable SED fit; i.e. SED models
significantly under-estimate the fluxes at MIR and submm wavelengths
even if they give good fit to the optical--NIR SEDs. The possible reasons 
are 1) the contamination from AGN (EROJ164023), 2) 
the uncertainty in the optical identification (N2 850.1), 
and 3) 
the low signal-to-noise ratio for faint objects
(SMMJ131225.7+424350 with $S_{850\mu\mathrm{m}}=2.4 \pm 0.74$ mJy). 
For SMMJ123629.13+621045.8, it seems that none of these reasons is 
suitable. 
Note that 
SMMJ123629.13+621045.8 is one of the reddest object ($R-K$=6.7). 
We may need to extend the parameter range to fit such an 
extremely red SED. 
Accordingly, we exclude these 4 galaxies in the following analysis. 

We present the fitting results of SMMJ02399-0136 for both L1 and L2. 
Since the relative contribution to MIR-submm flux from each 
component is uncertain due to the large beam size of telescopes, 
we assumed that the observed MIR-submm 
flux is dominated by either L1 or L2. 
If the observed dust emission is 
dominated by L2, 
the flux at rest-frame 2 $\mu$m is significantly 
under-estimated, while the flux at rest-frame 4 $\mu$m is properly 
reproduced. It is difficult to explain this discrepancy with 
the presence of hot dust component around the AGN. 
On the other hand, the SED of L1 component is reproduced 
well by the model. 
Therefore, we hereafter assume that observed dust emission 
originates mainly from the L1 component.

\section{Present-day scaling relations of submm galaxies}

\subsection{Present-day color-magnitude relation of submm galaxies}
The present-day colours and magnitudes are calculated under 
the assumption that effects of 
star-formation after the observed epoch are not significant 
for the resulting present-day colours and magnitudes.
This assumption is safe specifically 
for starbursts in the later evolutionary phase; at $t/t_0\sim 2$, 
about 60 \% of mass in the gas reservoir is already 
transformed into stars.

In Figure \ref{cm_uv}, we compare the predicted present-day 
colours ($U-V$) and magnitudes of submm galaxies with the 
colour-magnitude 
(CM) relation of elliptical galaxies at $z=0$.
In this figure, a solid curve with crosses 
indicates the present-day colour and magnitude of a starburst at $z=3$ 
with $M_T=10^{12}$ M$_\odot$, which corresponds to the formation 
of the brightest elliptical galaxy.
If the predicted present-day magnitude
is lower than that indicated by this line at a given $U-V$, 
present-day descendants of submm galaxies (or simply end-products) 
would be more massive than brightest elliptical galaxies. 

In the case of IMF$_{1.35}$, more than half of 
old ($t/t_0 \ge 2$) submm galaxies 
are more massive than the brightest elliptical galaxies. 
Moreover, the resulting $U-V$ colours of all 
sample galaxies are bluer than the CM relation. 
In order to redden the end-products, a significant amount 
of metal rich stars or old stars (i.e. red stars) 
should be additionally supplied. 
Since the stellar mass is already 
comparable to that of the brightest elliptical galaxies, 
such scenarios to supply red stars are not plausible. 
Therefore, we suggest that IMF$_{1.35}$ are not
appropriate for submm galaxies. 

On the other hand, in the case of IMF$_{1.10}$, the majority of 
old submm galaxies are less massive than the brightest elliptical galaxies. 
We derive the mean $V(B)$-band magnitude of $-22.03$ $(-20.97)$ 
mag for old submm galaxies.
Thus, the typical luminosity of the 
end-products of these submm galaxies is similar 
to those of $L^*$ elliptical galaxies which have 
$M_B^* = -20.75$ mag (Marinoni et al. 1999). 

For IMF$_{1.10}$, not only the magnitudes, but also the present-day 
colours of old submm galaxies are consistent with those of 
elliptical galaxies; 
i.e, old submm galaxies seem to follow the CM relation. 
This suggests that these submm galaxies can be real progenitors of 
elliptical galaxies.

Note that young ($t/t_0 < 2$) submm galaxies are systematically 
bluer than old ($t/t_0 \ge 2$) ones. This is because we assume that 
the star formation ceases at the observed epoch, 
irrespective of derived starburst ages, and therefore the 
mean metallicity of stars in young submm galaxies is
smaller than that of old ones. 

The observations of passively evolving elliptical galaxies 
suggest that the origin of CM relation is the 
systematic difference in the mean metallicity of stars 
rather than that in the mean age of stars 
(Kodama \& Arimoto 1997). Here we can check the 
origin of the CM relation independently of the previous 
studies. 
In Figure \ref{origin}, we show 
the present-day age and the luminosity-weighted 
metallicity as a function of present-day $V$-band magnitude. 
The scatter of age is considerable at all magnitudes. 
Such a large dispersion in age has recently been confirmed 
by Yamada et al. (2004) among 
elliptical galaxies in low density environments 
with a detailed study of 
H$\gamma_\sigma$ absorption index (Vazdekis \& Arimoto 1999). 
Therefore, three luminous submm galaxies ($M_V \le -21.5$)
with the age of $<8$ Gyr and $z \le 0.85$  
could be real progenitors of young field ellipticals. 
In Table 4, we give the mean characteristics 
of old submm galaxies at $z=0$ for each bright 
($M_V<-22$) and faint sample ($M_V \ge -22$). 
Although faint sample of old submm galaxies 
appears to be young in the mean value, we find no difference 
in the mean age when we include young submm galaxies (open circles 
in Figure \ref{origin}). 
Note that the present-day age depends mainly on the redshift, 
not on the starburst age, while the present-day magnitude 
depends on the starburst age. 
If the end-products of young submm galaxies remains 
to be faint ($M_V \ge -22$), 
there is no systematic difference in the present-day age 
between bright and faint sample. 
On the other hand, we find the systematic difference 
in the mean luminosity-weighted 
metallicity, $\Delta \log \langle Z_*/Z_\odot \rangle = 0.19$. 
Thus, the CM relation of elliptical galaxies could be 
caused by the systematic difference in the 
metallicity. However, we need more faint sample of old 
submm galaxies to derive firm conclusions on this issue. 

The analyses of the observed CM relation of elliptical 
galaxies suggest that the major star formation of 
elliptical galaxies occurred at $z>2$ 
(e.g. Bower et al. 1992; Kodama et al. 1998; Stanford et al. 1998). 
This suggests that the progenitor of elliptical galaxies 
would be found mostly at $z \ga 2$. 
On the other hand, the mean redshift of our sample 
is 1.6, while recent spectroscopic observations 
of submm galaxies suggest the median redshift of $z\sim 2.4$ 
(Chapman et al. 2003a). 
Our sample is likely to be biased towards 
low redshifts, since they should be bright enough in the optical/NIR 
photometric bands to perform the spectroscopic observations 
and also the SED fitting. 
Therefore, our sample occupies a lower tail of the 
redshift distribution of submm galaxies. 
Nevertheless, the present-day colours and 
magnitudes of the sample are consistent 
with the observed CM relation. 
This suggests that the physical 
mechanisms to establish the CM relation are still effective 
even at $z \la 2$. 

\subsection{Present-day size-magnitude relation of submm galaxies}
We compare the present-day $B$-band 
magnitudes and effective radii of submm galaxies
with the observed relation for elliptical galaxies, so-called 
the Kormendy relation. Since IMF$_{1.10}$ is more suitable
for submm galaxies as shown above, we mainly focus on this 
case. The effective radii of old 
submm galaxies range from $\sim 300$ pc to a few kpc. 
As shown in Figure \ref{kormendy}a, 
the predicted effective radii of less massive galaxies with 
$M_B \sim -19$ mag are consistent with the observed ones, 
while those of massive galaxies are an order of magnitude 
smaller than the observed ones. 

This discrepancy in $R_e$ for massive galaxies can be explained 
if the starburst region in submm galaxies are a multiple system 
rather than a unit system; i.e. several starburst regions are 
distributed within the potential of a whole galaxy, and these 
starburst regions 
eventually merge to form a more diffuse stellar system. 
Note that the recent observations 
of submm galaxies with the Hubble Space Telescope ({\it HST})
actually show the multiple 
structures (Chapman et al. 2003b). 
As we note in Section 2.1.2, the interpretation of 
the SED fitting remains the same if
each starburst region is triggered almost simultaneously, 
and the intrinsic bolometric surface brightness 
of each starburst region is similar to each other. 

The other possibility which may account 
for this discrepancy is that the stellar 
system expands, owing to dynamical response to 
gas removal as a galactic wind. 
Following Yoshii \& Arimoto (1987), we find that 
old submm galaxies could expand by no more than factor of $\sim$3, 
irrespective of the time-scale of gas removal. 
Moreover, this effect is expected to be more 
significant for less massive galaxies. However, the 
discrepancy of effective radius between submm galaxies and 
elliptical galaxies is larger for more massive submm galaxies. 
Therefore, it is difficult to explain the discrepancy of $R_e$ 
only by the dynamical response 
of a stellar system to the gas removal. 


The size-magnitude relation of submm galaxies seems to deviate 
from that of elliptical galaxies around $M_B \sim -19.5$ mag; 
i.e. the largest starburst regions in submm 
galaxies probably have the present-day magnitude of 
only $M_B \sim -19.5$ mag. 
The number of starburst regions can be 
estimated from the present-day magnitudes if 
submm galaxies with the present-day 
$M_B=-19.5$ mag are the largest 
starburst regions. 
Submm galaxies with the $L^*$ present-day luminosity 
are likely to have at least 
$\sim 3$ starburst regions within the size 
of effective radius $\sim$5 kpc. 
The clumpiness of submm galaxies may correlate with the mass 
of submm galaxies, since the discrepancy of the radius is larger 
for more massive submm galaxies. It is important to confirm 
the existence of the maximum scale in starburst regions, 
in order to understand the star formation process at 
high redshifts. To do this, we need 
high resolution imaging, e.g. 
with the Atacama Large Millimeter Array (ALMA). 

If submm galaxies consist of multiple starburst regions, 
we need to correct the derived $R_e$ for the multiplicity and 
for the effect of merging of these starburst regions, 
in order to compare the derived $R_e$ 
with the effective radius of elliptical galaxies. 
To estimate the correction factor for these effects, 
we simply assume that starburst regions in a submm galaxy
have a similar mass and SED to each other. We call this 
assumption as the `equality' condition of the starburst region.
We hereafter denote the stellar mass and the effective radius 
of each starburst region as $m_*$ and $r_e$, respectively. 
When a submm galaxy consists of $N$ starburst regions, 
we can write $r_e \sim R_e/\sqrt{N}$, since $m_*= M_*/N$ 
and $R_e \propto (N m_*)^{1/2}$. The gravitational energy 
per unit mass in a starburst region is 
$\frac{Gm_*}{r_e} \sim \frac{GM_*}{\sqrt{N} R_e}$.
If the gravitational energy per unit mass does not 
change during the assembly process of multiple starburst 
regions, and the total stellar mass is conserved, 
massive submm galaxies would eventually evolve into 
elliptical galaxies with the effective radius of 
$\sim \sqrt{N} R_e$; i.e. 
we need the correction factor $\sqrt{N}$ for effective 
radii in Figure \ref{kormendy}a.
In Figure \ref{kormendy}b, we show the size-magnitude 
relation corrected for the multiplicity. 
Note that the corrected size-magnitude relation has the 
similar slope with the observed size-magnitude relation 
of elliptical galaxies. The zero point is marginally 
consistent with the observation within the uncertainty, 
although it is systematically 
lower by a factor of $\sim 2$. The difference in the 
zero point might be explained by the expansion of 
stellar system, as a result of the gas removal from 
each starburst region at the end of its activity 
(e.g. Yoshii \& Arimoto 1987). 




\section{Evolution of submm galaxies}
The starburst activity can be characterized by 
the relative strength of self-gravity to feedback (TAH03). 
In massive and compact starbursts 
like nearby ULIRGs, the self-gravity of starburst region becomes 
strong enough to suppress the feedback effect. 
This is because the supernova rate 
depends mainly on the total baryonic mass of starburst regions, 
while the self-gravity increases 
with decreasing the size of starburst region for a given mass. 
First, we investigate the relative 
strength of self-gravity to feedback in submm galaxies, which 
can provide an insight into the evolution of submm galaxies. 

\subsection{Feedback versus self-gravity in submm galaxies}
TAH03 estimate the strength of feedback from the 
kinetic energy of gas per unit mass. We follow the same 
prescription as that in TAH03. 
The total kinetic energy of gas due to feedback 
with the typical velocity $V_g$ 
can be written as $ \frac{1}{2} M_g V_g^2 \simeq L_{\mathrm{kin}} 
t_{\mathrm{dyn}} $, 
where $L_{\mathrm{kin}}$ is the kinetic luminosity due to feedback, 
and $t_{\mathrm{dyn}} $ is the dynamical time-scale of the system.
Assuming that $ L_{\mathrm{kin}} = f_{\mathrm{kin}} L_{\mathrm{bol}}$ and 
$t_{\mathrm{dyn}} = t_0$, 
the kinetic energy per a unit mass can be written as  
\begin{equation}
\frac{1}{2} V_g^2 \simeq \frac{L_{\mathrm{kin}} t_{\mathrm{dyn}}}{M_g}
= f_{\mathrm{kin}} \frac{L_{\mathrm{bol}}}{\psi},
\end{equation}
where $\psi (= M_g/t_0)$ is the SFR. 
We use the relation 
\begin{equation}
\frac{L_{\mathrm{bol}}}{\varepsilon \mathrm{L}_\odot} = 
\frac{\psi}{1 \mathrm{M}_\odot \mathrm{yr}^{-1}} 
\left ( \frac{t}{t_0} \right )^\alpha
\end{equation}
where we find $\varepsilon = (1.7 \times 10^9$, $7.0 \times 10^9$) 
and $\alpha = (1.0, 0.42)$ for IMF$_{1.35}$ and IMF$_{1.10}$. 
Then, we derive the escape velocity, 
$V_\mathrm{esc}^2 \simeq 2GM(<R_e)/R_e$, where $M(<R_e)$ 
is the total mass within $R_e$. 
Here, we simply assume $f_{\mathrm{kin}} =0.01$ and 
$M(<R_e) = 2 M_* (<R_e)$ as in TAH03.
The limiting effect of feedback against self-gravity can be 
estimated by the comparison of $V_g$ with $V_\mathrm{esc}$. 
These velocities should be corrected when submm galaxies consist 
of $N$ starburst regions. 
For the equality condition of starbursts, 
we find the escape velocity of each starburst region 
$v^2_{\mathrm{esc}} \simeq V^2_{\mathrm{esc}}/\sqrt{N}$, 
since $m_* = M_*/N$ and $r_e \simeq R_e/\sqrt{N}$. 
On the other hand, we expect the velocity of gas due to feedback 
in each starburst $v_g \simeq V_g$, since 
the strength of feedback mainly depend on 
the efficiency of star formation, not on the mass scale. 
Therefore, we find $v_g^2/v_\mathrm{esc}^2 \simeq 
\sqrt{N} V_g^2/V_\mathrm{esc}^2 $. 
If starbursts are self-regulated, we expect 
$v_g^2/v_\mathrm{esc}^2 \ga 1$.


In Figure \ref{velmag}, we show $v_g^2/v_\mathrm{esc}^2$
corrected for the multiplicity, and find that most of old 
submm galaxies have $v_g^2/v_\mathrm{esc}^2 \ga 1$. 
This suggests that feedback is 
effective, and therefore the star formation in old 
submm galaxies is likely to be self-regulated. 

In the local universe, TAH03 suggest that UV-selected starburst 
galaxies (UVSBGs) are 
self-regulated, while ULIRGs are dynamically unstable. 
The difference between UVSBGs and ULIRGs can be seen as a 
systematic difference in the intrinsic 
bolometric surface brightness. 
In Figure \ref{lbol}, we show the bolometric luminosity $L_{bol}$ 
and the effective radius $R_e$ of submm galaxies, along with
those of nearby starburst galaxies. 
An arrow in Figure \ref{lbol} indicates 
the correction due to the multiplicity, assuming 
the equality condition of starbursts and $N=3$. 
The intrinsic bolometric surface brightness of old submm galaxies 
are systematically lower than those of nearby ULIRGs
which have $>$10$^{13}$ L$_\odot$ kpc$^{-2}$; instead, it 
seems to follow the relation of UVSBGs. 
Again, this may indicate that old submm galaxies 
are self-regulated starbursts. 
Thus, the effect of feedback could be very important for the 
evolution of submm galaxies.

\subsection{Role of feedback and origin of the CM relation}
The evolutionary trend of the feedback effect in submm galaxies 
can be seen in Figure \ref{velage}, 
which shows $v_g^2/v_\mathrm{esc}^2$ 
of submm galaxies (corrected for the multiplicity) 
as a function of starburst age $t/t_0$. 
For old submm galaxies, 
$v_g^2/v_\mathrm{esc}^2$ is found to be almost constant ($\sim 1$). 
Thus, feedback in submm galaxies is nearly balanced 
with the self-gravity for $t/t_0 \ga 1$. 
This means that the starburst activity in submm galaxies 
could easily cease at any starburst age. 

Even so, 
the star formation activity would not completely cease
when starburst regions reside in a large scale 
gravitational potential well, probably due to dark matter, 
since the gas can fall back again to starburst regions. 
Note that 
such dark matter haloes are necessary to explain the surface 
brightness profile at X-ray from nearby elliptical galaxies 
(e.g. Trinchieri, Fabbiano \& Canizares 1986; Matsushita et al. 1998). 
This also means that the mixing of gas could occur in 
the scale of dark matter halo; i.e. the chemical evolution in 
multiple starburst regions may be related to each other. 

We suggest that massive submm galaxies consist of 
multiple self-regulated starbursts in the dark matter halo.
In such systems, how the starburst activity 
ceases is determined not by the 
properties of the starburst itself, but by the 
gravitational potential of the dark matter halo. 
This means that both the resulting stellar mass and 
the stellar metallicity are controlled by 
the gravitational potential of the dark matter halo. 
This may be important to explain the tightness of 
the CM relation of elliptical galaxies.

\section{Conclusions} 
We analyse submm galaxies, including EROs and host galaxies of 
GRBs, by using the evolutionary SED model of starburst galaxies.
This model allows us to investigate the intrinsic properties of 
submm galaxies as starbursts, since the chemical 
evolution is consistently taken into account. 
We determine the evolutionary phase of submm galaxies, 
and derive the stellar mass and the metallicity of stars. 
Then, we predict the colour, magnitude and size of 
present-day descendants of 
submm galaxies, based on the results of the SED fitting. 
This prediction is more reliable for old submm galaxies 
($t/t_0 \ga 2$), since 
the large fraction of gas would have been already 
used to form stars, and therefore the effect of star formation 
after the observed epoch is not significant. 

We derive the present-day scaling relations of submm galaxies; 
i.e. the colour-magnitude and the size-magnitude relations at $z=0$. 
These scaling relations can provide new clues on the evolutionary 
link between submm galaxies and elliptical galaxies. 
By comparing them with the observed scaling relations of 
elliptical galaxies, we have reached the following conclusions:
\begin{itemize}
\item
The predicted present-day colours and 
magnitudes of submm galaxies suggest 
that the IMF of submm galaxies is flatter than IMF$_{1.35}$; 
otherwise the present-day magnitude of 
submm galaxies become brighter than the brightest elliptical galaxies.
\item
With IMF$_{1.10}$, the predicted colours and magnitudes of 
old submm galaxies are 
consistent with the observed CM relation of elliptical 
galaxies. The mean present-day magnitude of submm 
galaxies is similar to that of $L^*$ elliptical galaxies. 
This implies that these 
submm galaxies are quite likely to evolve into present-day 
elliptical galaxies after the starburst event. 
\item
The derived effective radii of less massive 
submm galaxies ($M_B \sim-19$ mag at $z=0$) 
are consistent with the observed 
size-magnitude relation of elliptical galaxies. 
The small size of massive submm galaxies, compared with 
the size of ellipticals with similar mass, suggests 
that they consist of multiple starburst regions. 
When the multiplicity of starburst regions is taken 
into account, the resulting effective 
radii become consistent with the observation 
within the uncertainty. 
\end{itemize}

Furthermore, 
we find that starbursts in submm galaxies are self-regulated; 
i.e. feedback is nearly balanced with the self-gravity. 
This mechanism may be
important to explain the tightness of 
the colour-magnitude relation of elliptical galaxies.

\section*{Acknowledgements}
TT would like to thank to N.\ Shibazaki, V.\ Vansevi\v cius, 
T.\ Matsumoto, M.\ Rowan-Robinson and G. White 
for their encouraging supports.
We are grateful to the referee S.\ Chapman 
for providing valuable data which significantly 
improve the quality of the paper. 
This work was financially supported in part by a
Grant-in-Aid for the Scientific Research (No.11640230, 13011201, 
13640230 $\&$ 14540220)
by the Japanese Ministry of
Education, Culture, Sports and Science.
This research has been supported in part by a Grant-in-Aid for the
Center-of-Excellence (COE) research.
Also, TT acknowledges the support of PPARC.

\newpage 

  \begin{figure*}
    \resizebox{7cm}{!}{\includegraphics{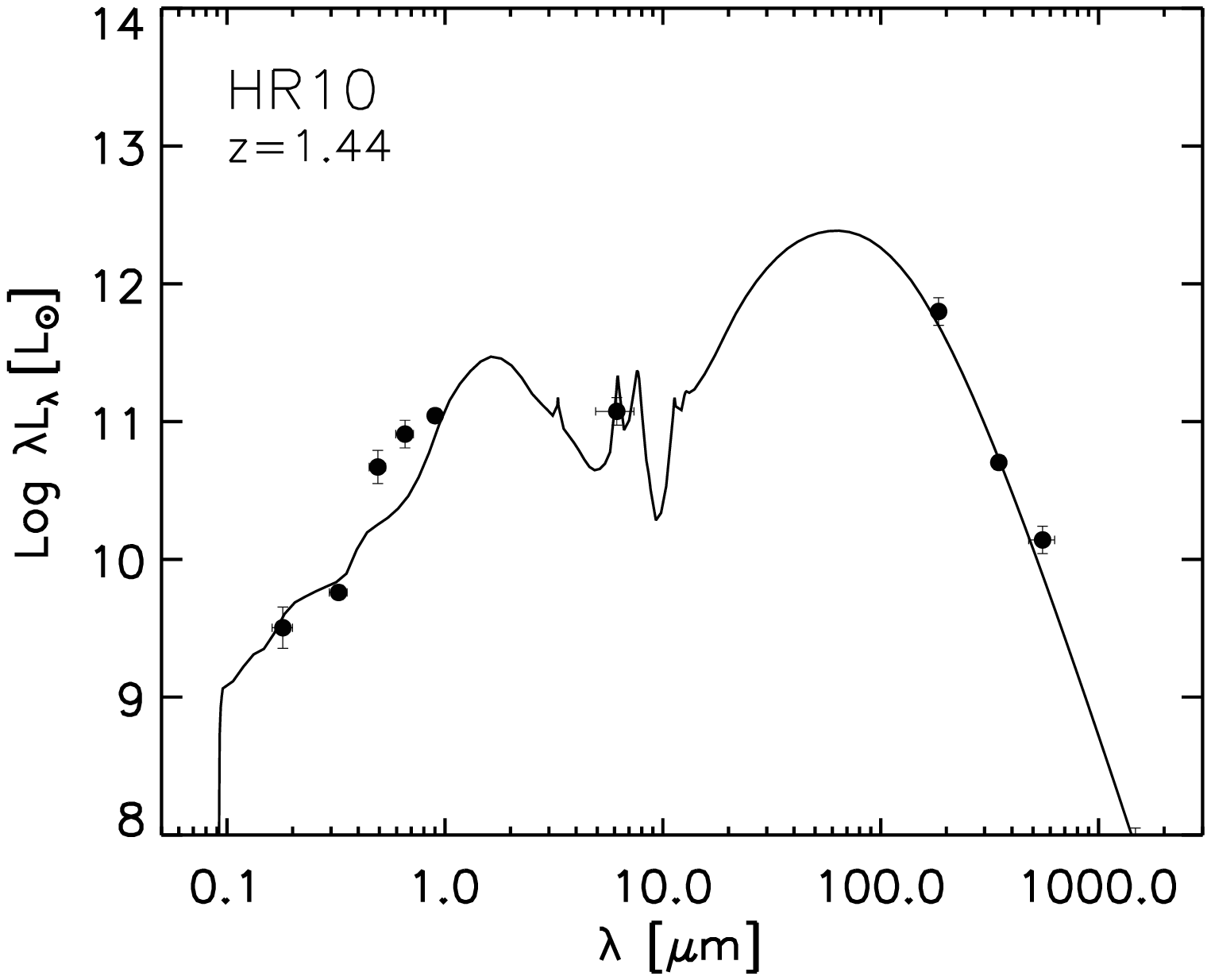}}
  \resizebox{7cm}{!}{\includegraphics{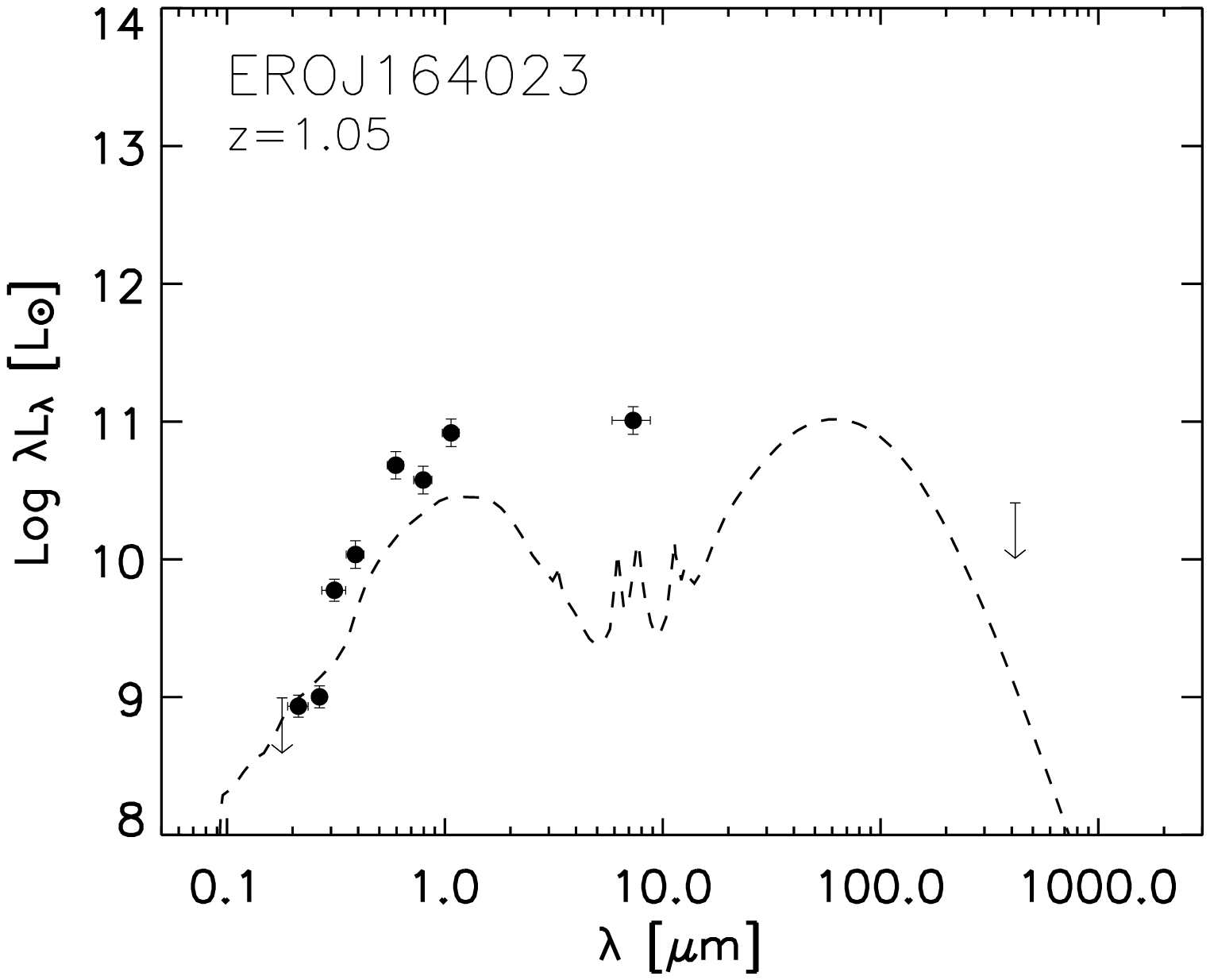}}
 \resizebox{7cm}{!}{\includegraphics{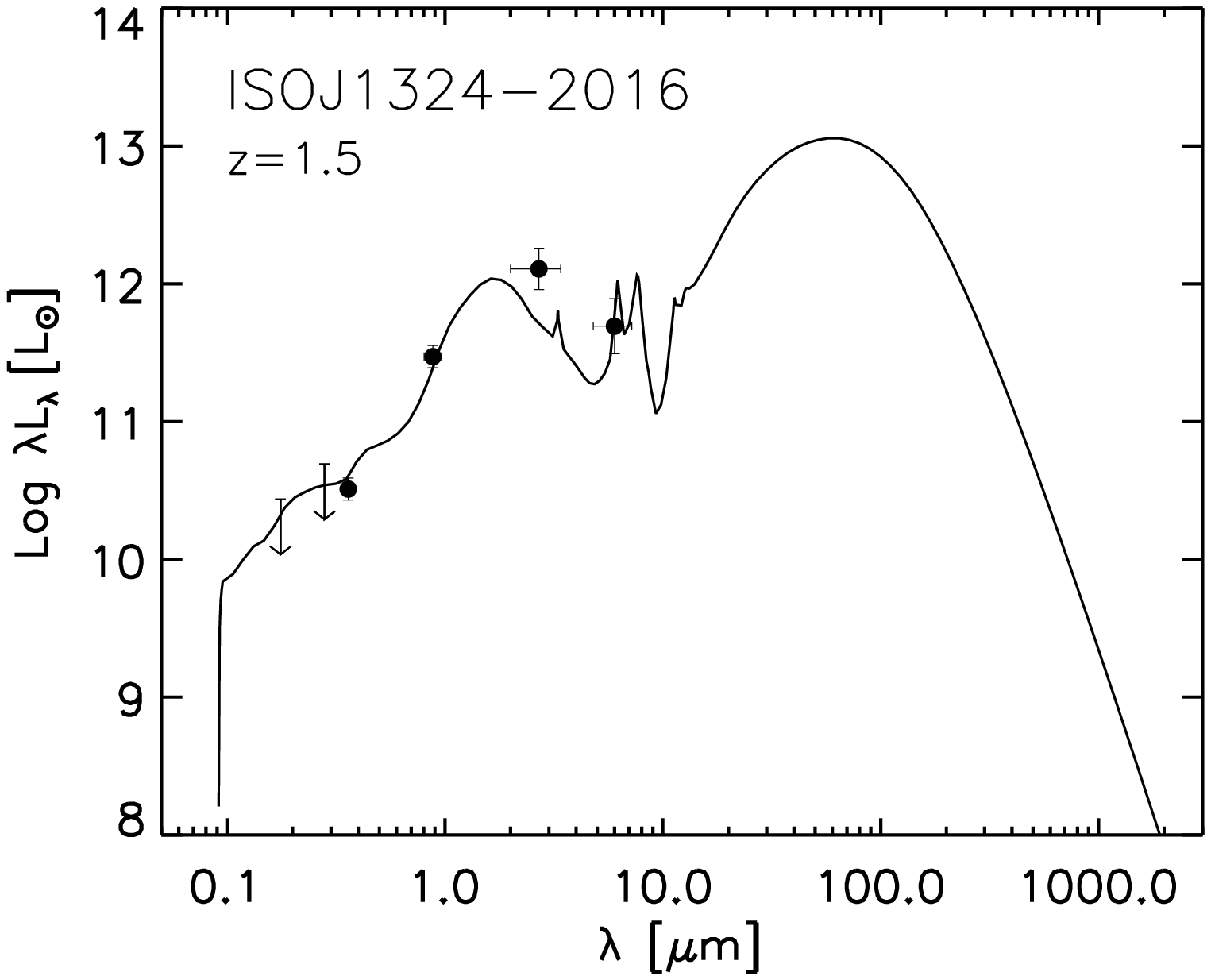}}
  \resizebox{7cm}{!}{\includegraphics{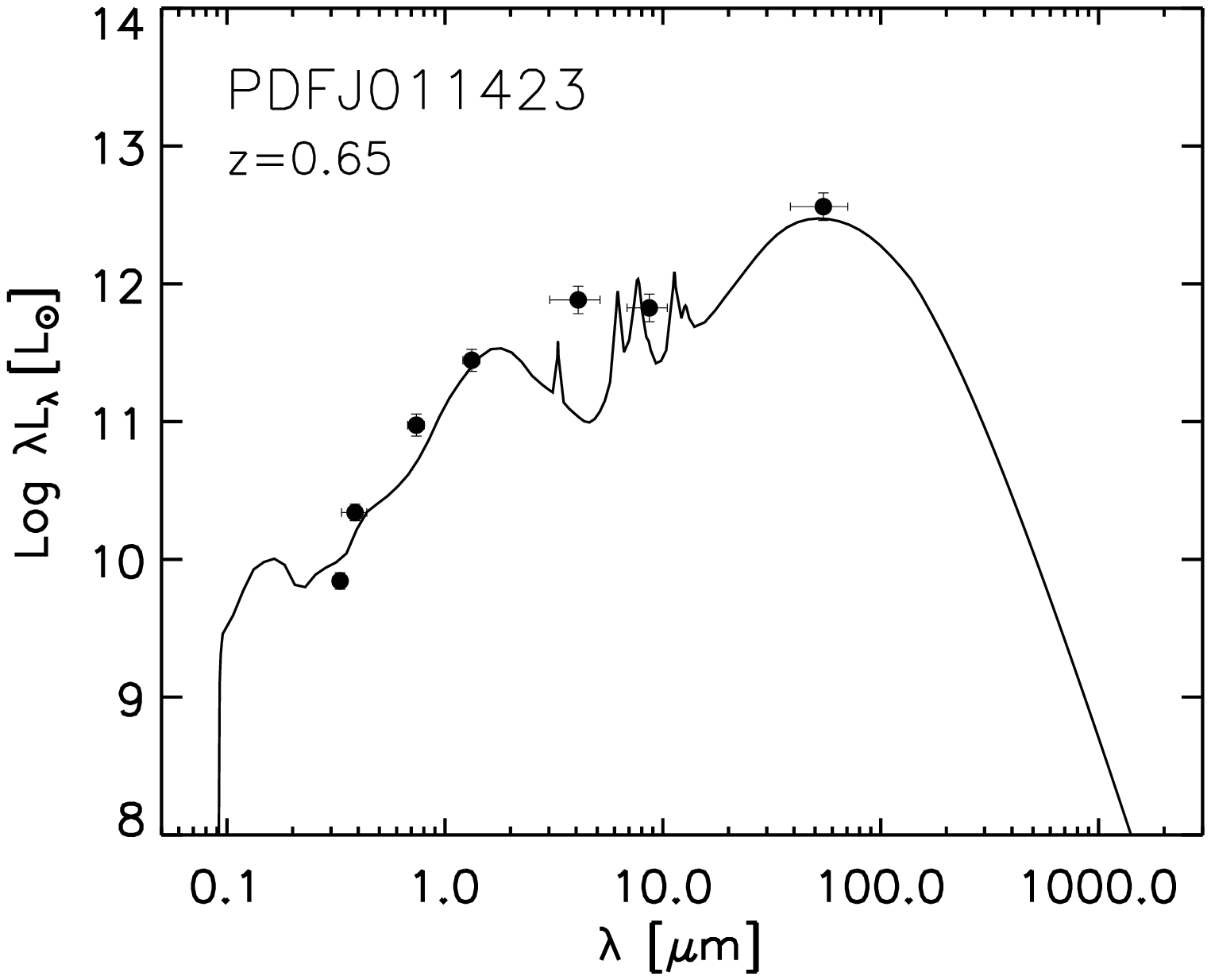}}
 \resizebox{7cm}{!}{\includegraphics{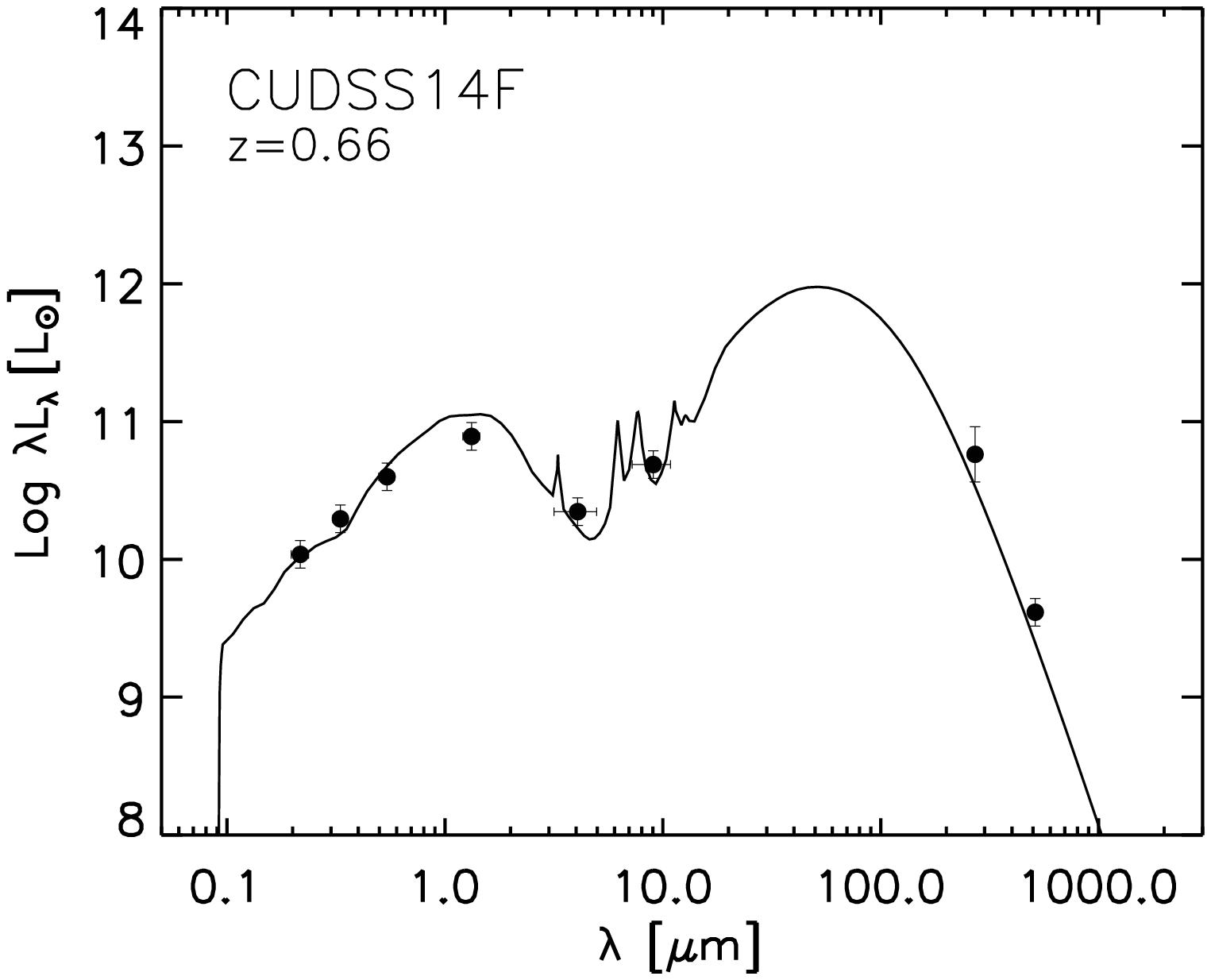}}
 \resizebox{7cm}{!}{\includegraphics{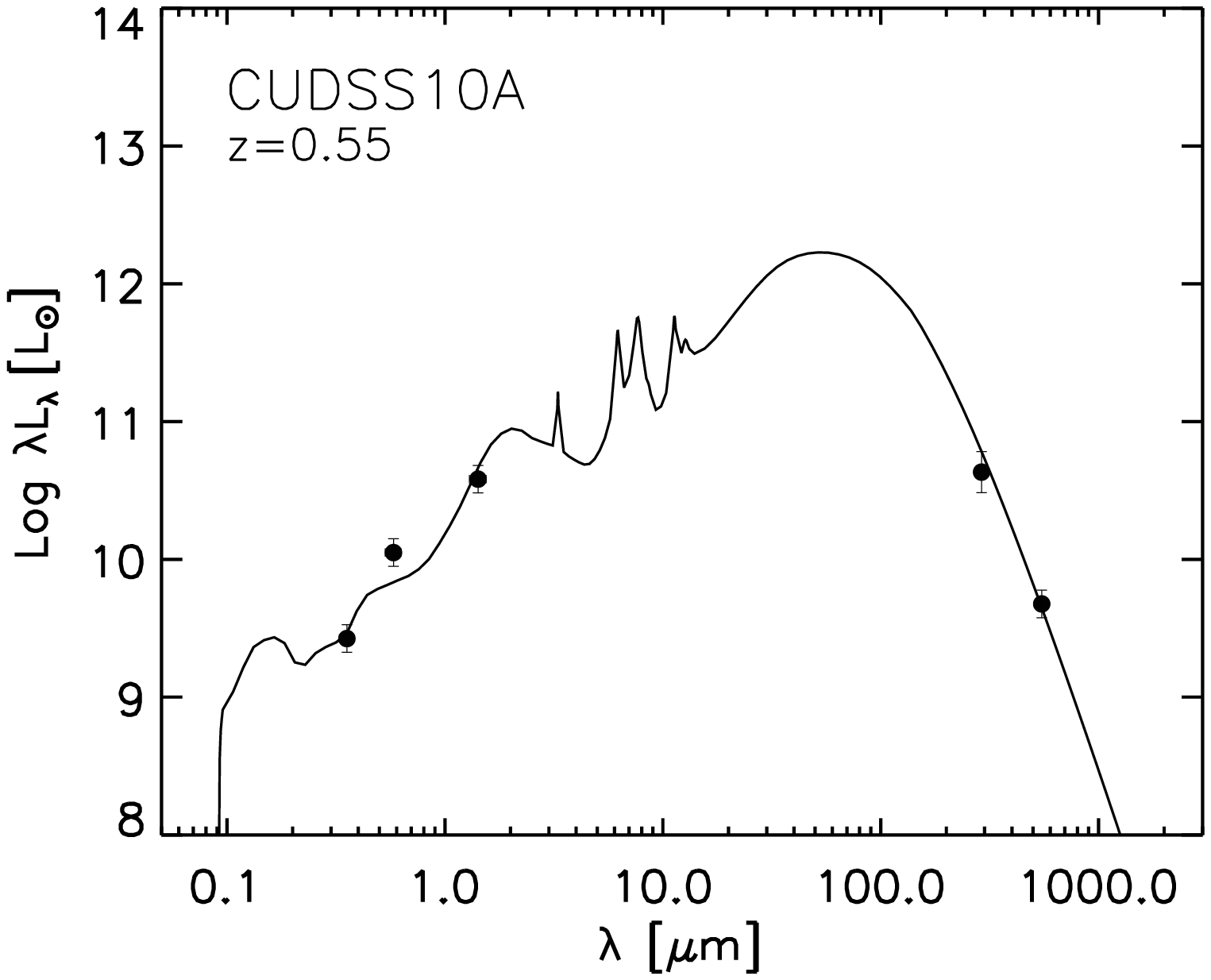}}
  \resizebox{7cm}{!}{\includegraphics{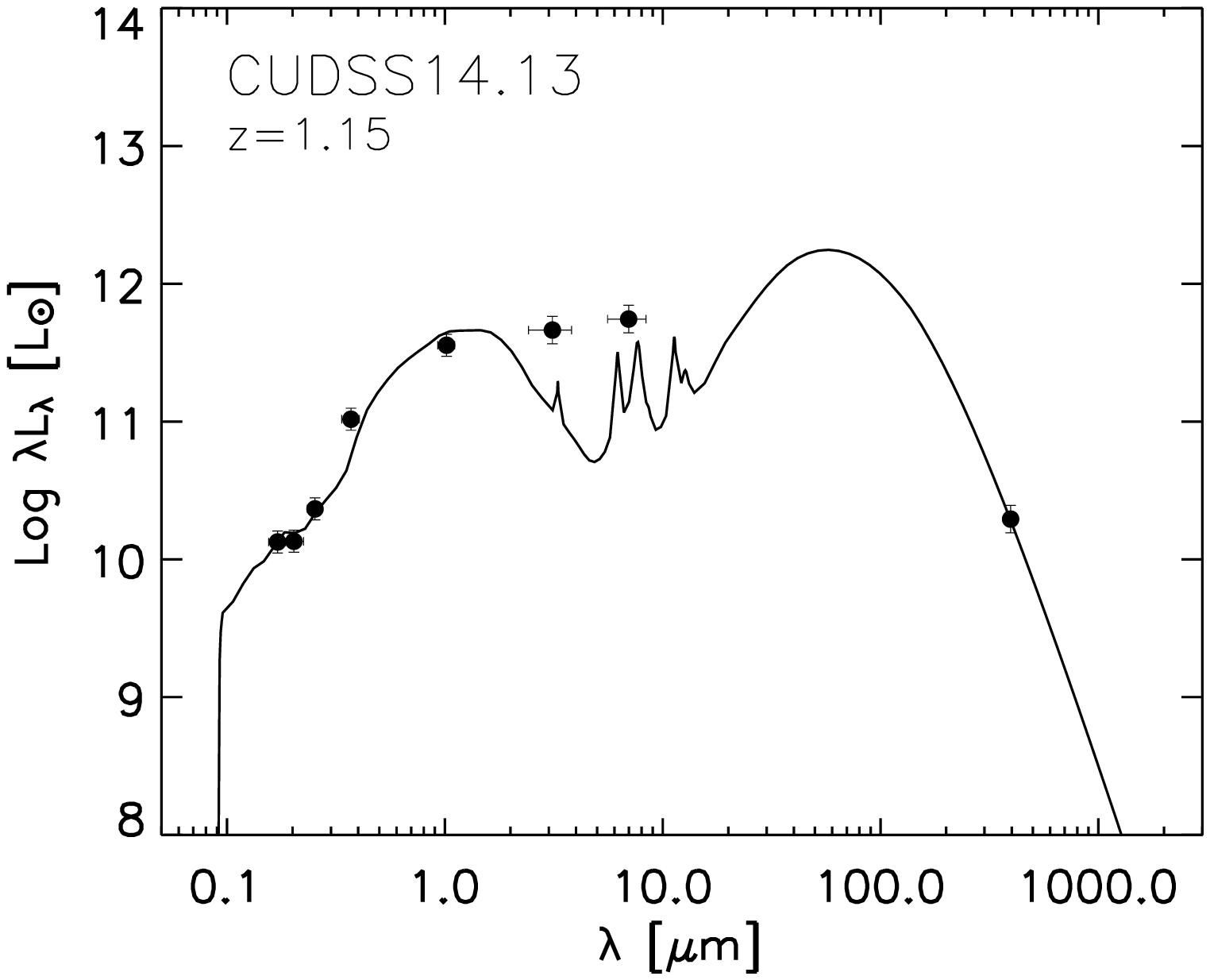}}
 \resizebox{7cm}{!}{\includegraphics{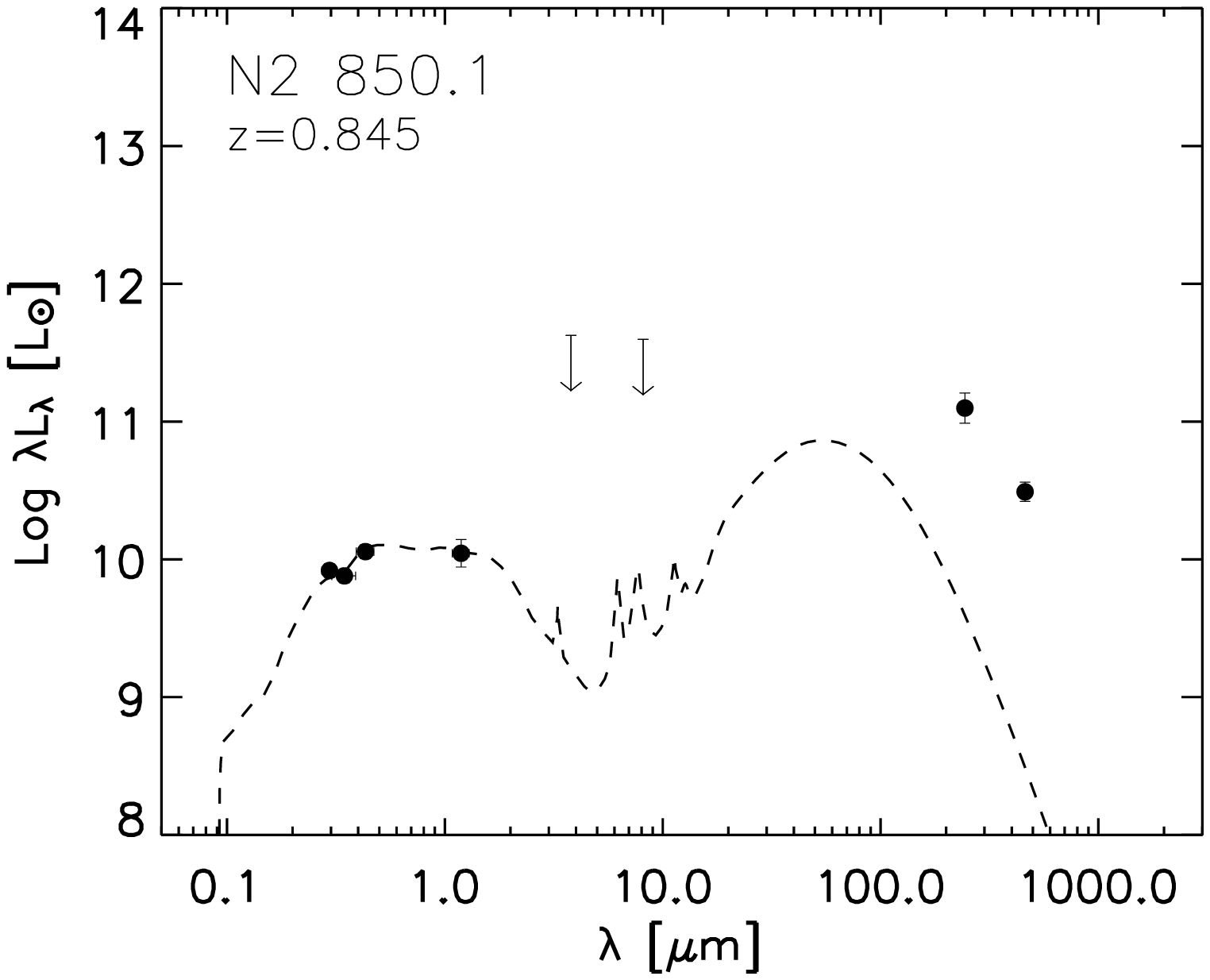}}
 \caption{Results of SED fitting  with the IMF of $x=1.10$ in the 
rest frame. 
The references for data are given in Table 1. 
Galaxies for which the SED model is 
indicated with dashed line are unused in the following 
analysis (Section 4 and 5). 
}
 \label{sed}
\end{figure*}

\addtocounter{figure}{-1}
\begin{figure*}
 \resizebox{7cm}{!}{\includegraphics{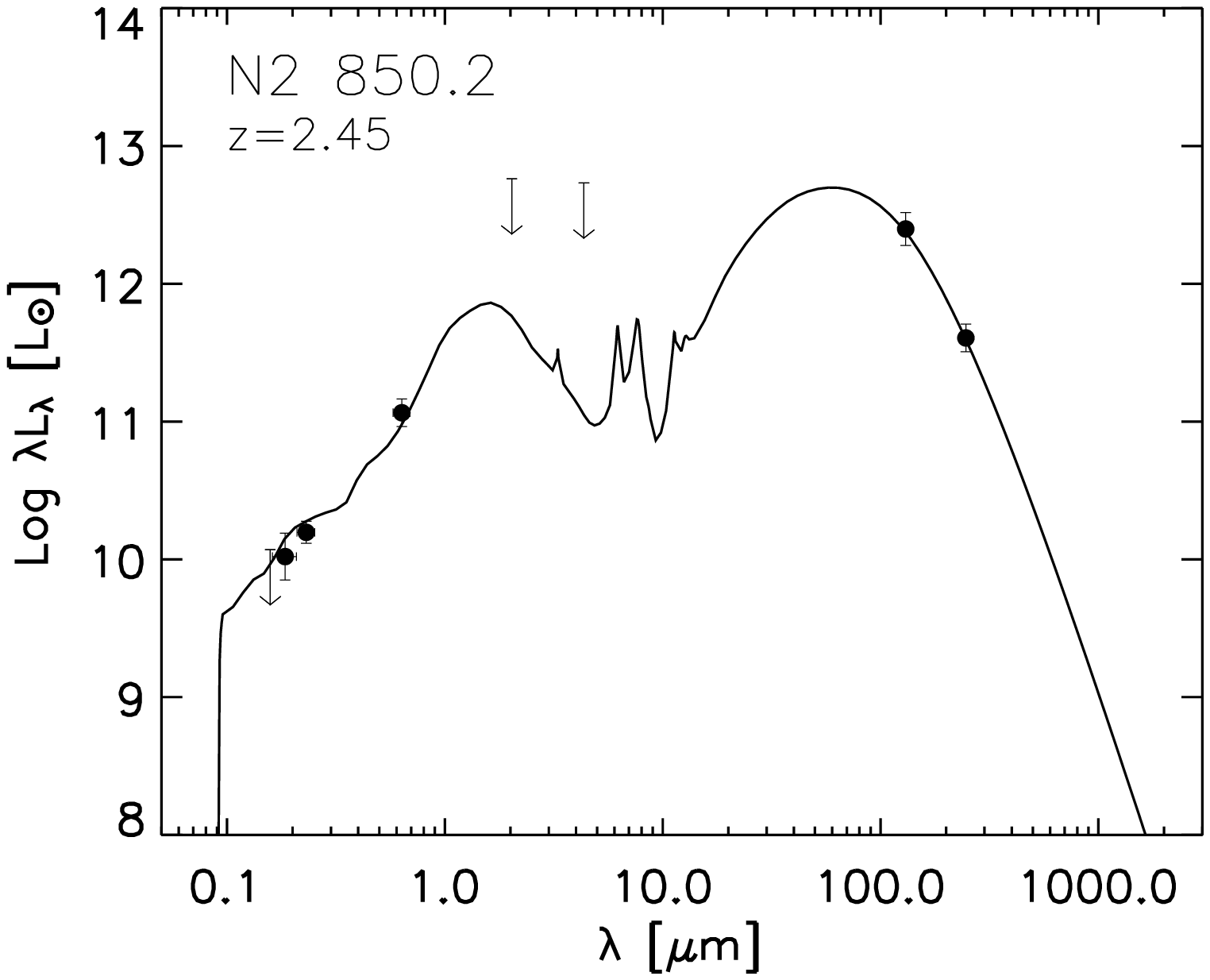}}
 \resizebox{7cm}{!}{\includegraphics{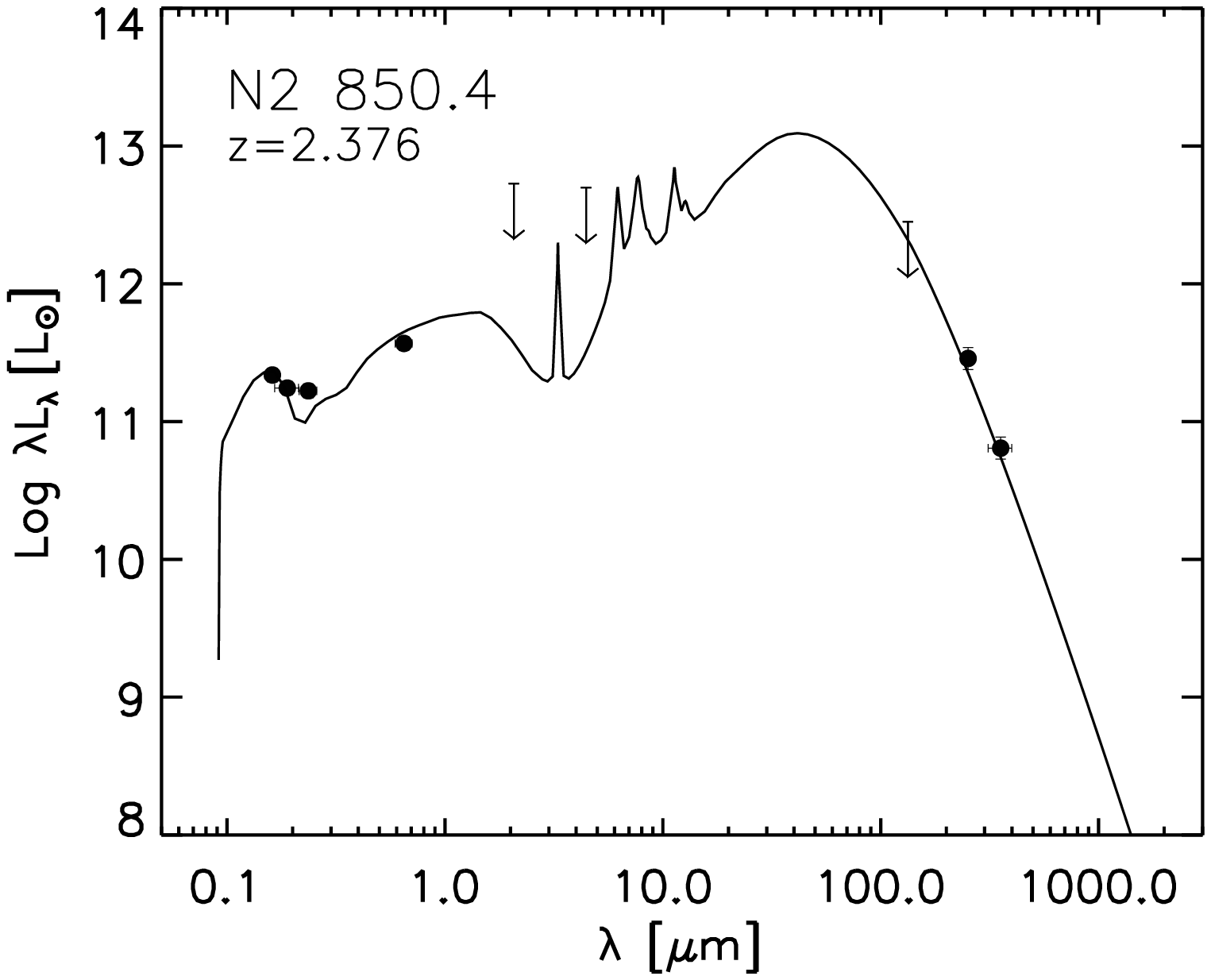}}
 \resizebox{7cm}{!}{\includegraphics{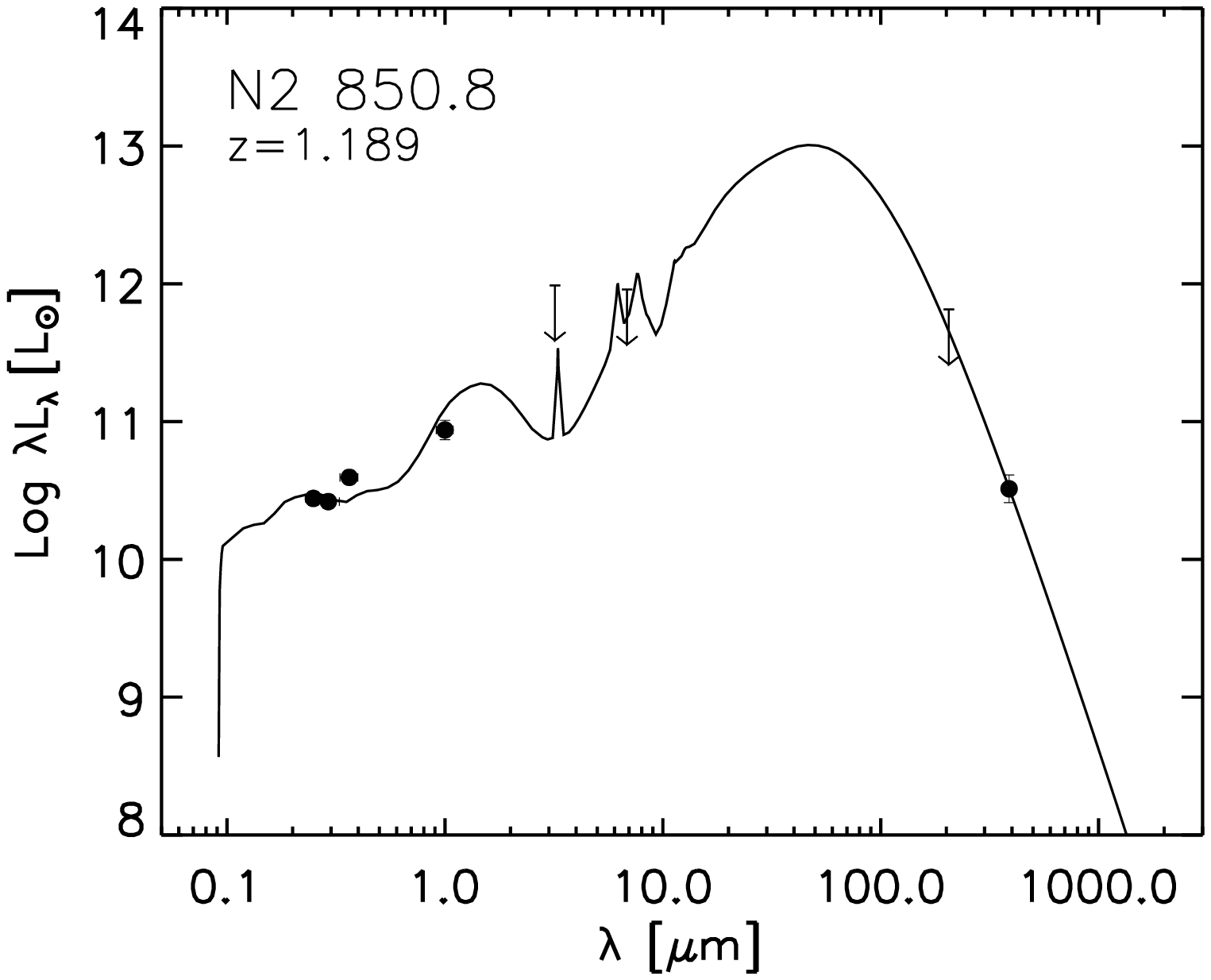}}
    \resizebox{7cm}{!}{\includegraphics{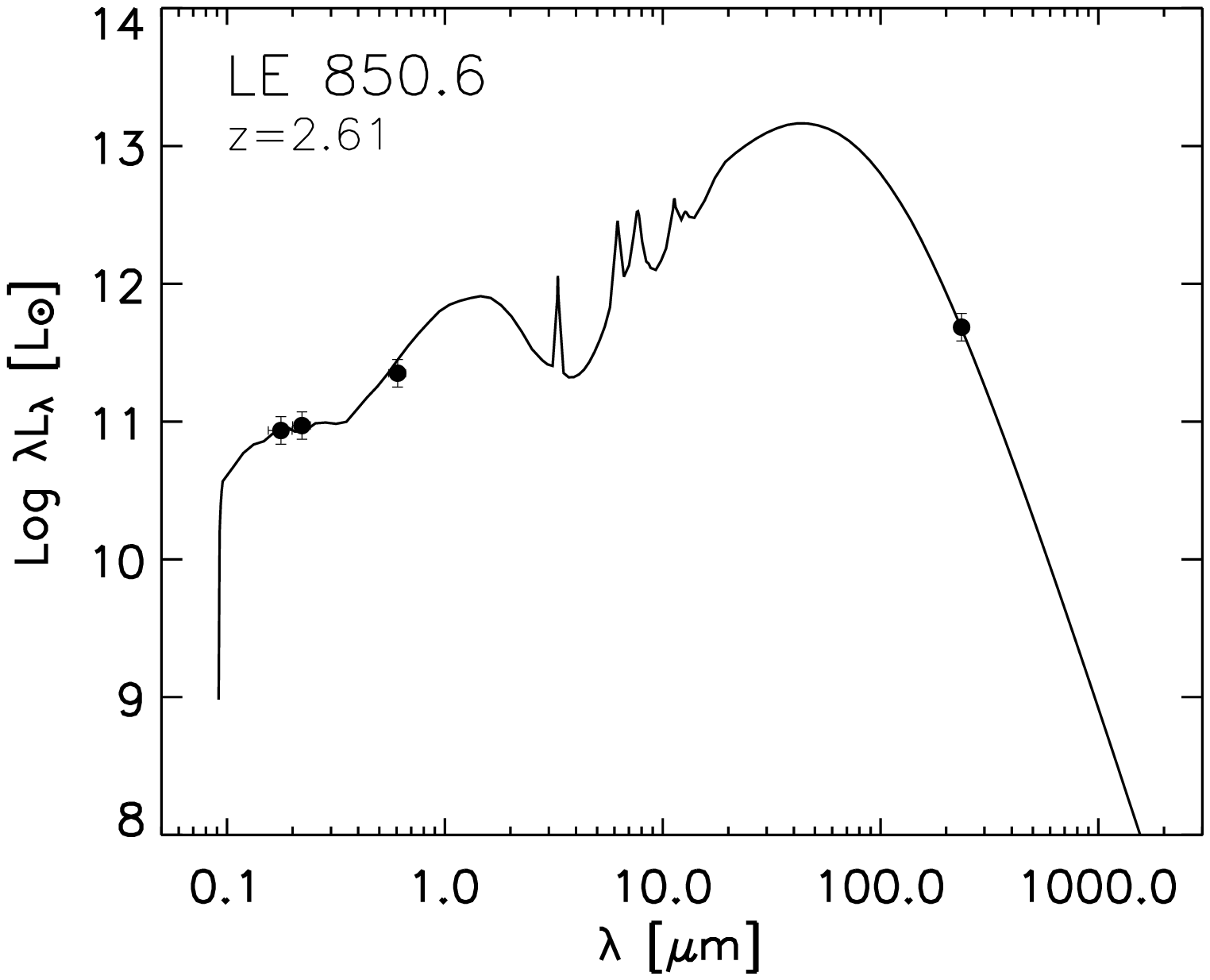}}
    \resizebox{7cm}{!}{\includegraphics{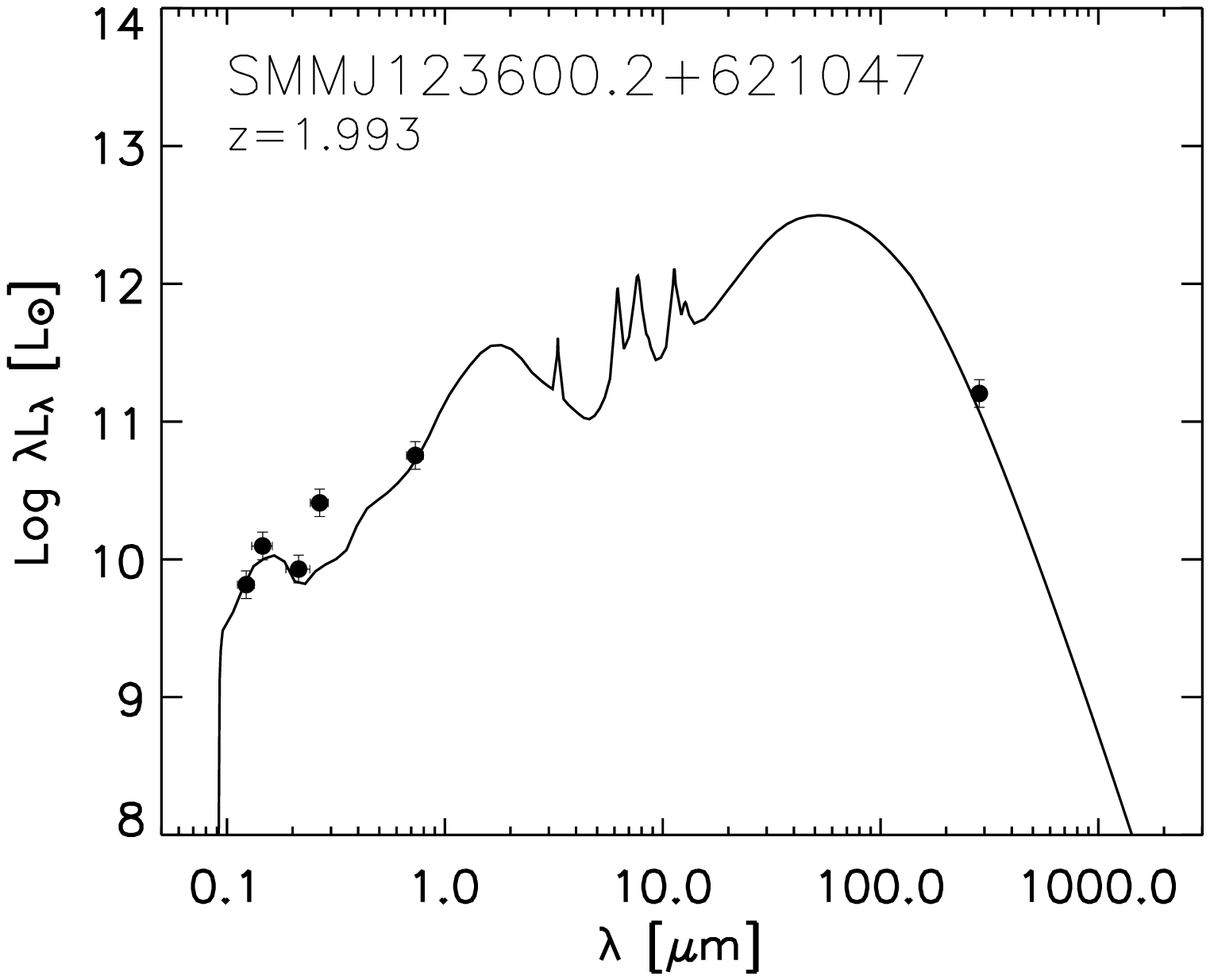}}
    \resizebox{7cm}{!}{\includegraphics{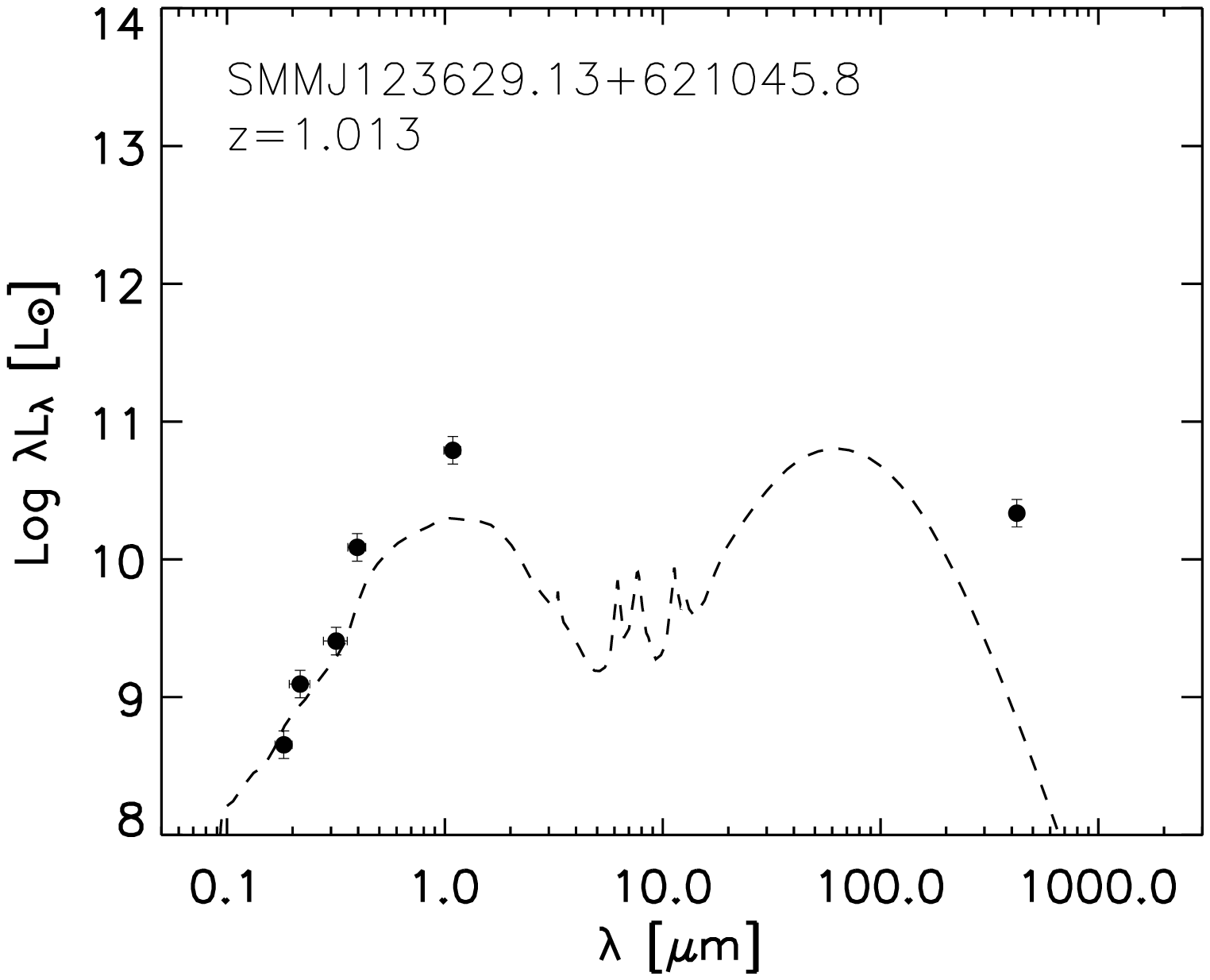}}
    \resizebox{7cm}{!}{\includegraphics{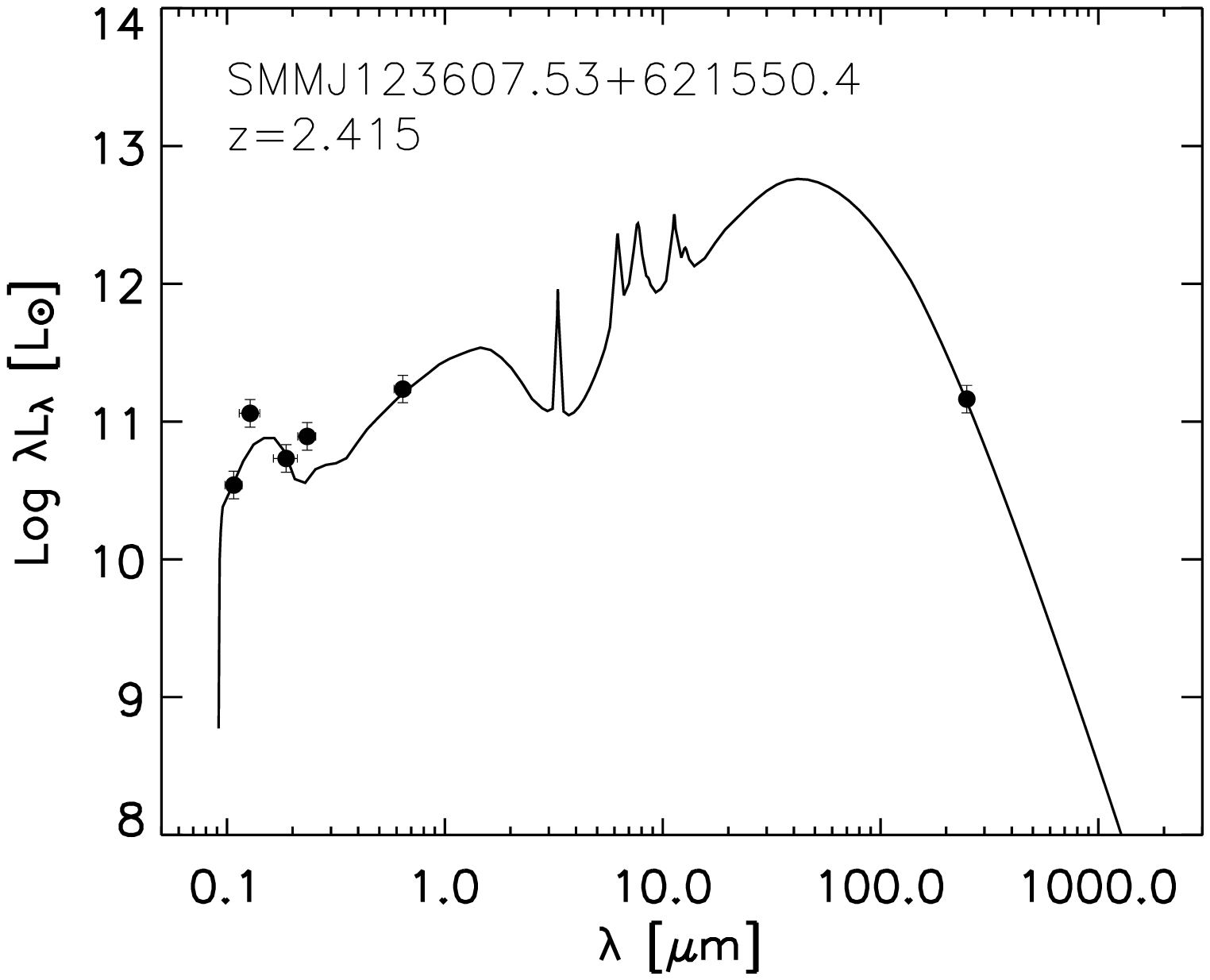}}
    \resizebox{7cm}{!}{\includegraphics{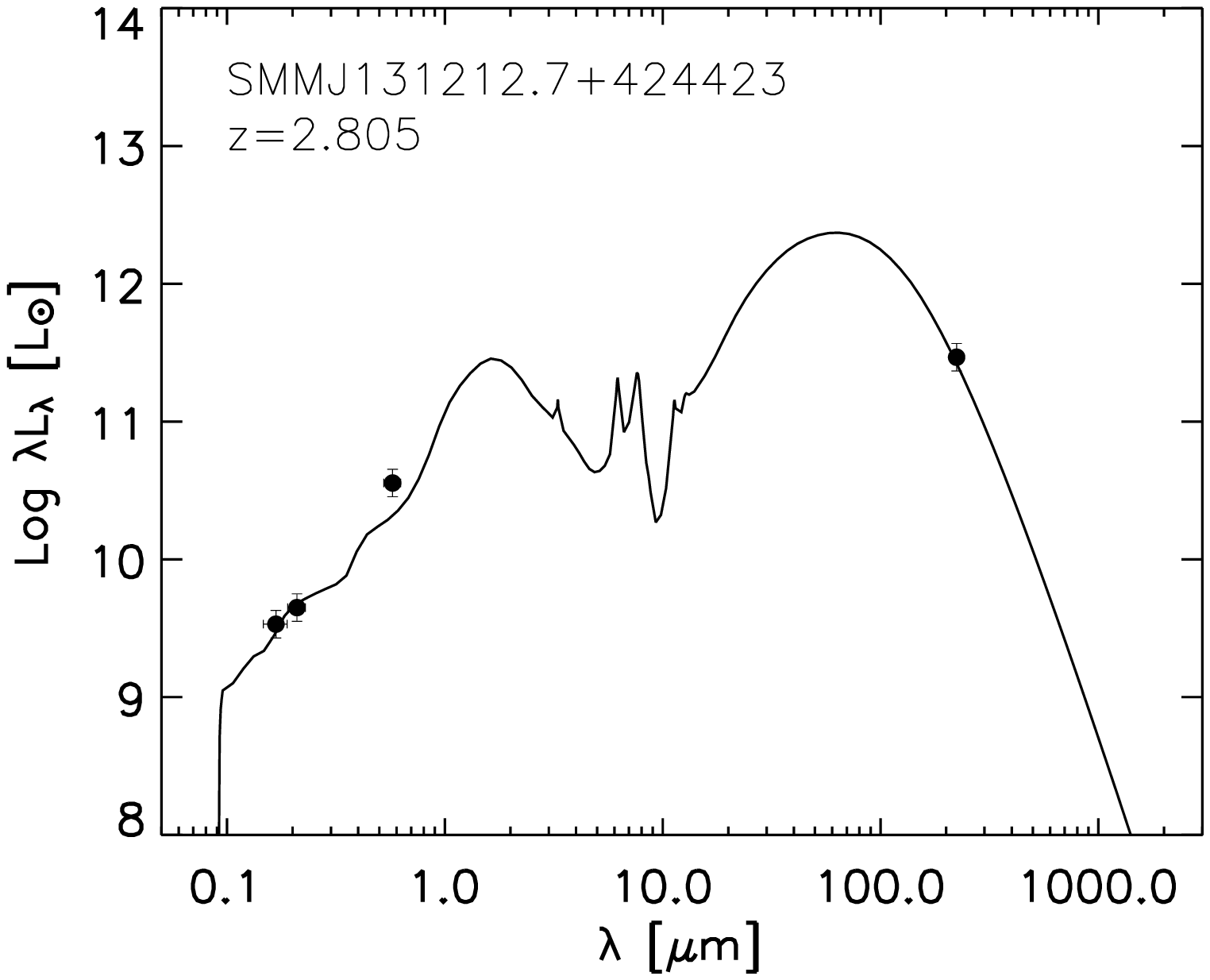}}
 \caption{{\it - continued.}
}
\end{figure*}

\addtocounter{figure}{-1}
\begin{figure*}
    \resizebox{7cm}{!}{\includegraphics{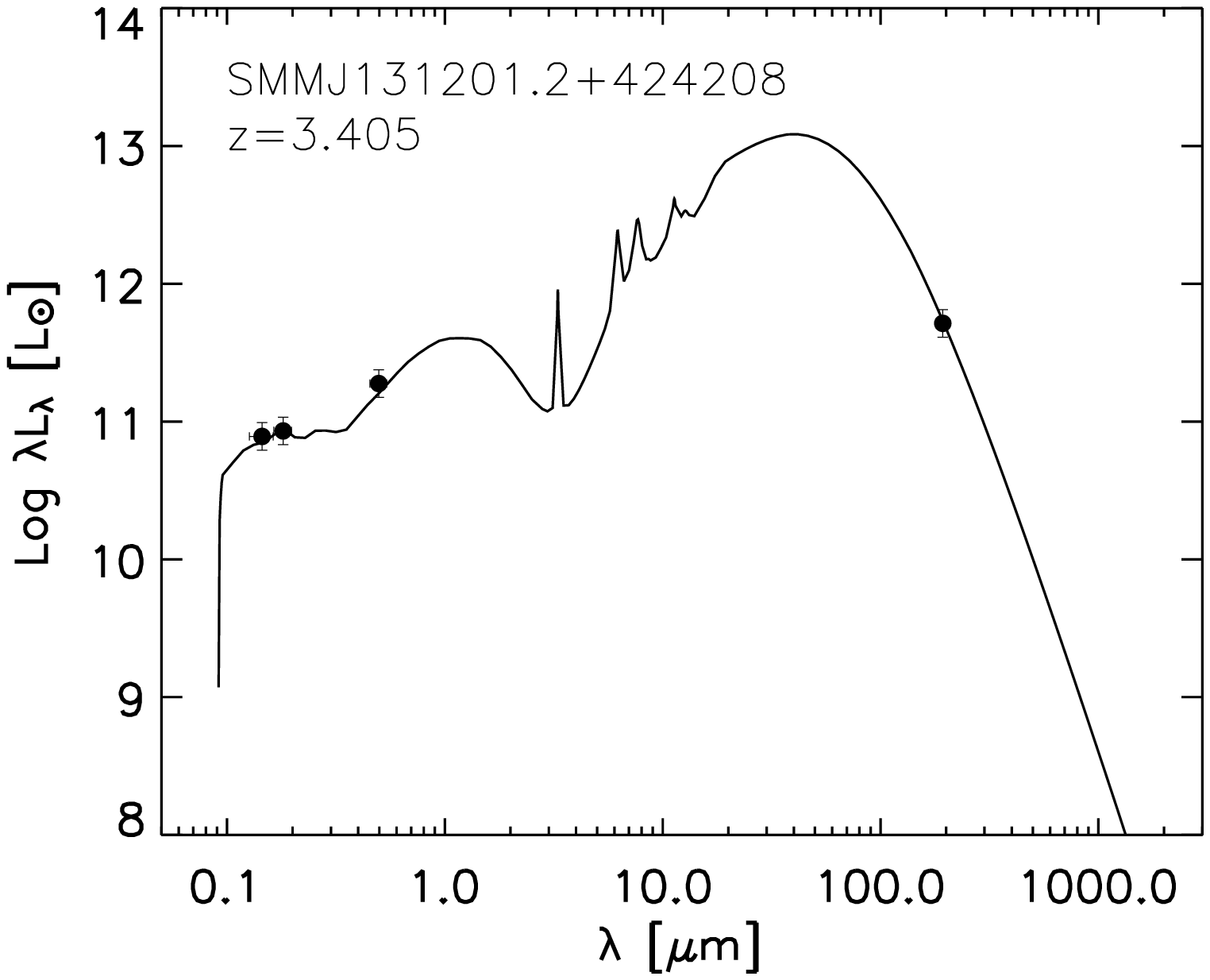}}
 \resizebox{7cm}{!}{\includegraphics{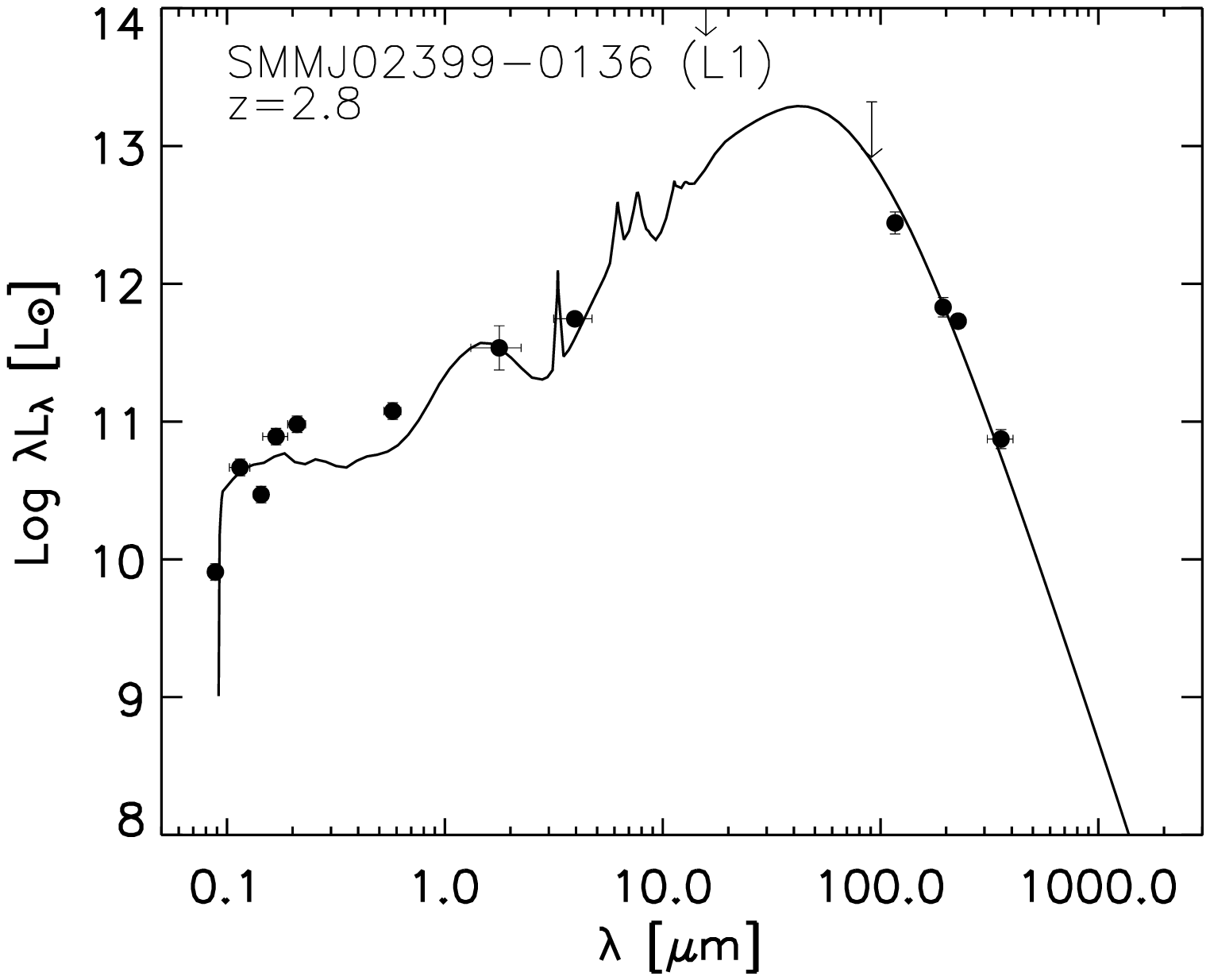}}
  \resizebox{7cm}{!}{\includegraphics{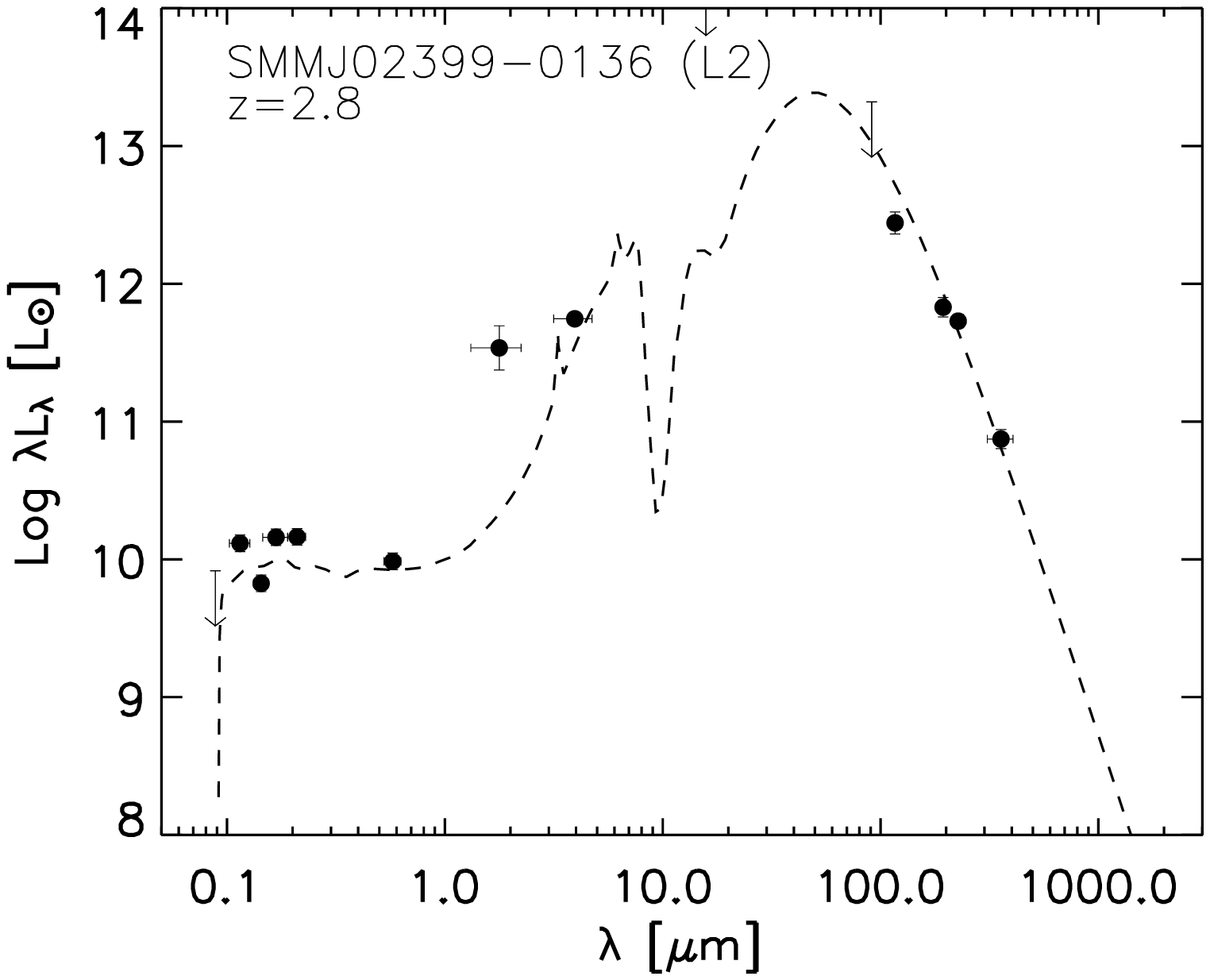}}
  \resizebox{7cm}{!}{\includegraphics{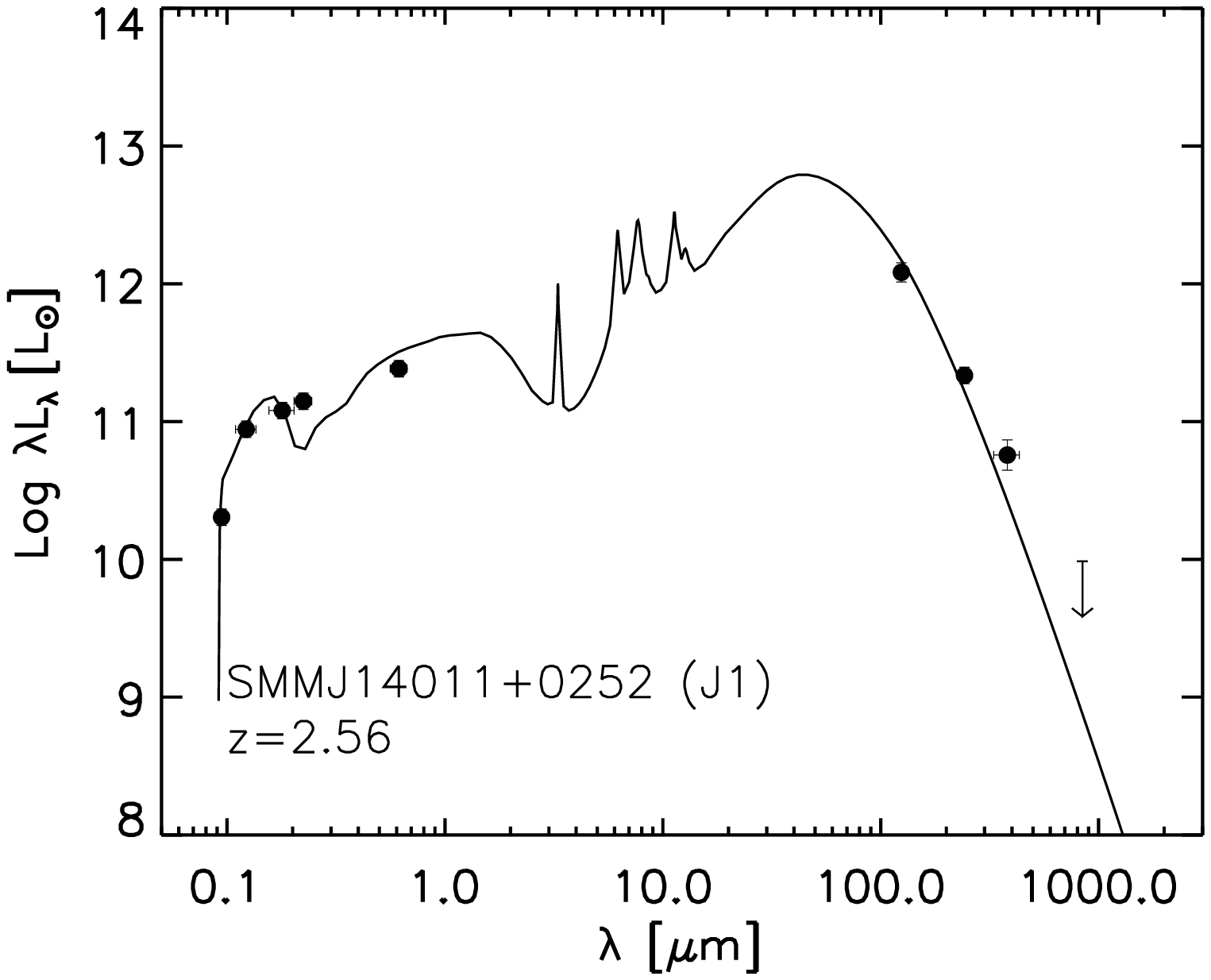}}
	 \resizebox{7cm}{!}{\includegraphics{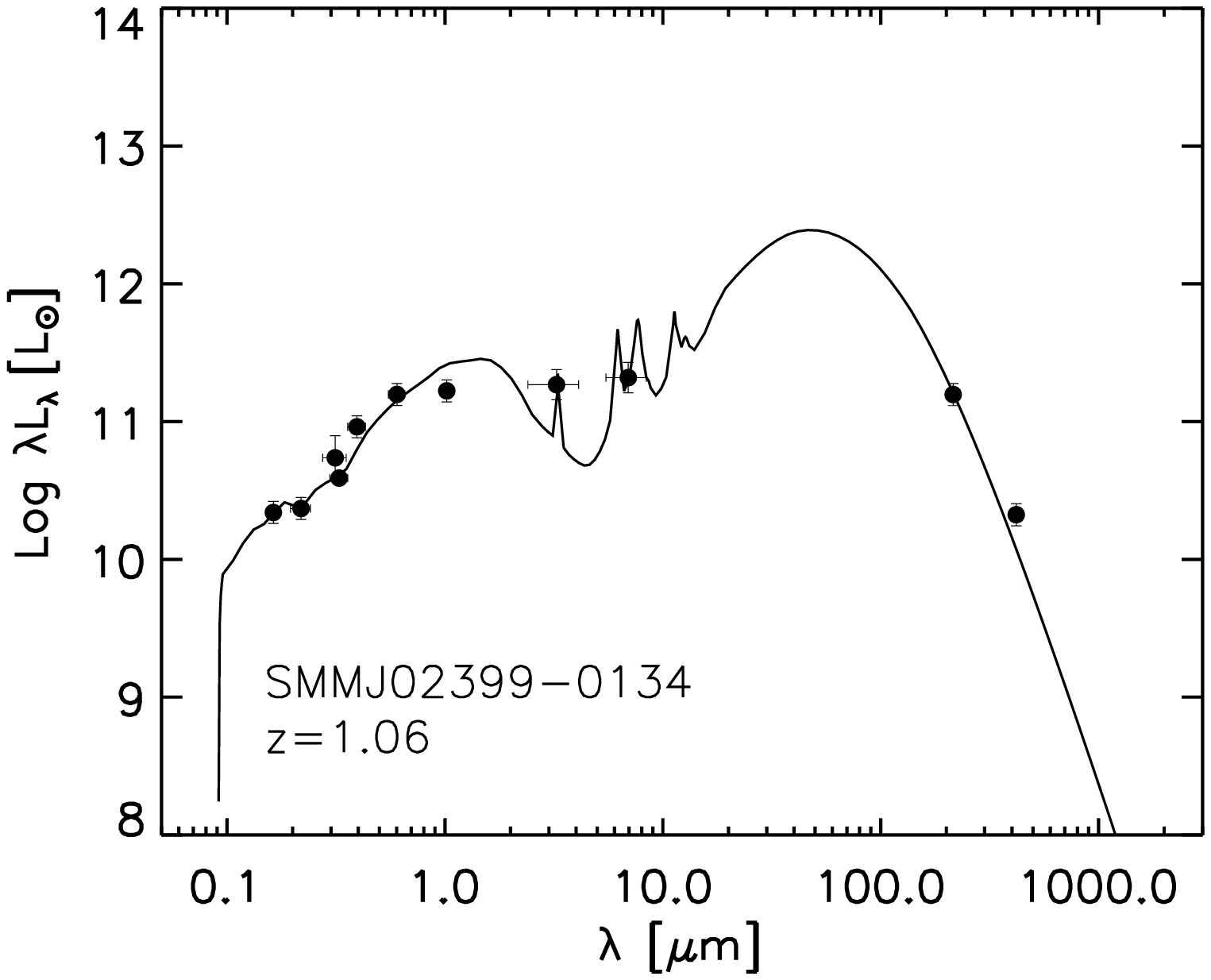}}
 \resizebox{7cm}{!}{\includegraphics{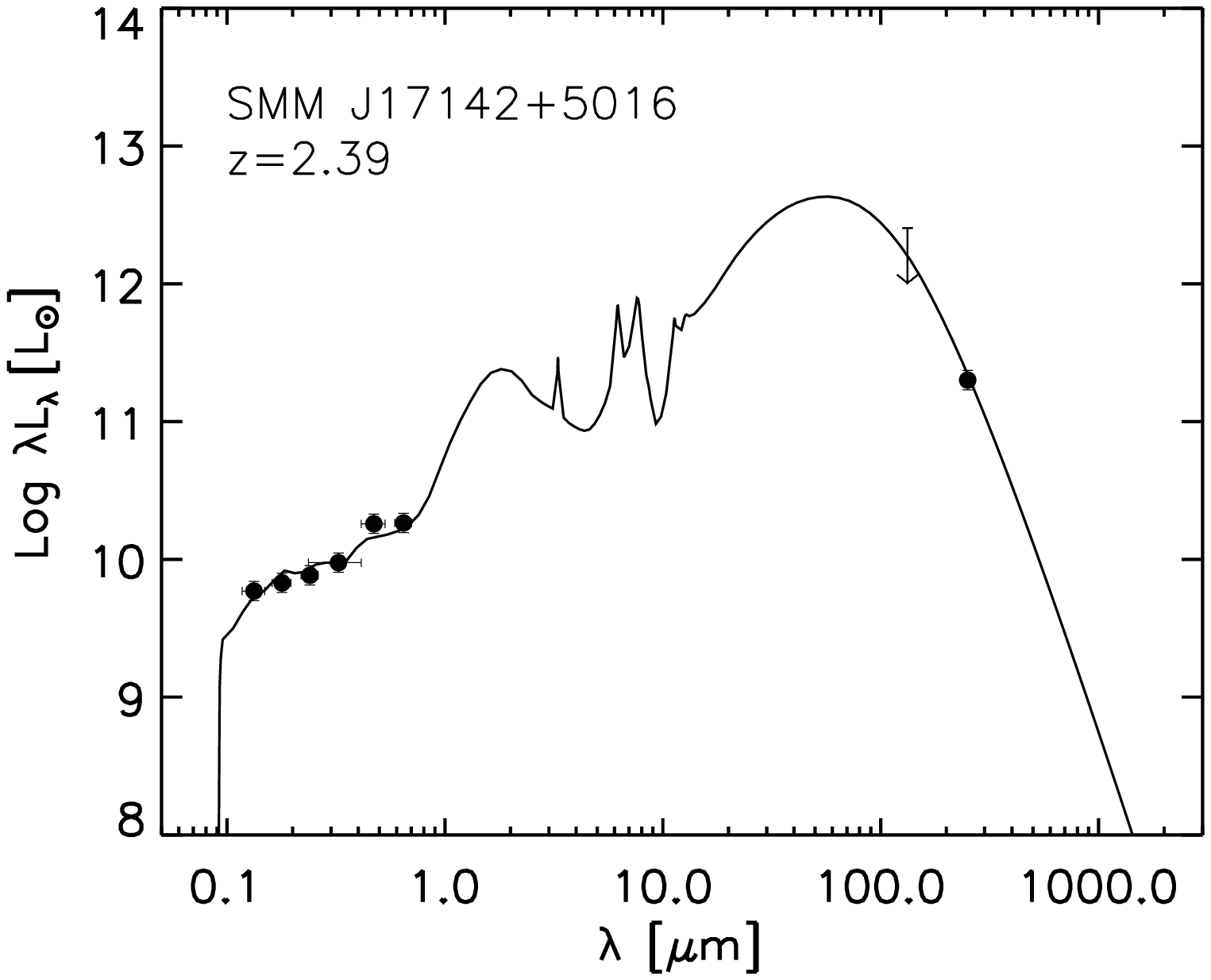}}
    \resizebox{7cm}{!}{\includegraphics{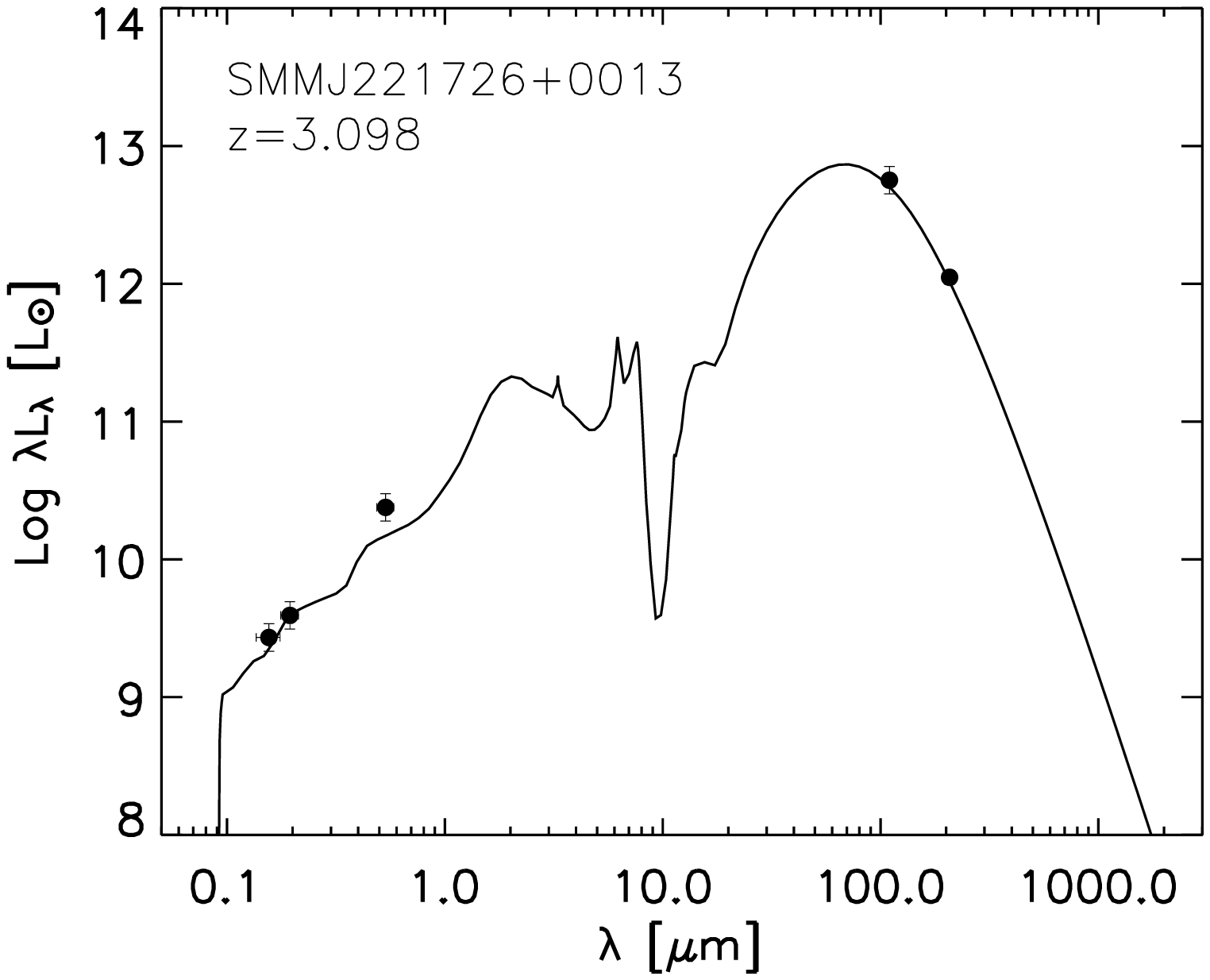}}
    \resizebox{7cm}{!}{\includegraphics{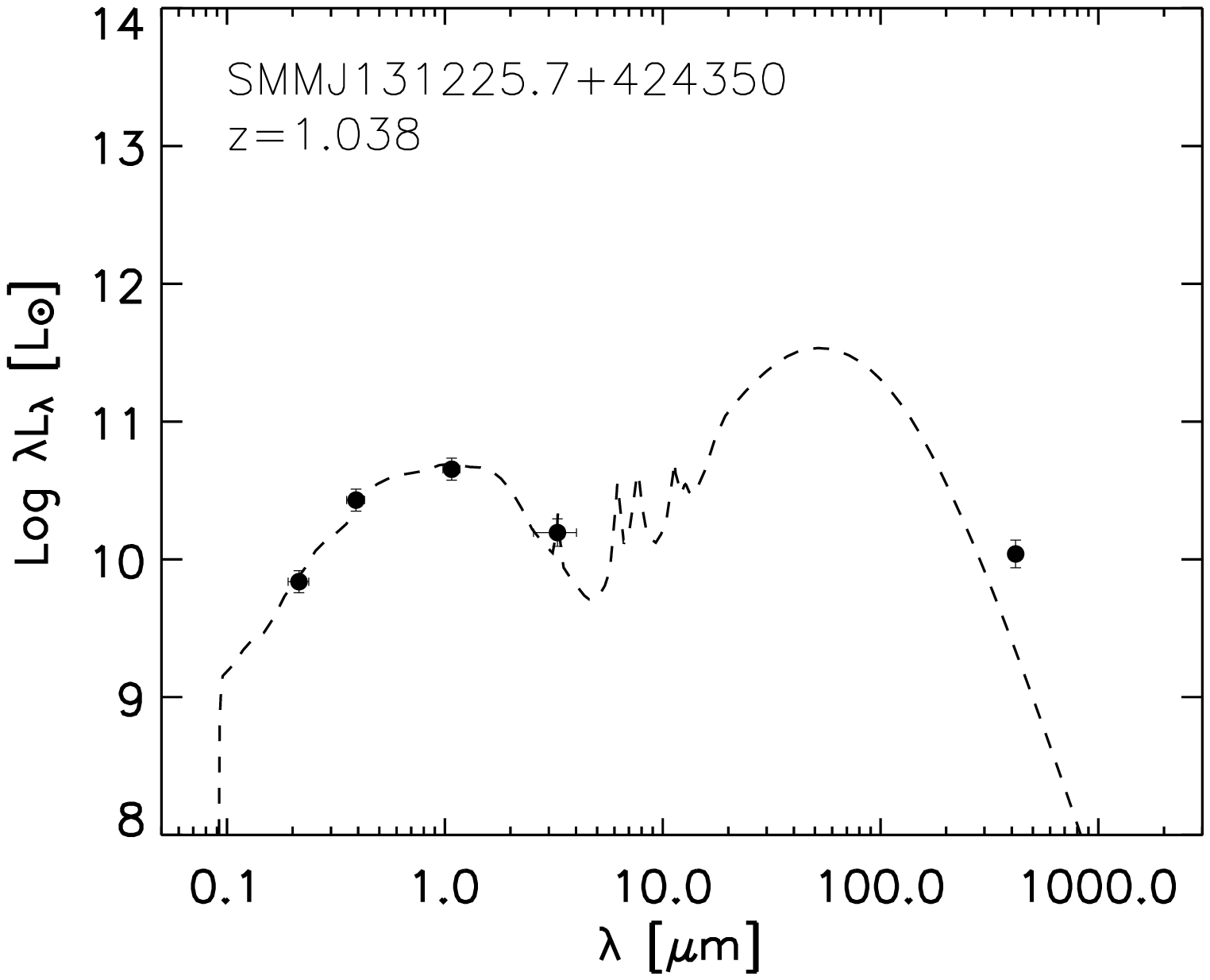}}
 \caption{{\it - continued.}
}
\end{figure*}

\addtocounter{figure}{-1}
\begin{figure*}
  \resizebox{7cm}{!}{\includegraphics{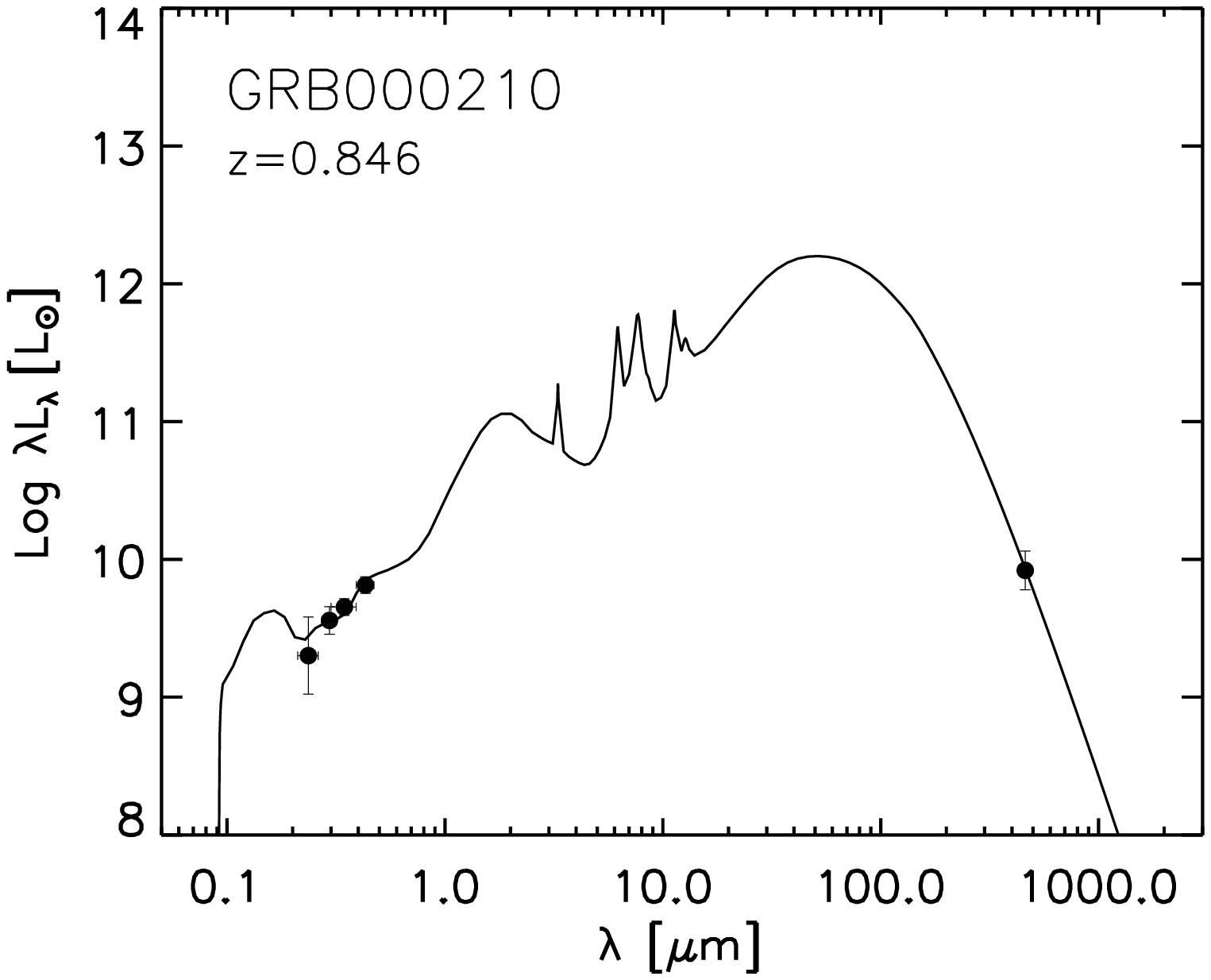}}
  \resizebox{7cm}{!}{\includegraphics{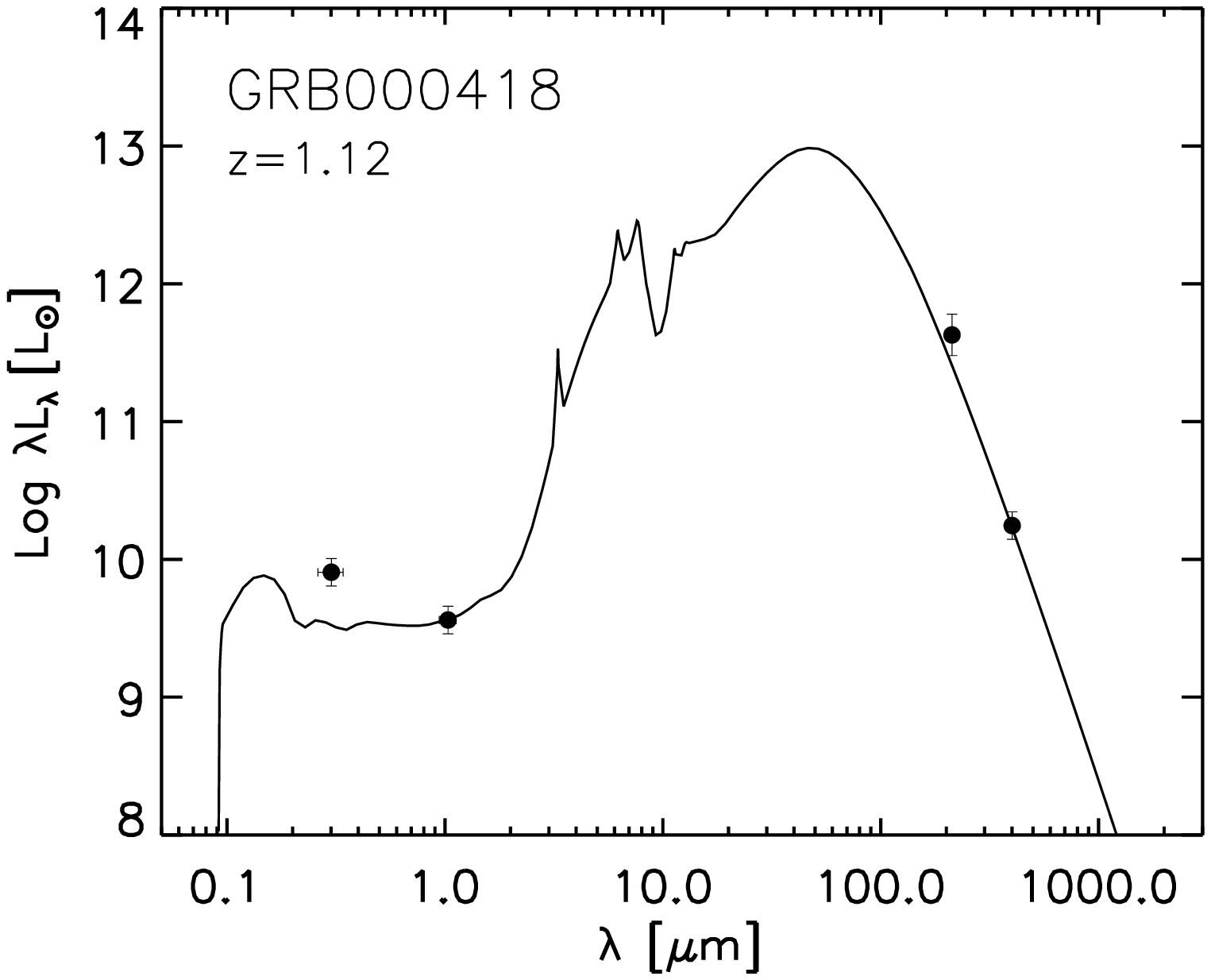}}
  \resizebox{7cm}{!}{\includegraphics{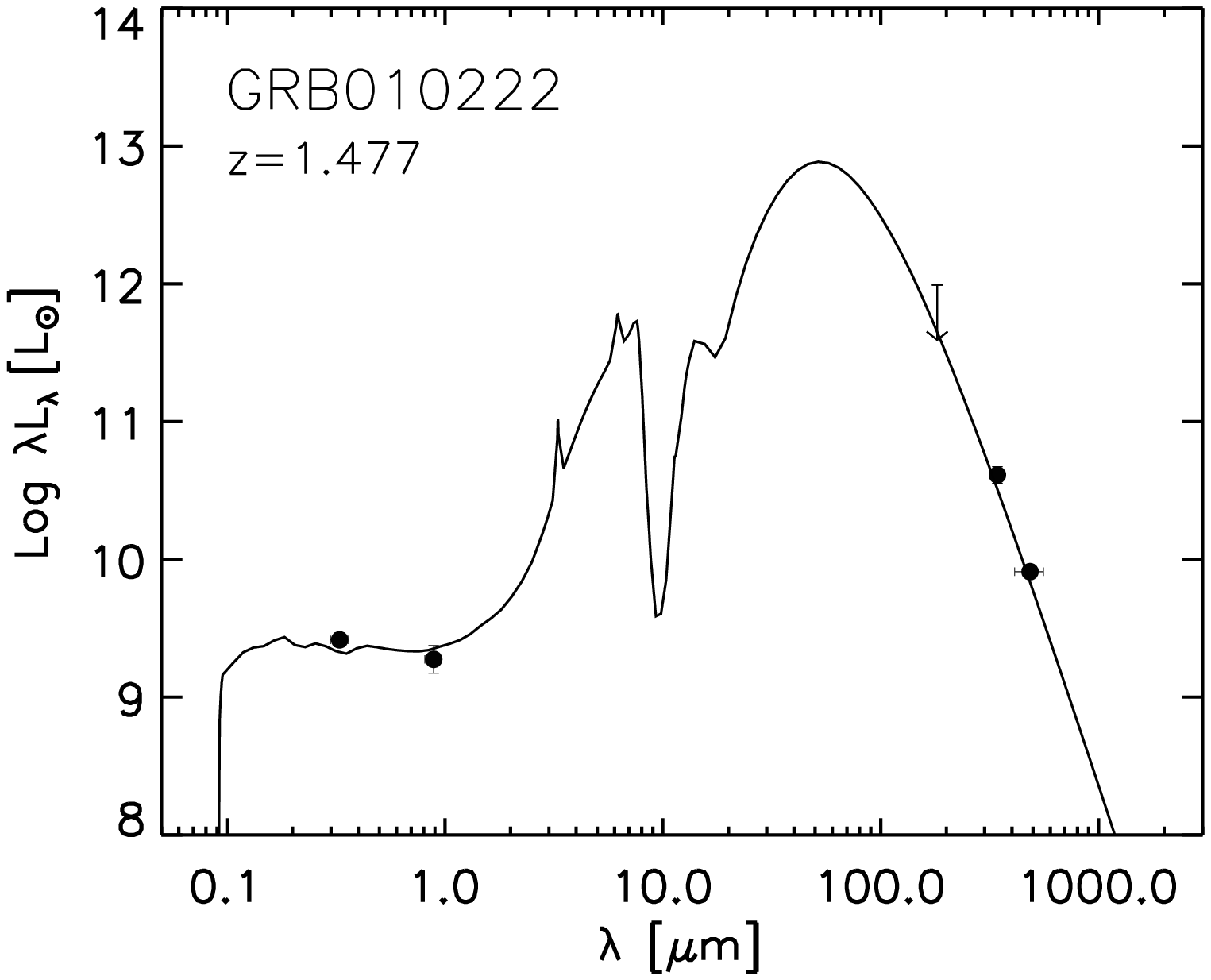}}
 \caption{{\it - continued.}
}
\end{figure*}

\begin{figure*}
 \begin{center}
 \resizebox{13cm}{!}{  \includegraphics{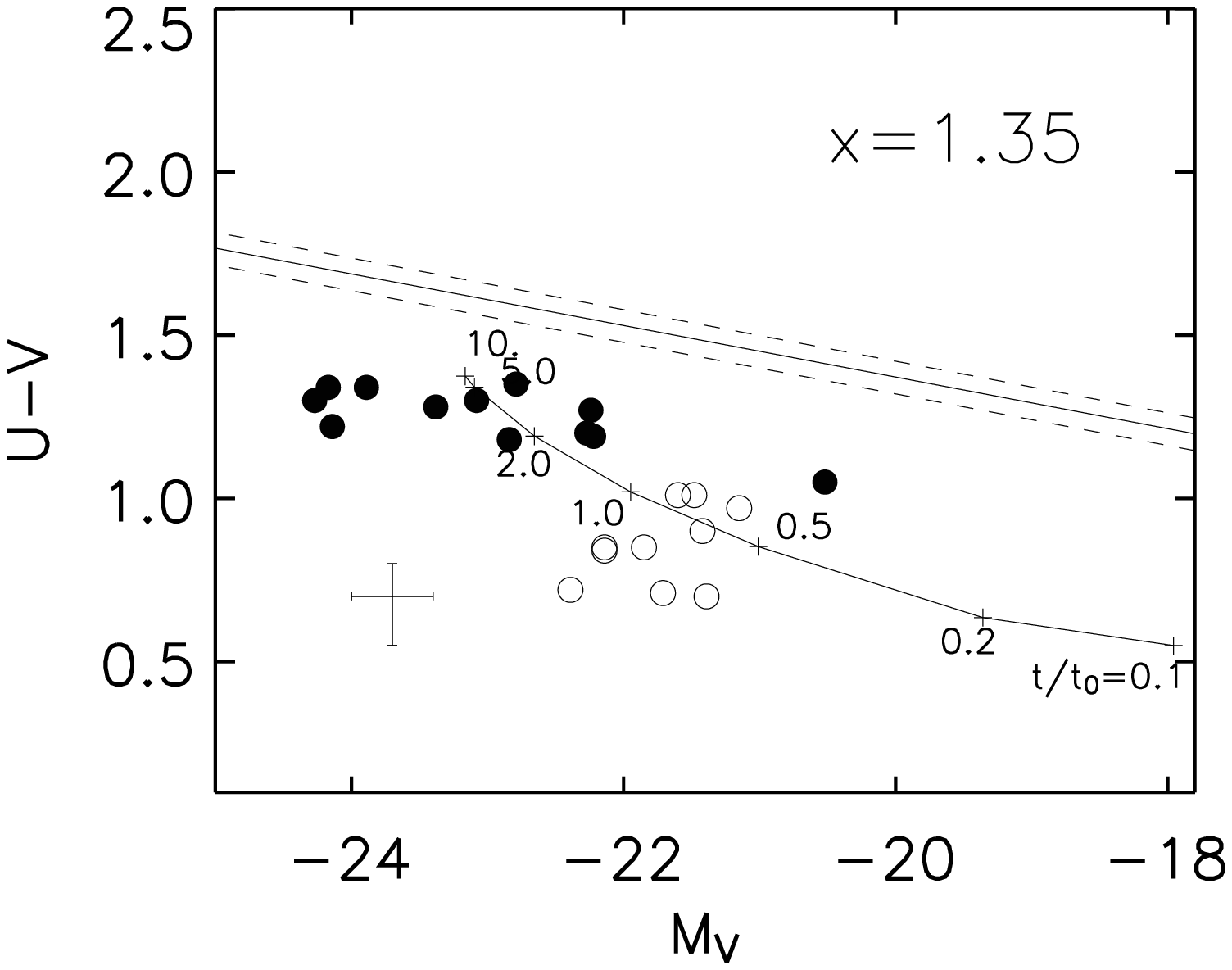}}
 \resizebox{13cm}{!}{  \includegraphics{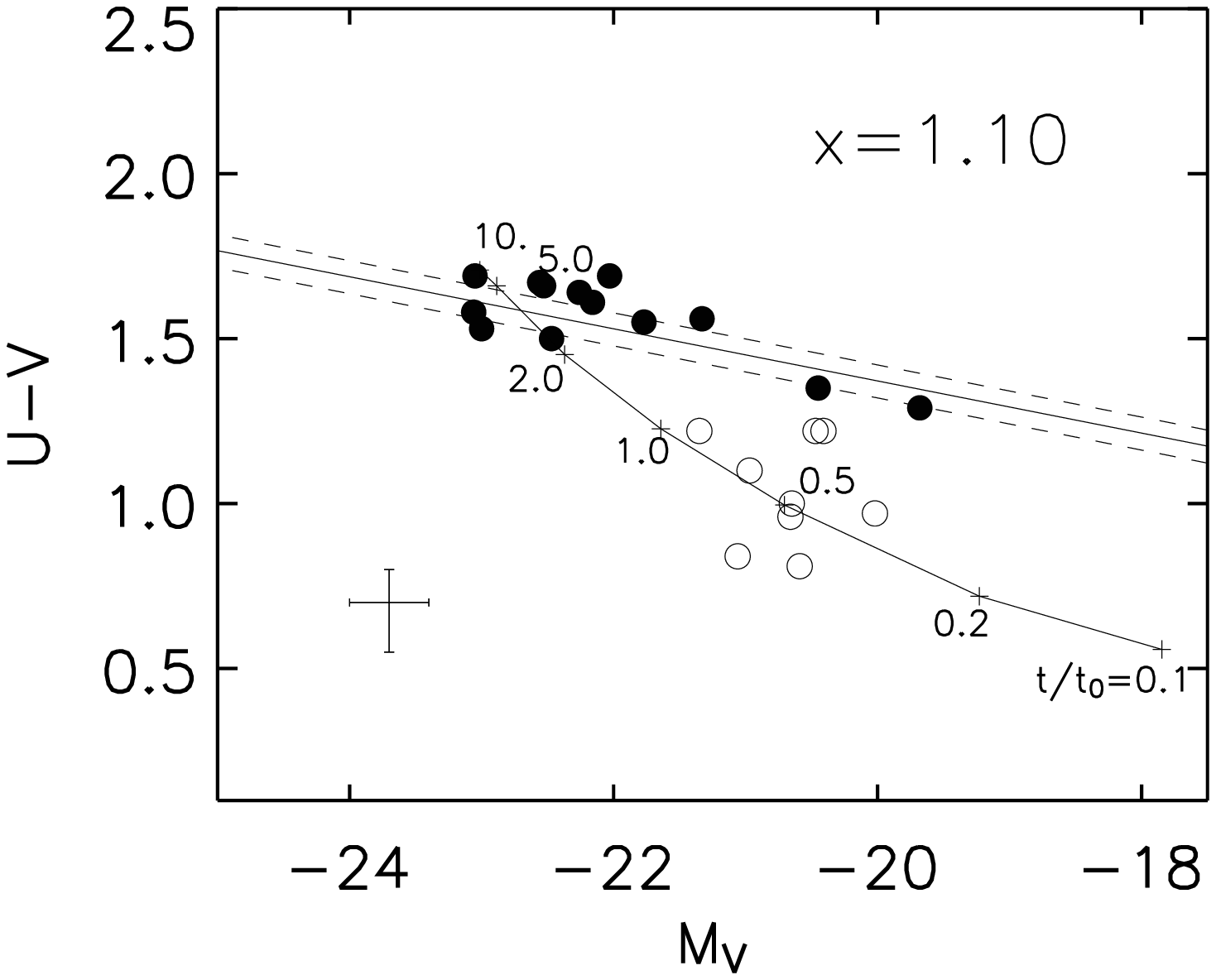}}
 \caption{
The predicted colours and magnitudes of submm galaxies 
at the present epoch. 
Open and solid circles indicate the sample galaxies with 
age of $t/t_0 \le 1.0$ and those old than $t/t_0 > 1.0$, respectively. 
The solid straight line is for the regression line to the CM relation
observed in the Coma cluster derived by Bower, Lucey \& Ellis (1992), and
dashed lines indicate the observed scatter $\pm 0.05$ mag.
The depicted loci with starburst age indicate 
the colours and magnitudes of starburst galaxies formed at $z\sim 3$ 
with the mass of gas reservoir $M_T=10^{12}$ M$_\odot$, in 
which the star formation ceases at indicated starburst age. 
Note that a starburst galaxy with $M_T=10^{12}$ M$_\odot$ becomes passively 
evolving galaxies with $M_V=-23$, i.e., typical brightest 
elliptical galaxy when 
$x=1.10$ and the star formation ceases at the starburst age of $t/t_0=5.0$
We show the typical errors at the lower left corner. 
}
 \label{cm_uv}
 \end{center}
 \end{figure*}

\begin{figure*}
 \begin{center}
 \resizebox{7cm}{!}{  \includegraphics{figures/age_MV.eps}}
\hspace{0.5cm}
 \resizebox{7cm}{!}{  \includegraphics{figures/zstar_MV.eps}}
 \caption{
The present-day age and luminosity-weighted stellar metallicity 
of submm galaxies. 
The symbols are the same as those in 
Figure \ref{cm_uv}. The IMF of $x=1.10$ is adopted. 
Left-hand panel: 
the present-day ages are plotted as a function of 
present-day $M_V$.
The present-day ages are calculated from 
the look-back time plus the starburst age assuming $t_0=100$ Myr;
right-hand panel: 
the luminosity-weighted stellar metallicities are plotted 
as a function of present-day $M_V$. 
}
 \label{origin}
 \end{center}
 \end{figure*}

\begin{figure*}
  \resizebox{13cm}{!}{\includegraphics{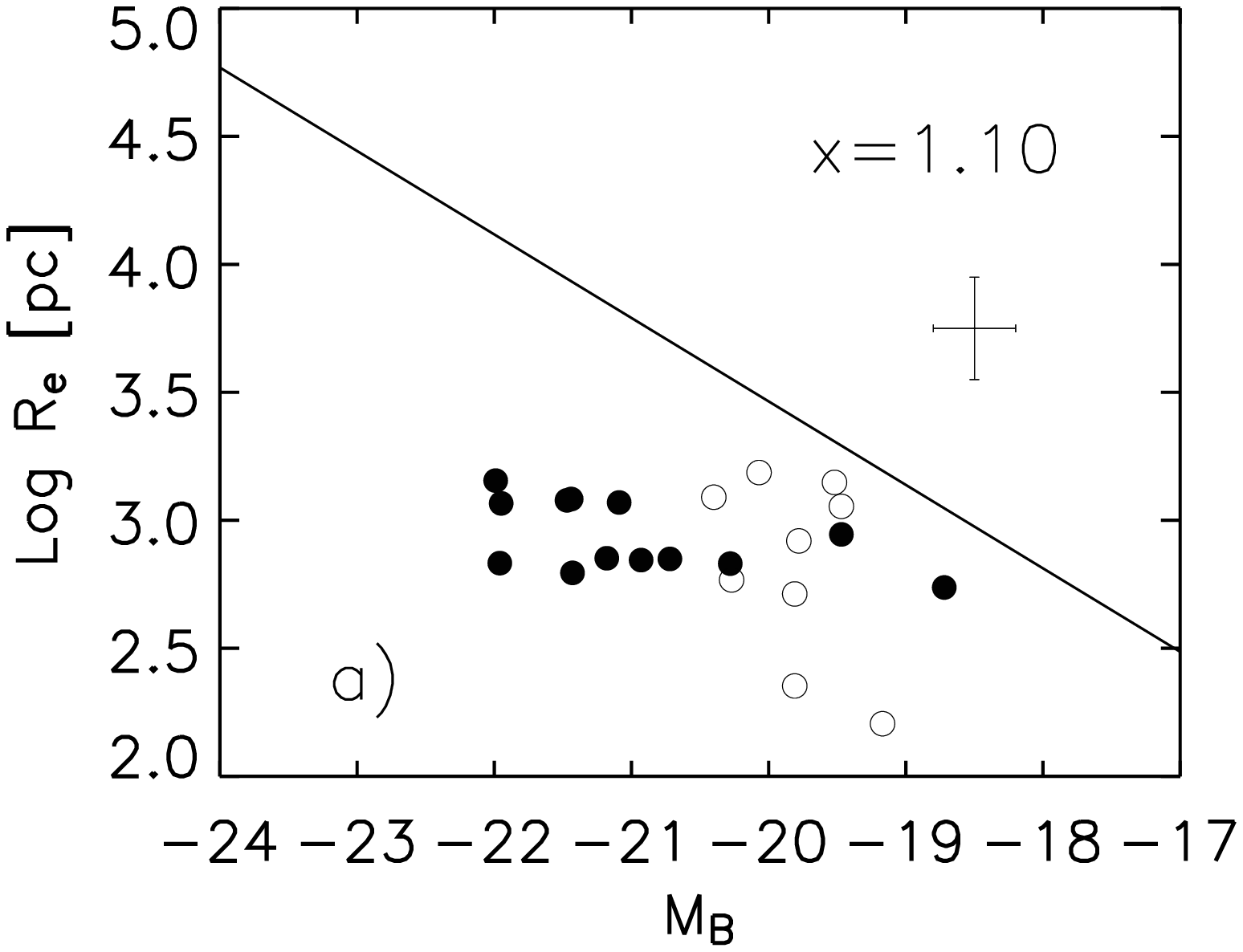}}
 \resizebox{13cm}{!}{\includegraphics{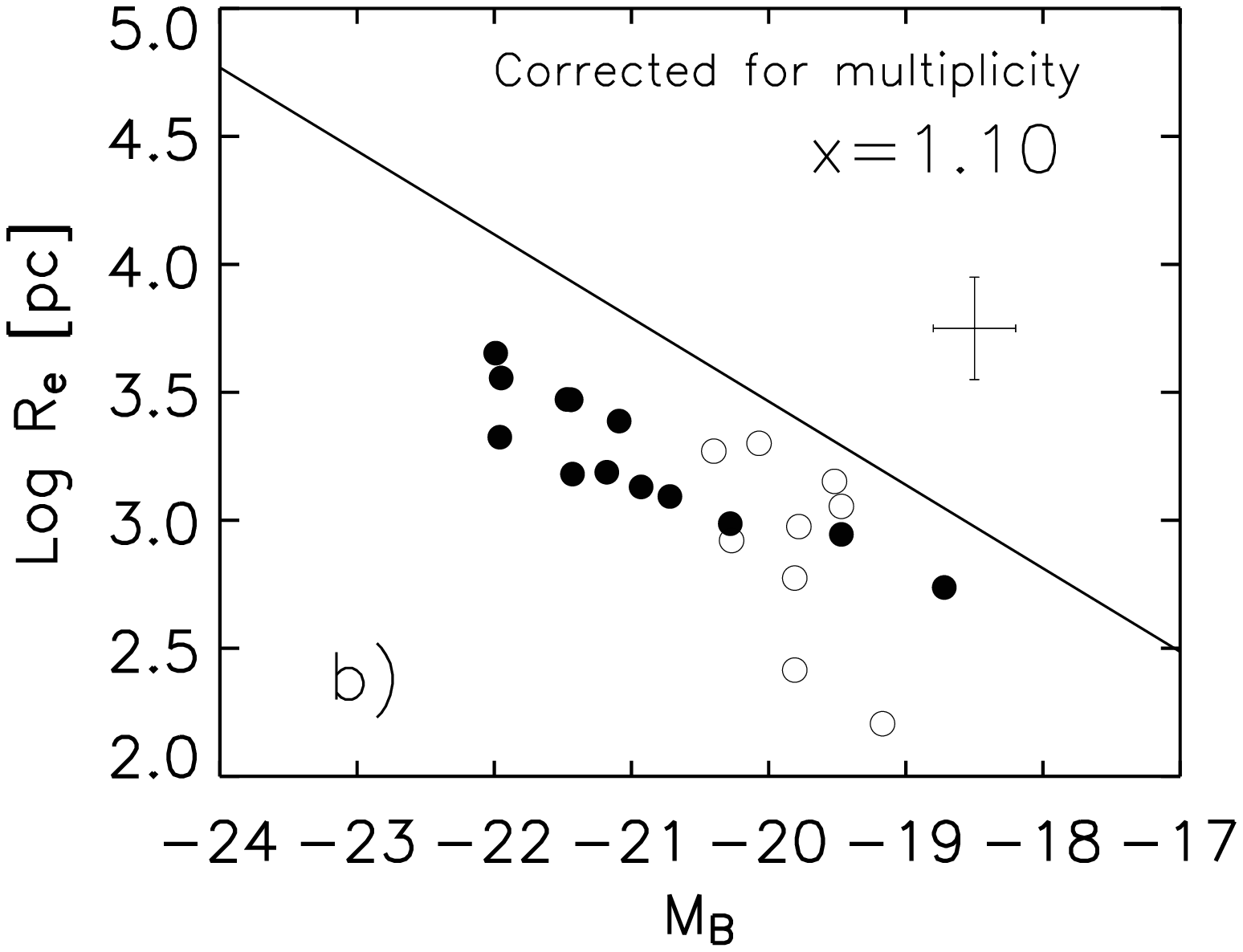}}
 \caption{a) Comparison of $B$-band magnitudes and effective radii 
with the observed size-magnitude relation of elliptical galaxies 
(solid line) obtained from the sample of Bender et al. (1992). 
The symbols are the same as 
those in Figure \ref{cm_uv}. The typical errors are shown. 
b) Same as a), but corrected for 
the multiplicity (see text in detail). 
}
 \label{kormendy}
\end{figure*}

\begin{figure*}
 \begin{center}
 \resizebox{12cm}{!}{\includegraphics{figures/velocity_mag_corrected.eps}}
 \caption{
The comparison of the gas velocity by feedback with the 
escape velocity as a function of the present-day $B$ magnitude with 
the IMF of $x=1.10$. The symbols are the same as those in 
Figure \ref{cm_uv}. 
The squared velocity ratio is corrected for 
the multiplicity. See text in detail. 
}
 \label{velmag}
 \end{center}
 \end{figure*}

\begin{figure*}
  \resizebox{13cm}{!}{\includegraphics{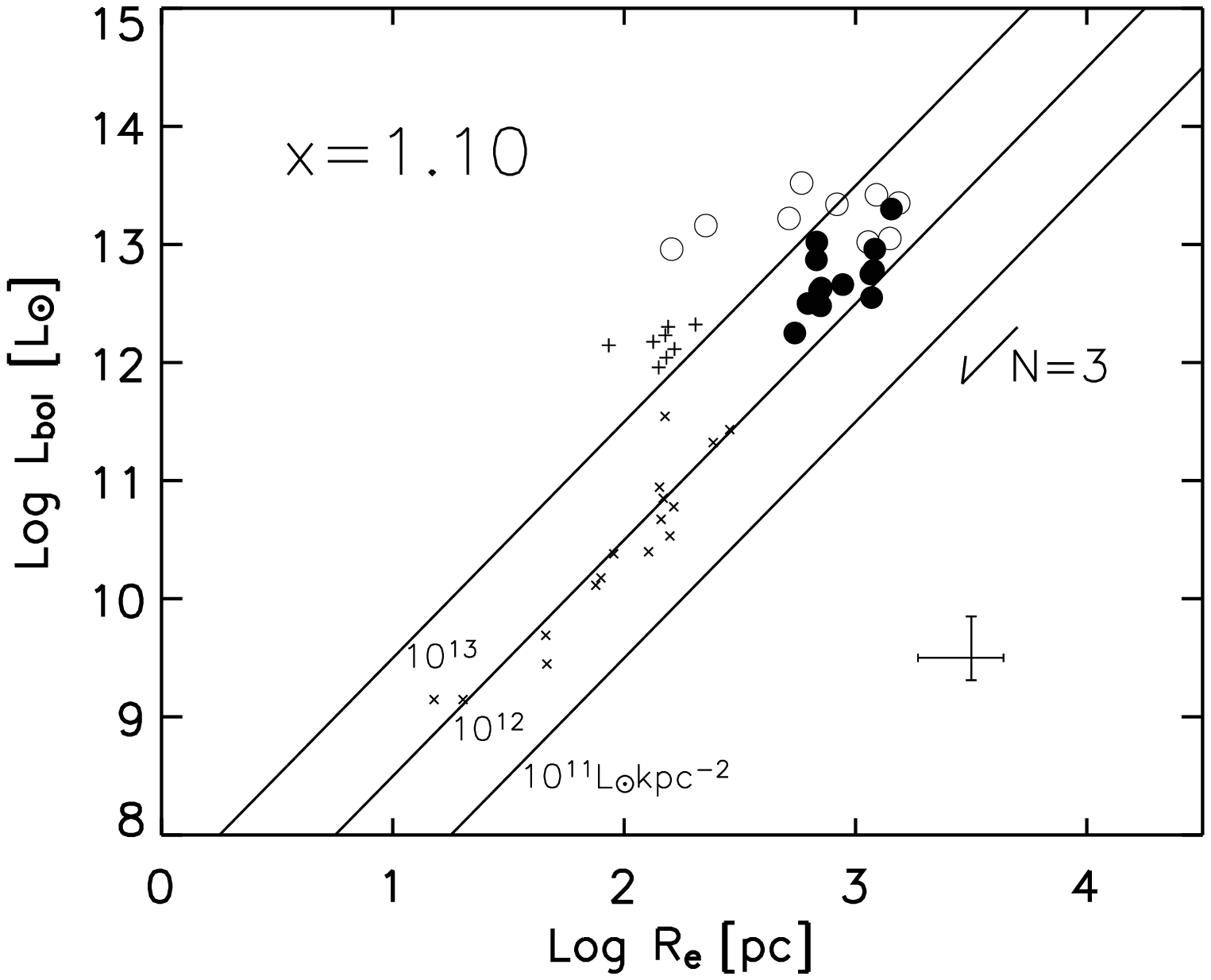}}
 \caption{The bolometric luminosities of the sample galaxies plotted
against intrinsic effective radii derived from the SED fitting. 
Crosses ($\times$) and pluses (+) indicate UVSBGs and ULIRGs, 
taken from TAH03. 
The other symbols are the same as those in Figure \ref{cm_uv}. 
We indicate the correction factor for the multiplicity with an 
arrow for the case of $N=3$.
The typical errors are shown.
}
 \label{lbol}
\end{figure*}

\begin{figure*}
 \resizebox{12cm}{!}{\includegraphics{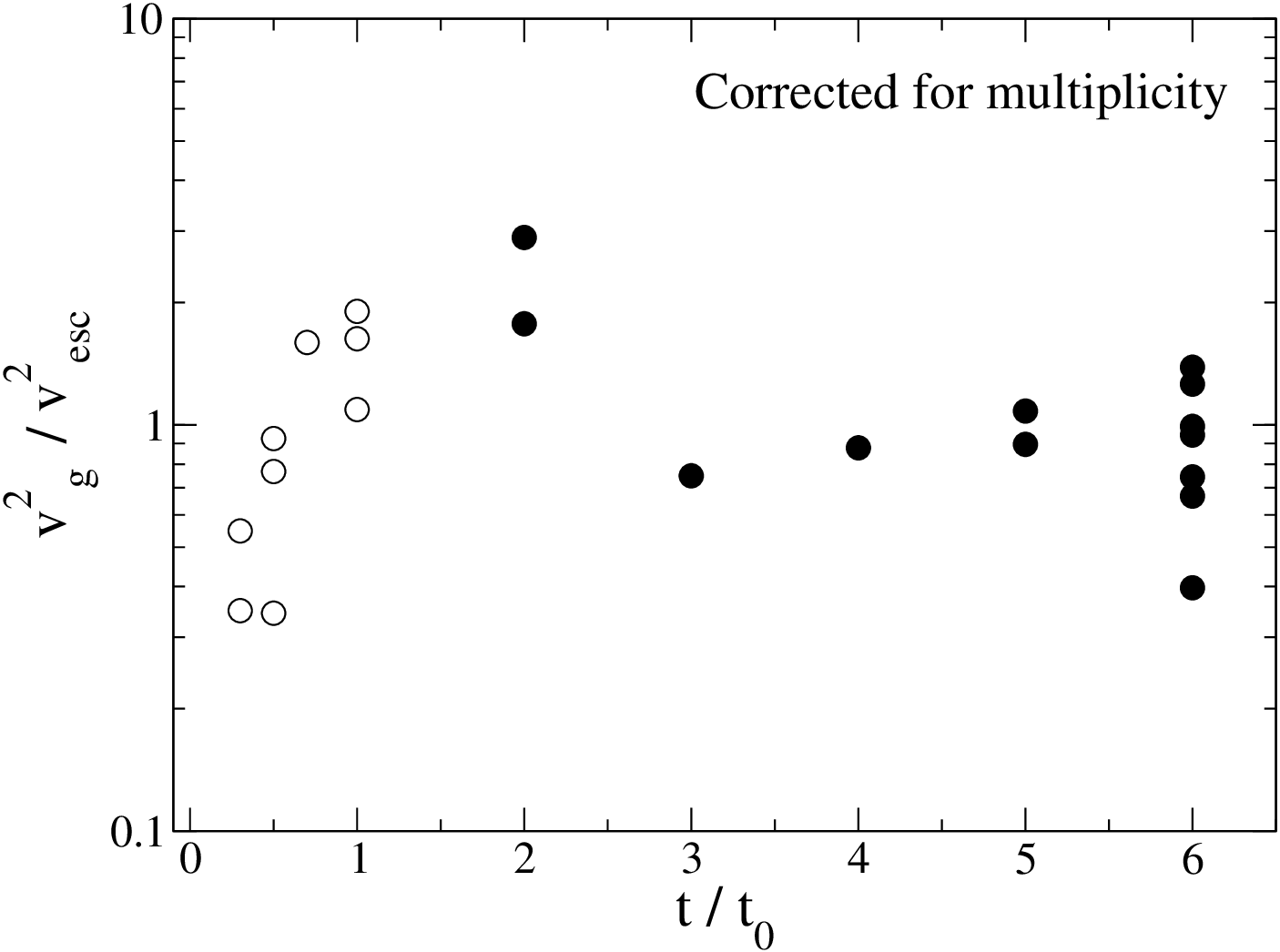}}
 \caption{
The squared velocity ratio 
as a function of the starburst age with the IMF of 
$x=1.10$. The symbols are the same as those in Figure \ref{cm_uv}. 
The squared velocity ratio is corrected for 
the multiplicity. See text in detail.
}
 \label{velage}
\end{figure*}



\begin{landscape}
\begin{table}
\begin{center}
Table 1.\hspace{4pt}
Summary of observations of the sample
\end{center}
\tabcolsep=1pt
\vspace{6pt}
\begin{tabular*}{22cm}{@{\hspace{\tabcolsep}
\extracolsep{\fill}}p{10pc}cccccccccccccccc}
\hline\hline\\ [-10pt]
\multicolumn{1}{c}{name} &
\multicolumn{1}{c}{$z$} &
\multicolumn{1}{c}{Amp.$^a$} &
\multicolumn{1}{c}{$U$} &
\multicolumn{1}{c}{$B$} &
\multicolumn{1}{c}{$V$} &
\multicolumn{1}{c}{$R$} &
\multicolumn{1}{c}{$I$} &
\multicolumn{1}{c}{$J$} &
\multicolumn{1}{c}{$H$} &
\multicolumn{1}{c}{$K$} &
\multicolumn{1}{c}{$S_{6.75 \mu \mathrm{m}}$} &
\multicolumn{1}{c}{$S_{15 \mu \mathrm{m}}$} &
\multicolumn{1}{c}{$S_{450 \mu \mathrm{m}}$} &
\multicolumn{1}{c}{$S_{850 \mu \mathrm{m}}$} &
\multicolumn{1}{c}{$S_{\mathrm{1.2-1.35mm}}^b$} &
\multicolumn{1}{c}{Ref.} \\
\multicolumn{1}{c}{} &
\multicolumn{1}{c}{} &
\multicolumn{1}{c}{} &
\multicolumn{1}{c}{[$\mu$Jy]} &
\multicolumn{1}{c}{[$\mu$Jy]} &
\multicolumn{1}{c}{[$\mu$Jy]} &
\multicolumn{1}{c}{[$\mu$Jy]} &
\multicolumn{1}{c}{[$\mu$Jy]} &
\multicolumn{1}{c}{[$\mu$Jy]} &
\multicolumn{1}{c}{[$\mu$Jy]} &
\multicolumn{1}{c}{[$\mu$Jy]} &
\multicolumn{1}{c}{[mJy]} &
\multicolumn{1}{c}{[mJy]} &
\multicolumn{1}{c}{[mJy]} &
\multicolumn{1}{c}{[mJy]} &
\multicolumn{1}{c}{[mJy]} &
\multicolumn{1}{c}{} \\
\hline \\[-8pt]
  HR10$^c$             & 1.44 &...&...  & 0.16 &...   &...   &0.52$^d$& 6.4  & 14.8 & 27.7 &...  & 0.203 & 32.3 & 4.89 & 2.13$^e$ & 1, 2\\
  EROJ164023       & 1.05 &1.4&$<$0.10& 0.13 & 0.19 & 1.32$^f$& 3.00   & 20.4 & 21.3 & 63.0 &...  & 0.530 &...   & $<$6 &...   & 3\\ 
  ISOJ1324-2016       & 1.50 &...&... &$<$0.98&...   &$<$2.8& 3.0   &...    &...   &67.  & 0.89 &0.76   &...   &...  &...    & 4 \\
  PDFJ011423$^g$      & 0.65 &...&... &...    & 3.02 & 11.1 &...    & 91.4  &...   & 485. & 4.1  & 7.6   &...   &...  &   ... & 5, 6 \\ 
  CUDSS14F           & 0.660 &...&3.01&...    & 8.32 &...   &  27.5  &...    &...   &132.&0.115  & 0.562& 20.  &2.7  &...    &  7, 8  \\
  CUDSS10A             &0.550&...&... &...    & 1.74 &...   &  12.0  &...    &...   &100. &...   &...    &  23.  &4.8  &...    & 7, 8   \\
  CUDSS14.13         & 1.15  &... &0.97&1.2   & 2.5  &...   &  16.4  &...    &...   &155.&0.613  & 1.653& ...  & 3.3 & ...   &   9, 10, 11, 12   \\
  N2 850.1             &0.845&...&... &   ... & 1.92 & 3.86 & 2.06   &...    &...   &10.3 & $<$1 & $<$2  & 23.  &11.2  &...    & 13, 14   \\
  N2 850.2             & 2.45 &...&... &...   &$<$0.16& 0.39 & 0.21   &...    &...   & 7.9 & $<$1 & $<$2  & 35.  &10.7 &...    & 13, 14, 15   \\
  N2 850.4             &2.376&...&... &...    & 3.99 & 4.48 & 3.75   &...    &...   &27.2 & $<$1 & $<$2  &$<$34.& 8.2 & 2.59$^h$  & 13, 14, 16   \\
  N2 850.8             &1.189&...&... &...    & 2.78 & 5.79 & 3.09   &...    &...   &35.2 & $<$1 & $<$2  &$<$43.& 5.1 &...    & 13, 14, 16   \\
  LE 850.6$^i$          &2.61 &...&... &..     &...   & ... & 2.0    &...    &...   &13.1 &...   & ...   &...   &11.  &...    & 13, 15  \\
  SMMJ123600.2+621047  & 1.993 & ... &0.12 &0.28 &...  & 0.28   &  1.0   & ... & ... &  6.4 & ...  & ... & ...  &  7.0 & ... &  15, 17\\
  SMMJ123629.13+621045.8& 1.013 & ... &0.045&0.15 &...  & 0.44   &  2.6   & ... &  ... & 36.8 & ... & ... &  ... &  5.0 & ... &  15 \\
  SMMJ123607.53+621550.4 & 2.415 & ... &0.41 &1.6  &...  & 1.1   &   2.0   & ... &  ... & 12.2 & ... & ... &  ... &  4.0 & ... &  15\\
  SMMJ131212.7+424423 & 2.805 & ... &...  &...  &...  & 0.048  &  0.080 & ... &  ... & 1.8 &  ... & ... &  ... &  5.6 & ... &  15, 17\\
  SMMJ131201.2+424208   & 3.405 & ... &...  &...  &...  & 0.70   &  0.96  & ... & ... &  5.8 &  ... & ... &  ... &  6.2 & ... &  15, 17 \\
  SMMJ02399-0136$^j$ L1& 2.80 &2.5&0.153$^k$&1.14&0.905& 2.79& 4.28   &...&...   &14.7 &0.13  & 0.47  & 69.  & 26.0 &  5.7$^e$& 18, 19   \\
  SMMJ02399-0136$^j$ L2& 2.80 &2.5&$<$0.12$^k$&0.320&0.205&0.518&0.653&...&...   &1.19 &0.13  & 0.47  & 69.  & 26.0 &  5.7$^e$& 18, 19 \\
  SMMJ14011+0252 J1$^l$& 2.56 &2.8&0.531& 3.00 &...   &6.00  &  8.77  &...    &...   &41.5 &...   &...    &42.  &14.6 & 6.06$^e$& 20   \\
  SMMJ02399-0134$^m$& 1.06 &2.5&4.45$^n$& 6.36 &...   & 21.3 &   45.  &  118. &...   &212. &0.750 & 1.80  &42.  & 11.0&...    & 19, 21 \\
  SMMJ17142+5016      & 2.39 &...&... &0.0875$^d$&0.135$^d$&...   & 0.205$^d$& 0.343$^o$& 0.956$^o$&1.33 &...&...&...& 5.6 &...    & 22, 23  \\
  SMMJ221726+0013  & 3.098 & ... &...  &...  &...  & 0.031  & 0.055  & ... &  ... & 0.92 & ... & ... & 45. &  17. & ... & 15, 24  \\
  SMMJ131225.7+424350 & 1.038&...&... & 0.77  &...   & .... & 5.5    & ...    & ...  &27. &0.027  & ... & ...  & 2.4 & ... &   25 \\
  GRB000210          & 0.846 &...&... & 0.370 & 0.833& 1.22 &  2.20  &...    &...   &...  &...   &...    &...   &3.0  &...    & 26, 27   \\
  GRB000418          & 1.119 &...&... &...    &...   & 1.1  &...     &...    &...   &1.7  &...   &...    &41.   &3.2  &...    & 27, 28   \\
  GRB010222          & 1.477 &...&... &...    &...   &...   &0.228$^d$&...   &...   &0.445$^p$&...   &...    &$<$37.8&3.74&1.05$^h$& 29   \\
\hline \\
\end{tabular*}
\begin{flushleft}
Notes: 
(a) Amplification by gravitational lens. Flux densities in this table have not been corrected for gravitational amplifications 
(b) Flux at millimetre wavelength, obtained with JCMT/SCUBA (1.35 mm) or IRAM/MAMBO (1.2 mm) 
(c) Additional photometry: $S_{12 \mu \mathrm{m}}$=0.85 mJy and $S_{3.6\mathrm{cm}}$=35 $\mu$Jy  
(d) Photometry with {\it HST}/WFPC2: F300W for $U$, F450W for $B$, F606W for $V$, F814W for $I$ 
(e) Observed with JCMT/SCUBA at 1.35 mm
(f) Originally taken with F702W band of {\it HST}/WFPC2
(g) Additional photometry: $S_{90 \mu \mathrm{m}}$=260 mJy
(h) Observed with IRAM/MAMBO at 1.2 mm
(i) Additional photometry at $R$-band (S. Chapman 2004, private communication)
(j) Additional photometry: $S_{350 \mu \mathrm{m}}<$323 mJy, $S_{750 \mu \mathrm{m}}$=28 mJy
(k) Originally taken with F336W band of {\it HST}/WFPC2
(l) Additional photometry: $S_{3 \mathrm{cm}} <$1.8 mJy 
(m) Additional photometry: $S_{0.675 \mu \mathrm{m}}$=15.8 $\mu$Jy 
(n) F336W band of {\it HST}/WFPC2
(o) Photometry with {\it HST}/NICMOS: F110W for $J$, F160W for $H$ 
(p) This flux is contaminated to some extent by the afterglow \\
REFERENCES. -- (1) Dey et al. 1999 
(2) Elbaz et al. 2002 
(3) Smith et al. 2001 and references therein 
(4) Pierre et al. 2001 
(5) Afonso et al. 2001 
(6) Georgakakis et al. 1999 
(7) Lilly et al. 1999 
(8) Eales et al. 1999
(9) Webb et al. 2003 
(10) Clements et al. 2003 %
(11) Flores et al. 1999a %
(12) Flores et al. 1999b %
(13) Ivison et al. 2002 
(14) Fox et al. 2002 
(15) Chapman et al. 2004, private communication
(16) Chapman et al. 2003b
(17) Chapman et al. 2003a
(18) Ivison et al. 1998 
(19) Smail et al. 2002 
(20) Ivison et al. 2001
(21) Soucail et al. 1999 
(22) Smail et al. 2003 
(23) Keel et al. 2002
(24) Chapman et al. 2003c %
(25) Sato et al. 2004 %
(26) Piro et al. 2002 
(27) Berger et al. 2003
(28) Berger et al. 2001 
(29) Frail et al. 2002
\end{flushleft}
\vspace{6pt}
\par\noindent
 \label{obs_tab}
\end{table}
\end{landscape}

\begin{landscape}
\begin{table}
\begin{center}
Table 2a.\hspace{4pt}
The results of SED fitting with the IMF of $x=1.35$
\end{center}
\tabcolsep=1pt
\vspace{6pt}
\begin{tabular*}{20cm}{@{\hspace{\tabcolsep}
\extracolsep{\fill}}p{10pc}cccccccccccccccc}
\hline\hline\\ [-10pt]
\multicolumn{1}{c}{name} &
\multicolumn{1}{c}{$t/t_0$} &
\multicolumn{1}{c}{$\Theta$} &
\multicolumn{1}{c}{$E.C.^b$} &
\multicolumn{1}{c}{$z$} &
\multicolumn{1}{c}{$\log M_T^c$} &
\multicolumn{1}{c}{$\tau_V$} &
\multicolumn{1}{c}{$\log$ SFR} &
\multicolumn{1}{c}{$\log M_*$} &
\multicolumn{1}{c}{$\log M_D$} &
\multicolumn{1}{c}{$\log L_{bol}$} &
\multicolumn{1}{c}{$\log R_e^d$} &
\multicolumn{1}{c}{$M_V^e$} &
\multicolumn{1}{c}{$M_B^f$} &
\multicolumn{1}{c}{$U-V^g$} &
\multicolumn{1}{c}{$t_{z=0}^h$} &
\multicolumn{1}{c}{$\log \langle Z_*/Z_\odot \rangle_{z=0}^i$} \\
\multicolumn{1}{c}{} &
\multicolumn{1}{c}{} &
\multicolumn{1}{c}{} &
\multicolumn{1}{c}{} &
\multicolumn{1}{c}{} &
\multicolumn{1}{c}{[$M_\odot$]} &
\multicolumn{1}{c}{} &
\multicolumn{1}{c}{[$M_\odot$ yr$^{-1}$]} &
\multicolumn{1}{c}{[$M_\odot$]} &
\multicolumn{1}{c}{[$M_\odot$]} &
\multicolumn{1}{c}{[$L_\odot$]} &
\multicolumn{1}{c}{[pc]} &
\multicolumn{1}{c}{} &
\multicolumn{1}{c}{} &
\multicolumn{1}{c}{} &
\multicolumn{1}{c}{[Gyr]} &
\multicolumn{1}{c}{} \\
\hline \\[-8pt]
 HR10               &   5.0&   0.3& SMC&    1.44&  11.94&   15.3&   2.74&  11.91&   9.20&  12.74&   2.76& -23.08& -22.11&   1.30&    9.0&  -0.09 \\
{\it  EROJ164023$^a$}  & {\it 6.0}&{\it 0.5}& {\it  MW} &{\it 1.05}&{\it 10.94}&{\it 4.3}&{\it 1.51}&{\it 10.93}& {\it 7.76}& {\it 11.59}& {\it 2.49}& {\it -20.72} & {\it -19.76}& {\it  1.27}& {\it 8.0}& {\it -0.06} \\
 ISOJ1324-2016      &   5.0&   0.3& SMC&    1.50&  12.42&   15.3&   3.22&  12.39&   9.68&  13.22&   3.00& -24.27& -23.30&   1.30&    9.2&  -0.09 \\
 PDFJ011423         &   6.0&   0.3& MW &    0.65&  12.21&   11.9&   2.78&  12.20&   9.03&  12.86&   2.91& -24.14& -23.20&   1.22&    6.2&  -0.06 \\
 CUDSS14F           &   2.0&   0.9& SMC&    0.66&  10.94&    5.2&   2.44&  10.68&   8.47&  12.21&   2.63& -20.52& -19.64&   1.05&    5.9&  -0.38 \\
 CUDSS10A           &   5.0&   0.3& MW &    0.55&  11.66&   18.0&   2.47&  11.63&   8.65&  12.45&   2.63& -22.84& -21.92&   1.18&    5.5&  -0.10 \\
 CUDSS14.13        &   6.0&   0.5& MW &    1.15&  12.02&    4.3&   2.58&  12.00&   8.84&  12.67&   3.03& -23.38& -22.41&   1.28&    8.3&  -0.06 \\
 {\it N2 850.1           }&{\it   0.5}&{\it   0.4}&{\it SMC}&{\it    0.84}&{\it  12.07}&{\it   32.5}&{\it   3.57}&{\it  11.00}&{\it   8.89}&{\it  13.07}&{\it   2.43}&{\it -21.66}&{\it -20.88}&{\it   0.81}&{\it    6.6}&{\it  -1.05} \\
 N2 850.2           &   5.0&   0.3& SMC&    2.45&  12.32&   15.3&   3.12&  12.29&   9.58&  13.12&   2.95& -23.89& -22.90&   1.34&   10.6&  -0.09 \\
 N2 850.4           &   0.5&   0.7& LMC&    2.38&  12.44&    8.9&   3.93&  11.36&   9.11&  13.44&   2.86& -22.14& -21.34&   0.84&   10.1&  -1.05 \\
 N2 850.8           &   0.7&   0.7& SMC&    1.19&  11.87&   10.8&   3.42&  11.03&   8.92&  13.03&   2.69& -21.42& -20.60&   0.90&    7.9&  -0.86 \\
 LE 850.6           &   0.3&   0.7& SMC&    2.61&  12.90&    8.9&   4.25&  11.44&   9.25&  13.66&   2.89& -22.39& -21.66&   0.72&   10.3&  -1.40 \\
SMMJ123600.2+621047 &   6.0&   0.3& MW &    1.99&  12.40&   11.9&   2.97&  12.39&   9.22&  13.05&   3.00& -24.17& -23.18&   1.34&   10.2&  -0.06 \\
{\it   SMMJ123629.13+621045.8$^a$}     &{\it   6.0}& {\it  0.5}&{\it SMC}&{\it    1.01}&{\it  10.50}&{\it    3.6}&{\it   1.07}&{\it  10.49}&{\it   7.60}&{\it  11.16}& {\it  2.27}&{\it -19.64}&{\it -18.68}&{\it   1.27}&{\it    7.8}&{\it  -0.06} \\
SMMJ123607.53+621550.4 &   1.0&   1.2& MW &    2.42&  11.80&    4.3&   3.39&  11.19&   8.80&  13.04&   3.00& -21.48& -20.60&   1.01&   10.2&  -0.68 \\
SMMJ131212.7+424423 &   5.0&   0.3& SMC&    2.81&  11.89&   15.3&   2.69&  11.86&   9.15&  12.69&   2.74& -22.79& -21.80&   1.35&   11.0&  -0.09 \\
SMMJ131201.2+424208 &   0.5&   0.9& LMC&    3.40&  12.35&    5.4&   3.84&  11.27&   9.02&  13.35&   2.92& -21.85& -21.05&   0.85&   10.9&  -1.05 \\
 SMMJ02399-0136 L1  &   0.5&   0.7& LMC&    2.80&  12.45&    8.9&   3.94&  11.38&   9.12&  13.45&   2.86& -22.14& -21.34&   0.85&   10.5&  -1.05 \\
{\it  SMMJ02399-0136 L2$^a$} &{\it 0.3}&{\it 0.3}& {\it LMC}&{\it 2.80}& {\it 12.69}& {\it  40.8}& {\it  4.04}&{\it  11.23}& {\it  8.91}& {\it 13.42}& {\it  2.42}& {\it -21.86}&{\it -21.12}& {\it  0.72}& {\it  10.5}&{\it  -1.40} \\
 SMMJ14011+0252 J1  &   1.0&   1.4& MW &    2.56&  11.85&    3.2&   3.44&  11.24&   8.86&  13.09&   3.10& -21.60& -20.73&   1.01&   10.3&  -0.68 \\
 SMMJ02399-0134     &   1.0&   1.0& SMC&    1.06&  11.55&    5.3&   3.15&  10.95&   8.83&  12.82&   2.80& -21.15& -20.30&   0.97&    7.5&  -0.69 \\
 SMMJ17142+5016     &   3.0&   0.4& LMC&    2.39&  11.72&   15.7&   3.01&  11.60&   9.10&  12.85&   2.73& -22.24& -21.28&   1.27&   10.4&  -0.23 \\
SMMJ221726+0013 &   2.0&   0.3& SMC&    3.10&  11.84&   46.7&   3.33&  11.58&   9.37&  13.09&   2.60& -22.22& -21.28&   1.19&   10.9&  -0.37 \\
{\it  SMMJ131225.7+424350$^a$} &{\it 3.0}&{\it 1.0}&{\it LMC}&{\it    1.04}&{\it  10.62}&{\it   2.5}&{\it   1.91}&{\it  10.50}&{\it   7.99}&{\it  11.75}&{\it   2.58}& {\it -19.74}& {\it -18.81}&{\it   1.18}&{\it    7.7}&{\it  -0.23} \\
 GRB000210          &   4.0&   0.3& SMC&    0.85&  11.55&   22.9&   2.60&  11.49&   8.96&  12.51&   2.55& -22.27& -21.33&   1.20&    7.0&  -0.15 \\
 GRB000418          &   0.3&   0.3& MW &    1.12&  12.51&   57.3&   3.86&  11.05&   8.59&  13.22&   2.33& -21.71& -20.98&   0.71&    7.7&  -1.39 \\
 GRB010222          &   0.3&   0.3& MW &    1.48&  12.44&   57.3&   3.79&  10.98&   8.52&  13.15&   2.30& -21.39& -20.66&   0.70&    8.6&  -1.40 \\
\hline \\
\end{tabular*}
\begin{flushleft}
Notes: 
(a) Submm galaxies unused in Section 4 and 5 
(b) extinction curve 
(c) initial mass of gas reservoir
(d) intrinsic effective radius
(e) present-day $V$-band absolute magnitude of submm galaxies
(f) present-day $B$-band absolute magnitude of submm galaxies
(g) present-day $U-V$ of submm galaxies
(h) age of submm galaxies at $z=0$ when $t_0$=100 Myr
(i) mean luminosity-weighted metallicity of stars
\end{flushleft}
\vspace{6pt}
\par\noindent
 \label{fit_tab}
\end{table}
\end{landscape}

\begin{landscape}
\begin{table}
\begin{center}
Table 2b.\hspace{4pt}
The results of SED fitting with the IMF of $x=1.10$
\end{center}
\tabcolsep=1pt
\vspace{6pt}
\begin{tabular*}{20cm}{@{\hspace{\tabcolsep}
\extracolsep{\fill}}p{10pc}cccccccccccccccc}
\hline\hline\\ [-10pt]
\multicolumn{1}{c}{name} &
\multicolumn{1}{c}{$t/t_0$} &
\multicolumn{1}{c}{$\Theta$} &
\multicolumn{1}{c}{$E.C^b$} &
\multicolumn{1}{c}{$z^c$} &
\multicolumn{1}{c}{$\log M_T^d$} &
\multicolumn{1}{c}{$\tau_V$} &
\multicolumn{1}{c}{$\log$ SFR} &
\multicolumn{1}{c}{$\log M_*$} &
\multicolumn{1}{c}{$\log M_D$} &
\multicolumn{1}{c}{$\log L_{bol}$} &
\multicolumn{1}{c}{$\log R_e^e$} &
\multicolumn{1}{c}{$M_V^f$} &
\multicolumn{1}{c}{$M_B^g$} &
\multicolumn{1}{c}{$U-V^h$} &
\multicolumn{1}{c}{$t_{z=0}^i$} &
\multicolumn{1}{c}{$\log \langle Z_*/Z_\odot \rangle_{z=0}^j$} \\
\multicolumn{1}{c}{} &
\multicolumn{1}{c}{} &
\multicolumn{1}{c}{} &
\multicolumn{1}{c}{} &
\multicolumn{1}{c}{} &
\multicolumn{1}{c}{[$M_\odot$]} &
\multicolumn{1}{c}{} &
\multicolumn{1}{c}{[$M_\odot$ yr$^{-1}$]} &
\multicolumn{1}{c}{[$M_\odot$]} &
\multicolumn{1}{c}{[$M_\odot$]} &
\multicolumn{1}{c}{[$L_\odot$]} &
\multicolumn{1}{c}{[pc]} &
\multicolumn{1}{c}{} &
\multicolumn{1}{c}{} &
\multicolumn{1}{c}{} &
\multicolumn{1}{c}{[Gyr]} &
\multicolumn{1}{c}{} \\
\hline \\[-8pt]
 HR10               &   6.0&   0.5& SMC&    1.44&  11.67&   14.8&   2.47&  11.64&   9.34&  12.63&   2.85& -22.26& -21.18&   1.64&    9.1&   0.35 \\
{\it EROJ164023$^a$  }&{\it   6.0}&{\it   0.9}&{\it SMC}&{\it    1.05}&{\it  10.36}&{\it    4.5}&{\it   1.16}&{\it  10.34}&{\it   8.04}&{\it  11.32}&{\it   2.45}&{\it -19.13}&{\it -18.06}&{\it   1.60}&{\it    8.0}&{\it   0.36} \\
 ISOJ1324-2016      &   4.0&   0.7& SMC&    1.50&  12.04&   14.5&   3.23&  11.96&   9.95&  13.30&   3.16& -23.06& -21.99&   1.58&    9.1&   0.25 \\
 PDFJ011423         &   6.0&   0.7& MW &    0.65&  11.81&    8.9&   2.60&  11.78&   9.20&  12.75&   3.07& -23.00& -21.95&   1.53&    6.2&   0.35 \\
 CUDSS14F           &   2.0&   1.6& SMC&    0.66&  10.70&    4.8&   2.25&  10.40&   8.65&  12.25&   2.74& -19.68& -18.72&   1.29&    5.9&   0.00 \\
 CUDSS10A           &   6.0&   0.5& MW &    0.55&  11.56&   17.4&   2.35&  11.53&   8.95&  12.50&   2.79& -22.47& -21.43&   1.50&    5.6&   0.34 \\
 CUDSS14.13        &   6.0&   0.9& LMC&    1.15&  11.60&    3.8&   2.39&  11.57&   9.13&  12.55&   3.07& -22.16& -21.09&   1.61&    8.3&   0.36 \\
{\it N2 850.1$^a$    }&{\it   2.0}&{\it   2.4}&{\it SMC}&{\it    0.84}&{\it   9.60}&{\it    2.1}&{\it   1.16}&{\it   9.31}&{\it   7.55}&{\it  11.16}&{\it   2.37}&{\it -16.80}&{\it -15.83}&{\it   1.32}&{\it    6.8}&{\it   0.01} \\
 N2 850.2           &   5.0&   0.7& SMC&    2.45&  11.86&   10.5&   2.85&  11.81&   9.66&  12.96&   3.08& -22.53& -21.44&   1.66&   10.6&   0.30 \\
 N2 850.4           &   0.7&   2.2& MW &    2.38&  11.90&    3.4&   3.47&  11.03&   9.06&  13.35&   3.19& -20.97& -20.07&   1.10&   10.1&  -0.50 \\
 N2 850.8           &   0.5&   1.0& SMC&    1.19&  11.87&   12.9&   3.37&  10.76&   9.04&  13.22&   2.71& -20.66& -19.81&   0.96&    7.9&  -0.70 \\
 LE 850.6           &   1.0&   1.4& LMC&    2.61&  11.87&    6.1&   3.49&  11.23&   9.40&  13.42&   3.09& -21.35& -20.40&   1.22&   10.4&  -0.31 \\
SMMJ123600.2+621047 &   6.0&   0.7& MW &    1.99&  11.83&    8.9&   2.63&  11.80&   9.23&  12.78&   3.08& -22.56& -21.47&   1.67&   10.2&   0.34 \\
{\it SMMJ123629.13+621045.8$^a$}   &{\it   6.0}&{\it   1.0}&{\it  SMC}&{\it    1.01}&{\it  10.16}&{\it    3.7}&{\it   0.96}&{\it  10.14}&{\it  7.84}&{\it  11.12}&  {\it 2.40}&{\it -18.65}&{\it -17.58}&{\it   1.59}& {\it   7.8}&{\it   0.37} \\
SMMJ123607.53+621550.4&   1.0&   2.0& MW &    2.42&  11.49&    4.2&   3.11&  10.84&   8.88&  13.02&   3.05& -20.41& -19.47&   1.22&   10.2&  -0.31 \\
 SMMJ131212.7+424423 &   6.0&   0.5& SMC&    2.81&  11.66&   14.8&   2.45&  11.63&   9.33&  12.61&   2.84& -22.03& -20.93&   1.69&   11.1&   0.34 \\
 SMMJ131201.2+424208 &   0.5&   1.4& LMC&    3.40&  11.99&    5.5&   3.49&  10.88&   9.02&  13.34&   2.92& -20.65& -19.78&   1.00&   10.9&  -0.70 \\
 SMMJ02399-0136 L1  &   0.3&   0.9& LMC&    2.80&  12.45&   11.3&   3.81&  10.96&   9.04&  13.52&   2.77& -21.06& -20.27&   0.84&   10.5&  -1.03 \\
{\it SMMJ02399-0136 L2$^a$  }&{\it   0.3}&{\it   0.4}&{\it LMC}&{\it    2.80}&{\it  12.43}&{\it   57.2}&{\it   3.79}&{\it  10.94}&{\it   9.01}&{\it  13.47}&{\it   2.40}&{\it -21.01}&{\it -20.22}&{\it   0.84}&{\it   10.5}&{\it  -1.03} \\
 SMMJ14011+0252 J1  &   1.0&   2.4& MW &    2.56&  11.52&    2.9&   3.13&  10.87&   8.91&  13.05&   3.15& -20.47& -19.52&   1.22&   10.3&  -0.31 \\
 SMMJ02399-0134     &   2.0&   1.6& LMC&    1.06&  11.11&    4.0&   2.67&  10.82&   8.92&  12.66&   2.94& -20.45& -19.47&   1.35&    7.6&   0.01 \\
 SMMJ17142+5016     &   3.0&   0.7& LMC&    2.39&  11.46&   16.4&   2.85&  11.31&   9.30&  12.87&   2.83& -21.33& -20.28&   1.56&   10.4&   0.15 \\
 SMMJ221726+0013   &   6.0&   0.3& SMC&    3.10&  12.08&   41.0&   2.87&  12.05&   9.75&  13.02&   2.83& -23.05& -21.96&   1.69&   11.3&   0.34 \\
{\it SMMJ131225.7+424350$^a$}&{\it   2.0}&{\it   2.0}&{\it SMC}&{\it    1.04}&{\it  10.26}&{\it    3.1}&{\it   1.82}&{\it   9.97}&{\it   8.21}&{\it  11.81}&{\it   2.62}&{\it -18.33}&{\it -17.35}&{\it   1.34}&{\it    7.6}&{\it   0.01} \\
 GRB000210          &   5.0&   0.7& MW &    0.85&  11.39&   12.4&   2.38&  11.35&   8.92&  12.48&   2.85& -21.77& -20.72&   1.55&    7.1&   0.33 \\
 GRB000418          &   0.3&   0.5& MW &    1.12&  12.13&   51.5&   3.49&  10.65&   8.58&  13.16&   2.35& -20.59& -19.81&   0.81&    7.7&  -1.03 \\
 GRB010222          &   0.5&   0.4& LMC&    1.48&  11.65&   67.8&   3.15&  10.54&   8.68&  12.96&   2.21& -20.02& -19.17&   0.97&    8.7&  -0.70 \\
\hline \\
\end{tabular*}
\begin{flushleft}
See notes in Table 2a
\end{flushleft}
\vspace{6pt}
\par\noindent
 \label{fit_tab}
\end{table}
\end{landscape}

\begin{table*}
\begin{center}
Table 3.\hspace{4pt}
Summary of SED fitting results
\end{center}
\tabcolsep=1pt
\vspace{6pt}
\begin{tabular*}{6cm}{@{\hspace{\tabcolsep}
\extracolsep{\fill}}p{5pc}ccc}
\hline\hline\\ [-10pt]
Age & $x=1.35$ & $x=1.10$ \\
\hline
$t/t_0 \le 1$ & 10 & 10 \\
$t/t_0 = 2$ &   2 & 2 \\
$t/t_0 = 3$ &   1 & 1 \\
$t/t_0 = 4$ &   1 & 1 \\
$t/t_0 = 5$ &   5 & 2 \\
$t/t_0 = 6$ &   3 & 6 \\
\hline 
Optical depth & & \\
\hline 
$\tau_V \le 3.$      & 0 & 1 \\
$3 < \tau_V \le 5.$  & 3 & 3 \\
$5 < \tau_V \le 10.$ & 6 & 4 \\
$10 < \tau_V \le 20.$& 9 & 9 \\
$\tau_V > 20.$       & 4 & 3 \\
\hline
\multicolumn{2}{l}{Extinction curve} \\
\hline
MW & 8 & 8 \\
LMC & 4 & 7 \\
SMC & 10 & 7 \\
\hline 
\hline 
\end{tabular*}
\begin{flushleft}
Note: we exclude EROJ164023, N2 850.1, SMMJ123629.13+621045.8, 
and SMMJ131225.7+424350, for which we find no reasonable SED fit. 
\end{flushleft}
\vspace{6pt}
\par\noindent
 \label{fit}
\end{table*}

\begin{table*}
\begin{center}
Table 4.\hspace{4pt}
The mean characteristics of old submm galaxies at $z=0$. 
\end{center}
\tabcolsep=1pt
\vspace{6pt}
\begin{tabular*}{16cm}{@{\hspace{\tabcolsep}
\extracolsep{\fill}}p{9pc}cccccc}
\hline\hline\\ [-10pt]
\multicolumn{1}{c}{} &
\multicolumn{1}{c}{ Sample } &
\multicolumn{1}{c}{$<z>$} &
\multicolumn{1}{c}{ $<M_V>$} &
\multicolumn{1}{c}{ $<U-V>$} &
\multicolumn{1}{c}{ $<t_{z=0}>$ [Gyr]} &
\multicolumn{1}{c}{$<\log \langle Z_*/Z_\odot \rangle >$} \\
\hline \\
Bright sample ($M_V<-22$)&9 & 1.7  & -22.57 & 1.62  & 9.0 (9.0)$^a$ & 0.33 \\  
Faint sample ($M_V>-22$) &4 &1.2  & -20.81 & 1.44  & 7.7  (9.0)$^a$ & 0.14 \\
\hline
All sample               &13 &1.6  & -22.03  & 1.56  & 8.64 & 0.27 \\
\hline \\[-8pt]
\hline \\
\end{tabular*}
\begin{flushleft}
Note: the mean values are given for old 
submm galaxies ($t/t_0 \ge 2$).
All the values are derived with the IMF of $x=1.10$. 
a) The mean of present-day age including young submm galaxies.
\end{flushleft}
\vspace{6pt}
\par\noindent
 \label{cmtab}
\end{table*}

\appendix

\section{Summary of observations}

\subsection{Extremely red objects} 
We collect extremely red objects (EROs) detected in the 
MIR -- submm wavelengths. In order to increase the sample, 
we included EROs detected at the MIR wavelengths by 
Infrared Space Observatory ({\it ISO}), but not have observed 
flux at submm wavelengths. 
The samples are HR10, EROJ164023, ISOJ1324-2016, and 
PDFJ011423. 

These EROs are bright in the NIR 
enough to perform spectroscopic observations, and therefore redshifts are
spectroscopically determined. From the spectra, 
EROJ164023, ISOJ1324-2016 and PDFJ011423 are suggested to be 
composite starburst-AGN galaxies 
(Smith et al. 2001; Pierre et al. 2001; Afonso et al. 2001), 
while there is no signature of AGN in the spectra of HR10 
(Dey et al. 1999).

\subsection{Submm-selected galaxies}

\subsubsection{Canada-UK Deep Submm Survey}
The secure optical identifications of six submm sources detected in 
the Canada-UK Deep Submm Survey (CUDSS) are reported in Lilly et al. (1999), 
one of which is a nearby spiral galaxy at $z=0.074$ and therefore excluded 
in this study. Two submm galaxies (CUDSS 14F and 14A) are associated 
with radio sources. 
CUDSS 14F is detected with $ISO$ at 6.75 and 15 $\mu$m, and therefore 
this source is most securely identified, despite the faintest submm flux 
among these sample. 
The redshifts of CUDSS 14F and 10A are already known to be $z=$0.660 and 
0.550, respectively. So far, no signature of AGN activity is found 
in these submm galaxies (e.g. Gear et al. 2000). 

Together with CUDSS 14F and 10A, we collect CUDSS14.13 from the new 
catalogue by Webb et al. (2003). The multi-band photometric data 
are reported in Clements et al. (2003). This source has been 
detected at X-ray wavelengths (Waskett et al. 2003) and likely 
contains an AGN. 

\subsubsection{SCUBA 8-mJy survey}
The SCUBA 8-mJy survey is the largest submm extragalactic 
survey undertaken to date, covering 260 arcmin$^2$. 
Ivison et al. (2002) performed the optical identification of 
these submm galaxies in the Lockman Hole and ELAIS N2 regions 
by using the 1.4-GHz imaging map; as a result, 18 out of 30 submm 
sources are reliably identified as radio sources. 
The spectroscopic redshift of some sources are available in  
Chapman et al (2003a) 
and also provided by S.\ Chapman (2004, private communication). 
From this survey, 5 sources (N2 850.1, N2 850.4, N2 850.8, 
and LE 850.6) satisfies the requirements for our sample. 
Note that the $R$-band data of LE 850.6 is provided by 
S.\ Chapman (2004, private communication). 


Although the radio map suggests 
N2 850.1 is associated with the bright, 
compact optical galaxy at $z=0.845$, Ivison et al. (2002) 
claim that this galaxy acts as a gravitational lens 
to amplify the background faint submm source, 
considering the unreasonable 450-/850-$\mu$m and submm/radio 
spectral indices as a submm source at $z < 1$ (see also 
Chapman et al. 2002). 
Assuming that the galaxy at $z=0.845$ is the 
true optical counterpart, we fit the SED of 
this submm source, but found no suitable SED model. 

\subsubsection{Hubble Deep Field and SA13 field}
The photometric data and the spectroscopic redshifts of 
sample in these fields (i.e. SMMJ123600.2+621047, 
SMMJ123629.13+621045.8, SMM J123607.53+621550.4, 
SMMJ131212.7+424423, and   SMMJ131201.2+424208) are 
taken from Chapman et al. (2003a) or 
provided by S.\ Chapman (2004, private communication) if 
not available in Chapman et al. (2003a). 

The deep SCUBA map of HDF is reported by Dunlop et al. (1998), 
which is re-analysed by Serjeant et al. (2003). 
No source is however collected from these reports, 
since these sources are too faint at submm wavelengths, 
compared to the other sample. Note that even the optical 
identification of the brightest source HDF850.1 is complicated 
by the possible gravitational lensing by a foreground galaxy 
(Dunlop et al. 2002).

\subsection{Lensed submm galaxies}
The effects of gravitational lens has been used to push 
below the confusion limit of the blank-field surveys. 
In the list of submm sources by Smail et al. (2002), 
three submm sources, SMM J02399-0136, SMM J14011+0252 
and SMM J02399-0124 have enough photometric data to perform 
the SED fitting, all of which have spectroscopic redshifts. 
We adopt the lens amplification of 
2.5, 2.8 and 2.5 for SMM J02399-0136, SMM J14011+0252 and 
SMM J02399-0134, respectively (Smail et al. 2002).

SMM J02399-0136 is one of the brightest submm sources with 
the 850-$\mu$m flux of 26 mJy.  
The optical counterpart of SMM J02399-0136 is 
a system of two interacting/merging galaxies, 
named L1 for the compact component and L2 for the disturbed 
relatively diffuse component. The spectroscopic redshifts ($z=2.80$)
measured for each component suggest that L1 component is 
physically associated with L2 component. 
The angular separation 
of L1 and L2 is about 3$''$ (22 kpc for the adopted cosmology). 
Each component is resolved by a 
1.3$''$ synthesized beam at 1.4-GHz, and appears to have 
different radio spectral indices. The UV spectrum of L1 exhibits 
high-ionization lines with the width of $\sim$1000 km s$^{-1}$ 
and therefore this component probably has an AGN in the center. 
The presence of an AGN is also confirmed by the detection in 
the hard X-ray by {\it Chandra} (Bautz et al. 2000).  
On the other hand, the presence of a large amount of gas 
(M(H$_2$)$ \sim 2 \times 10^{11}$ M$_\odot$) is inferred from 
the strong CO emission lines. This means that a significant 
fraction of FIR luminosity could arise from starburst activity. 
The rest-frame FIR-to-5-GHz flux ratio is similar to that 
seen in nearby starbursts. These observations suggest that 
SMM J02399-0136 is a system of composite starburst-AGN galaxy. 

The optical counterpart of 
SMM J14011+0252 is again an interacting/merging pair of 
galaxies at $z=2.56$, J1 and J2 (Ivison et al. 2001). 
The CO position by Downes \& Solomon (2003) 
differs significantly ($\ga 1''$) from those obtained by 
Frayer et al. (1999) and Ivison et al. (2001); as a result, 
the new position agrees with the optical position of the 
J1 complex within the 1 $\sigma$ errors\footnote{
Downes and Solomon 
(2002) also suggest that SMM J14011+0252 is gravitationally 
lensed not only by the foreground cluster, but also 
by an individual galaxy on the line of sight. 
Since this hypothesis is not yet confirmed well 
(e.g. the redshift of suspected lensing galaxy is 
unknown, individually), we think it is premature to 
take this effect into account. 
Therefore, we consider the amplification of source 
only by the cluster, and assumed that J1 complex is 
the true optical counterpart of SMM J14011+0252.
}.
The high resolution image with the HST shows that 
the morphology of J1 component is complex; i.e. 
main component with relatively regular shape is 
associated with an extended envelope. Interestingly enough, 
the CO and dust emission is spatially extended over few 
arcseconds ($\ga 10$ kpc). Therefore, this galaxy is 
powered by starbursts rather than an AGN, which is 
consistent with the spectral features at rest-frame 
1200 -- 2400 \AA~ (Ivison et al. 2000). 
Also, this system is undetected in hard X-ray 
observations with {\it Chandra} (Fabian et al. 2000). 


SMM J02399-0134 is identified with a ring 
galaxy at $z=1.06$, which is also relatively bright at 
7 and 15 $\mu$m (Soucail et al. 1999). 
The emission lines [OII] 3727 \AA\ and [NeV] 3426 \AA\ 
are detected, which are typical of starburst galaxies 
hosting a central AGN (Soucail et al. 1999). 
The presence of an obscured AGN is confirmed through 
a detection in the hard X-ray band by {\it Chandra} 
(Bautz et al. 2000). 


\subsection{Submm galaxies in the proto-cluster region}
If submm sources at high redshifts are progenitors of 
present-day elliptical galaxies, they should be found 
preferentially in over-density regions at high redshifts, i.e. 
in proto-clusters. 

Smail et al. (2003) report the 
spectroscopically confirmed, submm-selected companion to 
a high-$z$ radio galaxy 53W002 which has been shown to 
reside in an over-density of Ly-$\alpha$ detected galaxies. 
By using a 1.4-GHz map, this submm source, SMM J17142+5016, 
is identified with a narrow-line AGN at $z=2.390$. This 
galaxy itself is one of the brightest Ly-$\alpha$ emitters 
in this region, and associated with an extended ($>6''$) 
Ly-$\alpha$ halo.

SMMJ221726+0013 is a bright submm galaxy resides in a giant 
Ly-$\alpha$ halo (Chapman et al. 2003c) at $z=3.098$. This giant 
Ly-$\alpha$ halo `blob-1' ($\ga 100$ kpc) lies in an overdensity 
region discovered in the survey of Lyman-break galaxies 
(Steidel et al. 2000). Chapman et al. (2003c) identify the 
optical counterpart as an elongated galaxy J1. The optical-NIR 
photometric data of J1 are provided by S.\ Chapman (2004, private 
communication). No signature of AGN is found 
(e.g. Chapman et al. 2003c; Bower et al. 2004). 


\subsection{A submm-detected faint 6.7 $\mu$m galaxy}
The SCUBA observations in the field of very deep 6.7 $\mu$m survey 
(1 $\sigma$ sensitivity of 3 $\mu$Jy) 
with {\it ISO} result in the detection of 3 submm 
sources (Sato et al. 2002). 
We collect SMMJ131225.7+424350 for which 
the spectroscopic redshift is available (Sato et al. 2002, 2004).

\subsection{GRB host Galaxies}
Recent follow-up observations of $\gamma$-ray bursts (GRBs) 
at submm/radio wavelengths indicate that 
about 20\% of GRB host galaxies are ULIRGs (Berger et al. 2003). 
Here we analyse three GRBs with the spectroscopic redshift; 
i.e. GRB 000210, GRB 000418 and 
GRB 010222 at $z=0.846$, 1.119 and 1.477, respectively. 
All GRB hosts are detected not only at submm, but 
also at radio wavelengths, and therefore the 
optical identification is reliable 
(Frail et al. 2002; Berger et al. 2003).

\section{Error estimates of the SED fitting} 
In Figure \ref{chi2a}, we show contour maps of $\Delta \chi^2$ 
for the sample used in Section 4, 
in which $\Delta \chi^2 =1.0,$ 2.71, and 6.63 correspond to the 
probability of 68.3 \%, 90 \%, and 99 \%, respectively, when projected 
to each axis, i.e. $t/t_0$ and $\Theta$. In most cases, the 
contour is elongated along the axis for the starburst age. 
However, note that the range of the starburst age within the 
confidence level of 68.3 \% is reasonably small.

\begin{figure*}
  \resizebox{6cm}{!}{\includegraphics{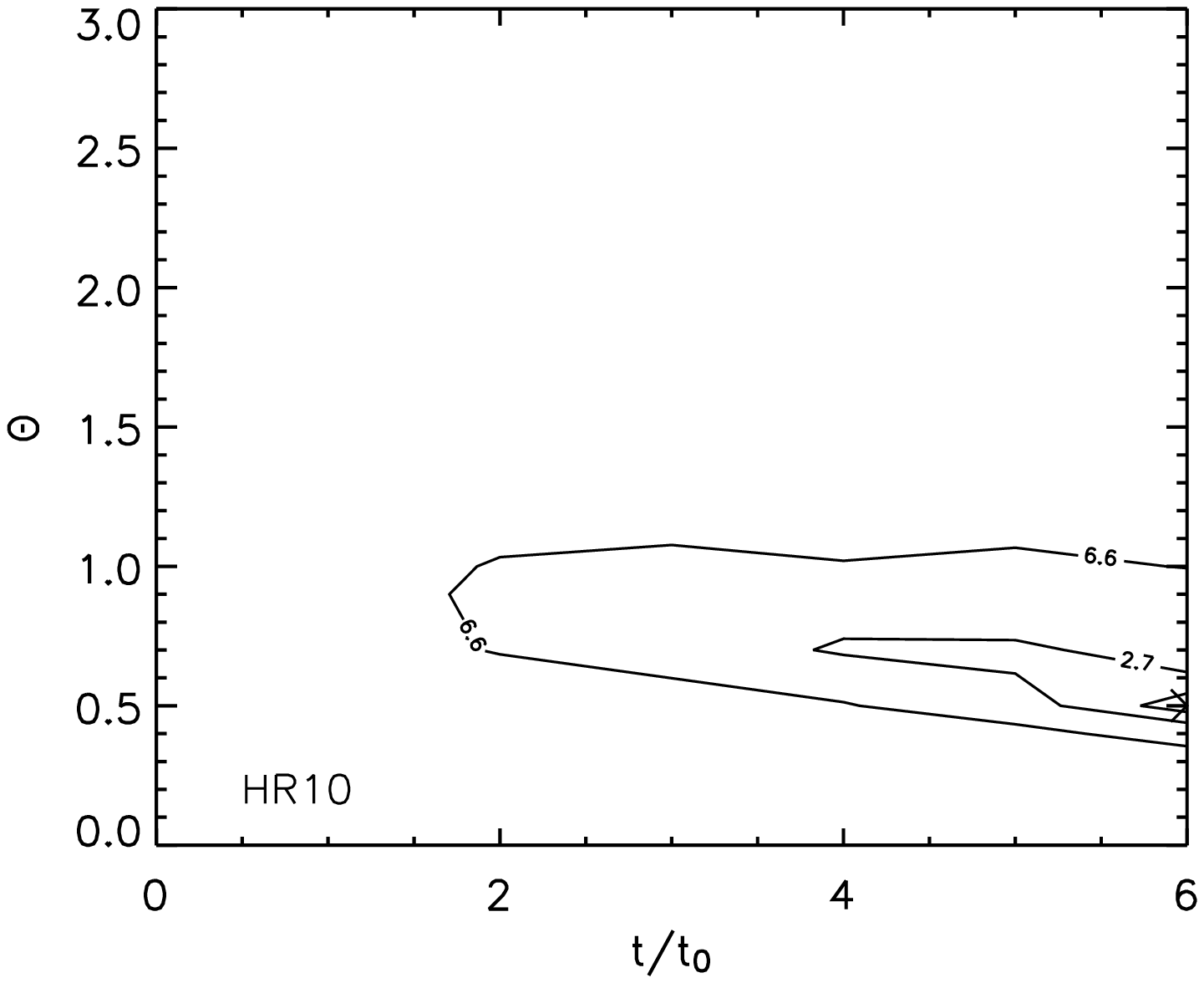}}
  \resizebox{6cm}{!}{\includegraphics{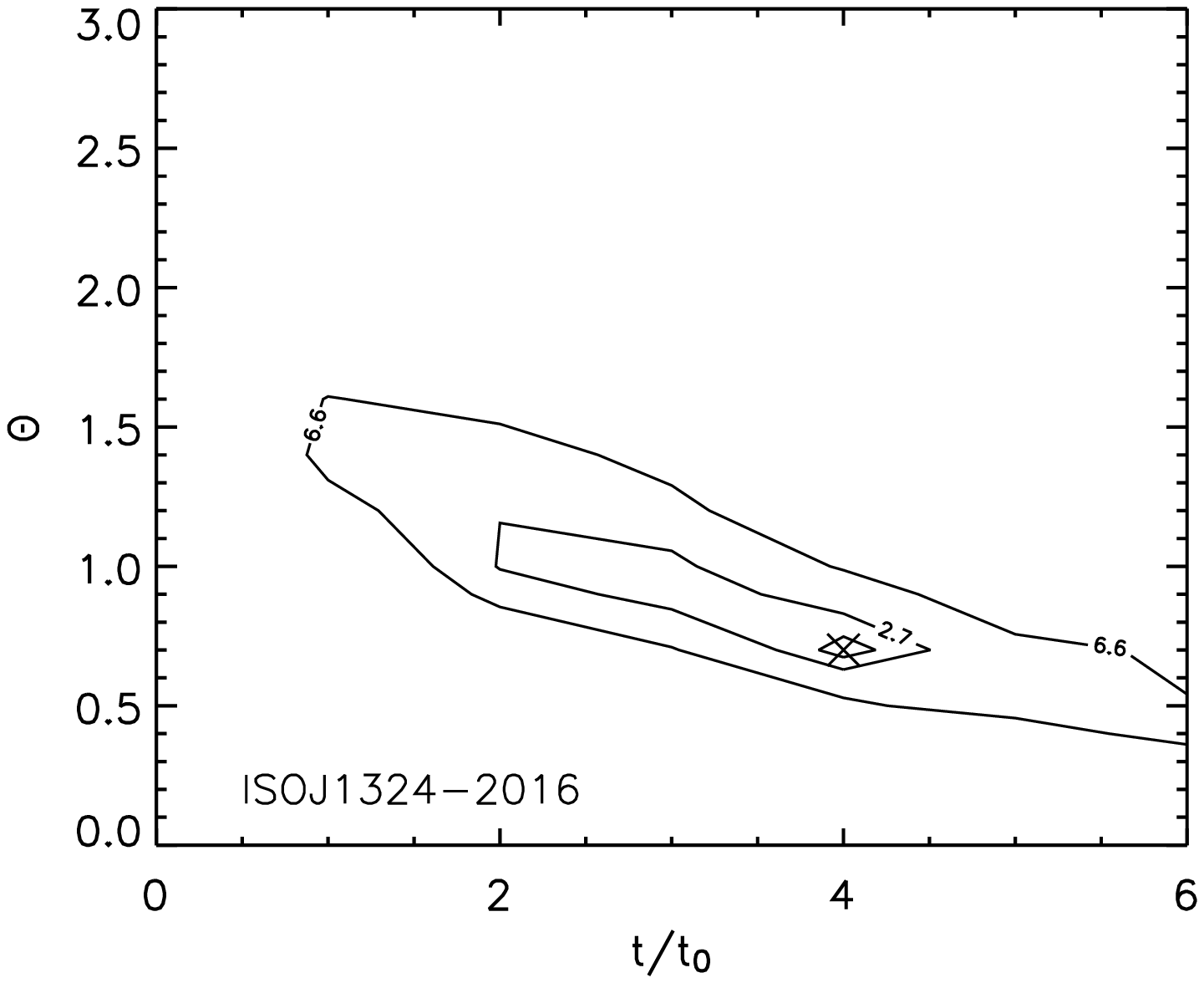}}
  \resizebox{6cm}{!}{\includegraphics{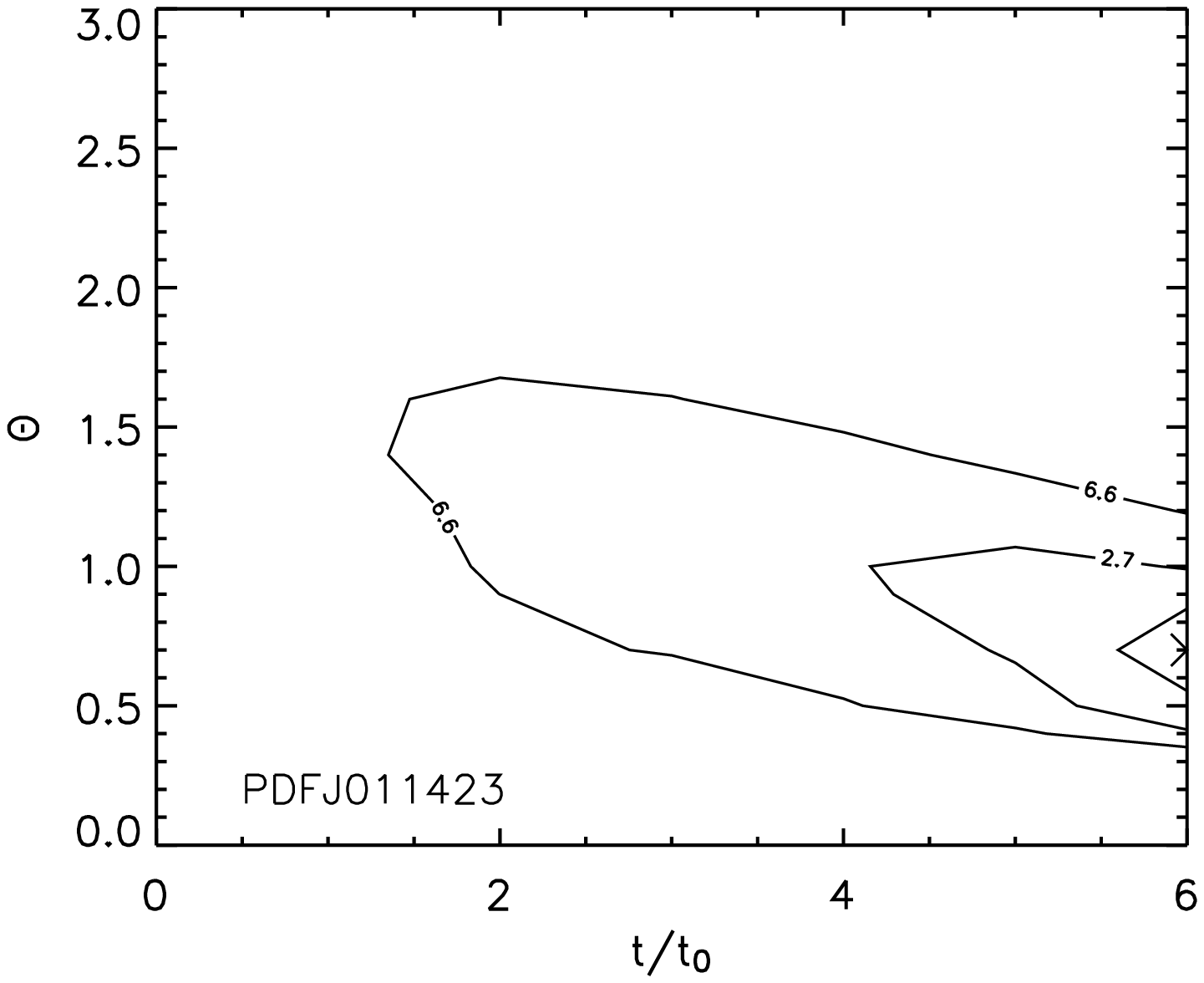}}
  \resizebox{6cm}{!}{\includegraphics{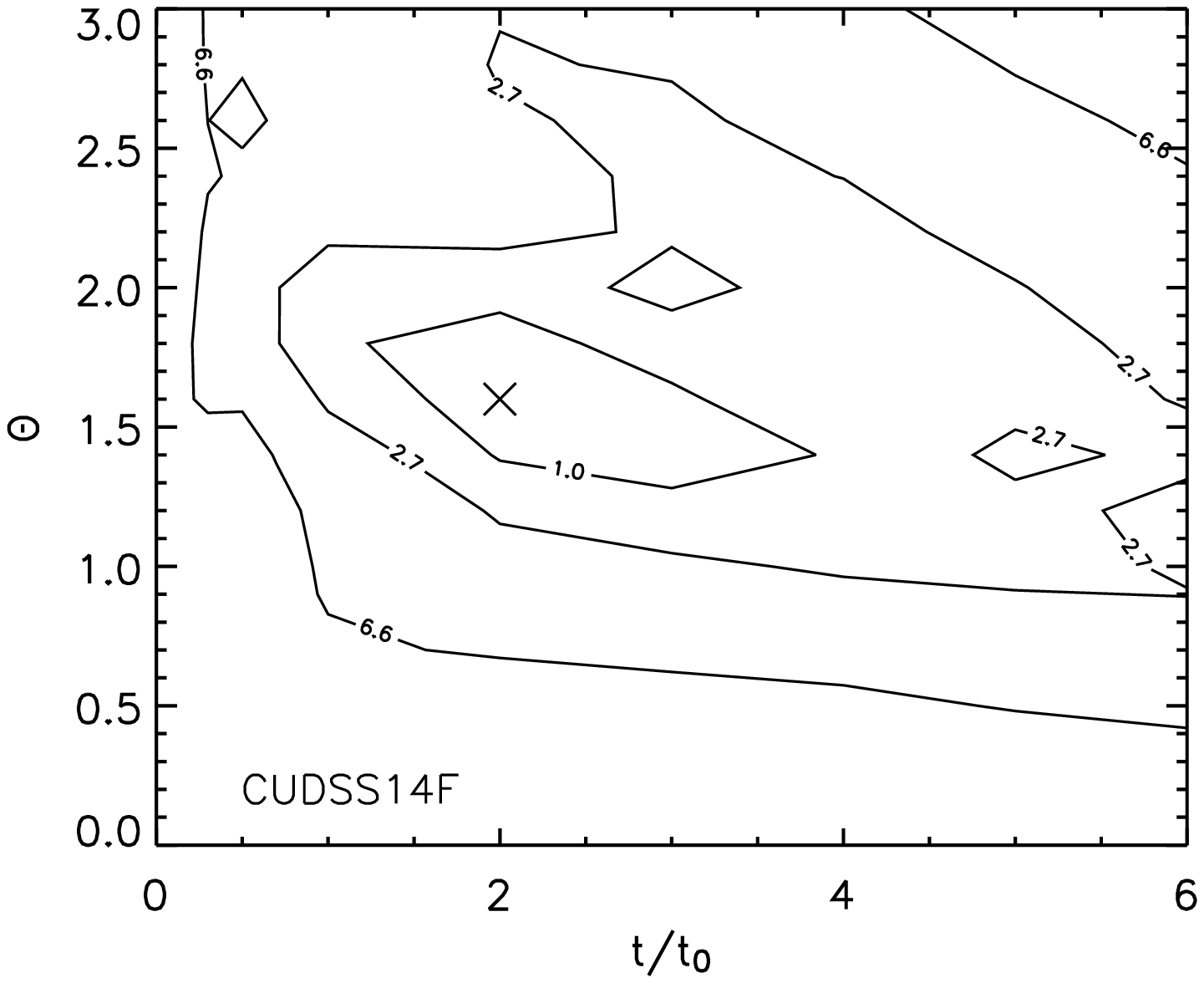}}
  \resizebox{6cm}{!}{\includegraphics{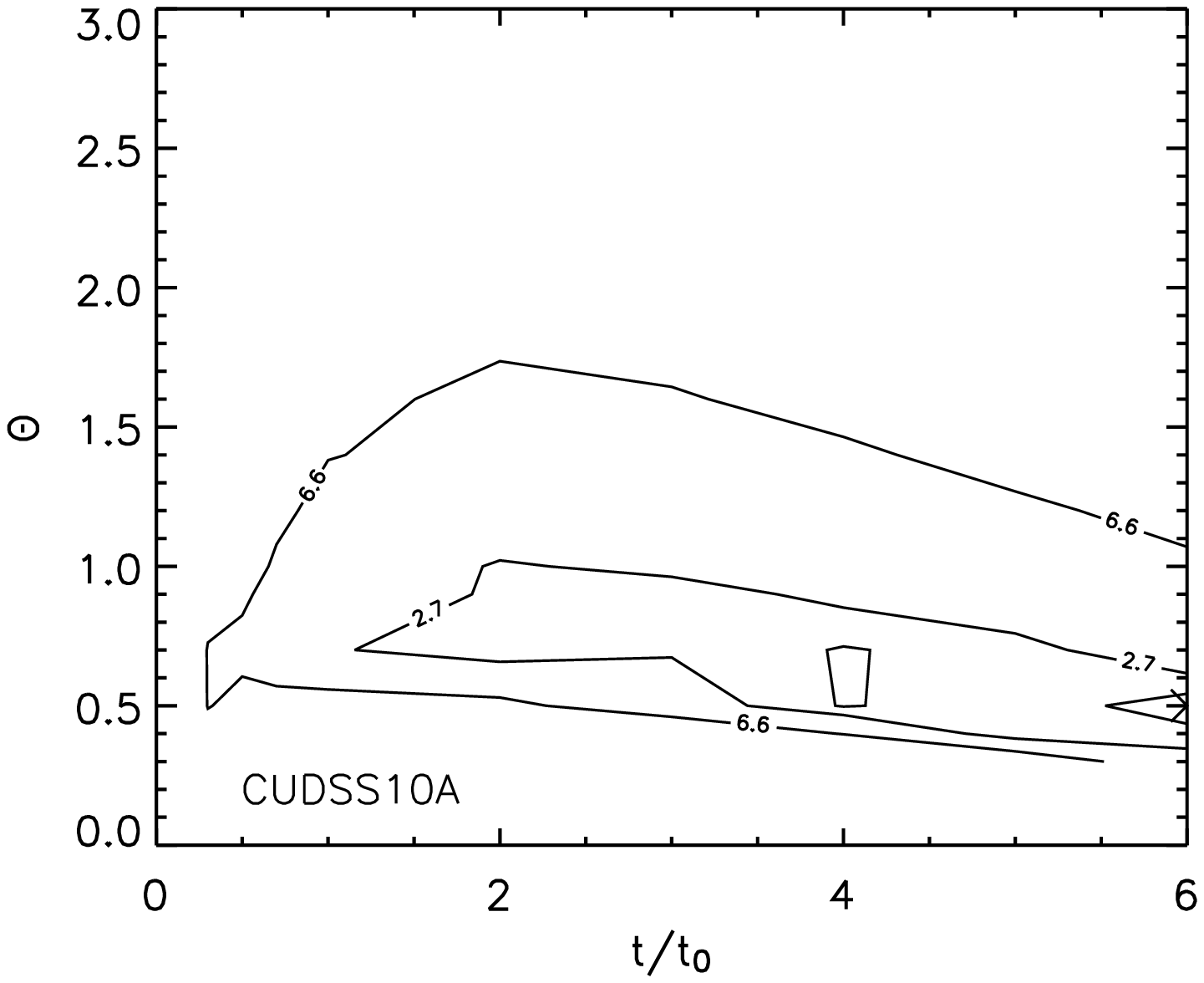}}
  \resizebox{6cm}{!}{\includegraphics{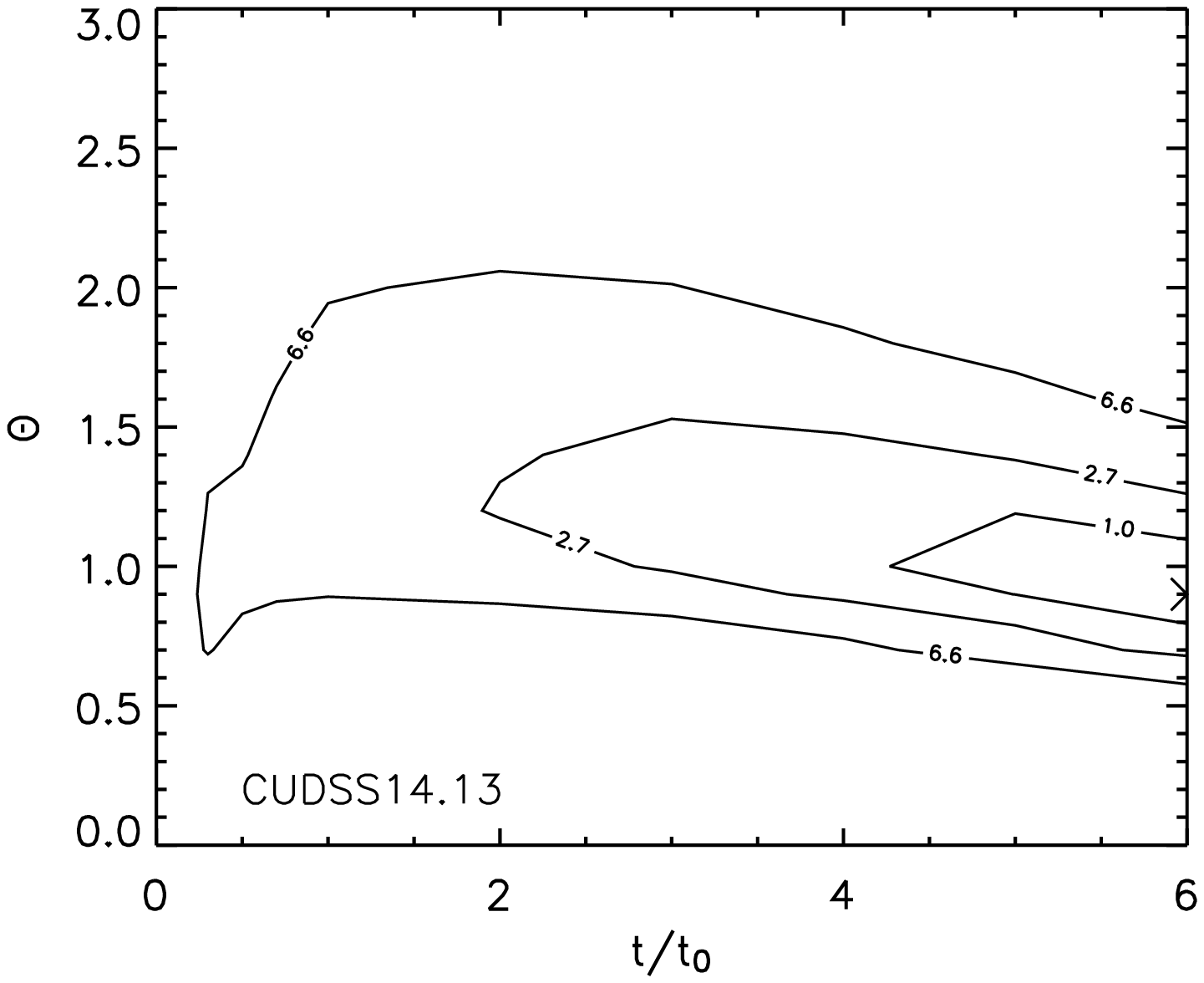}}
  \resizebox{6cm}{!}{\includegraphics{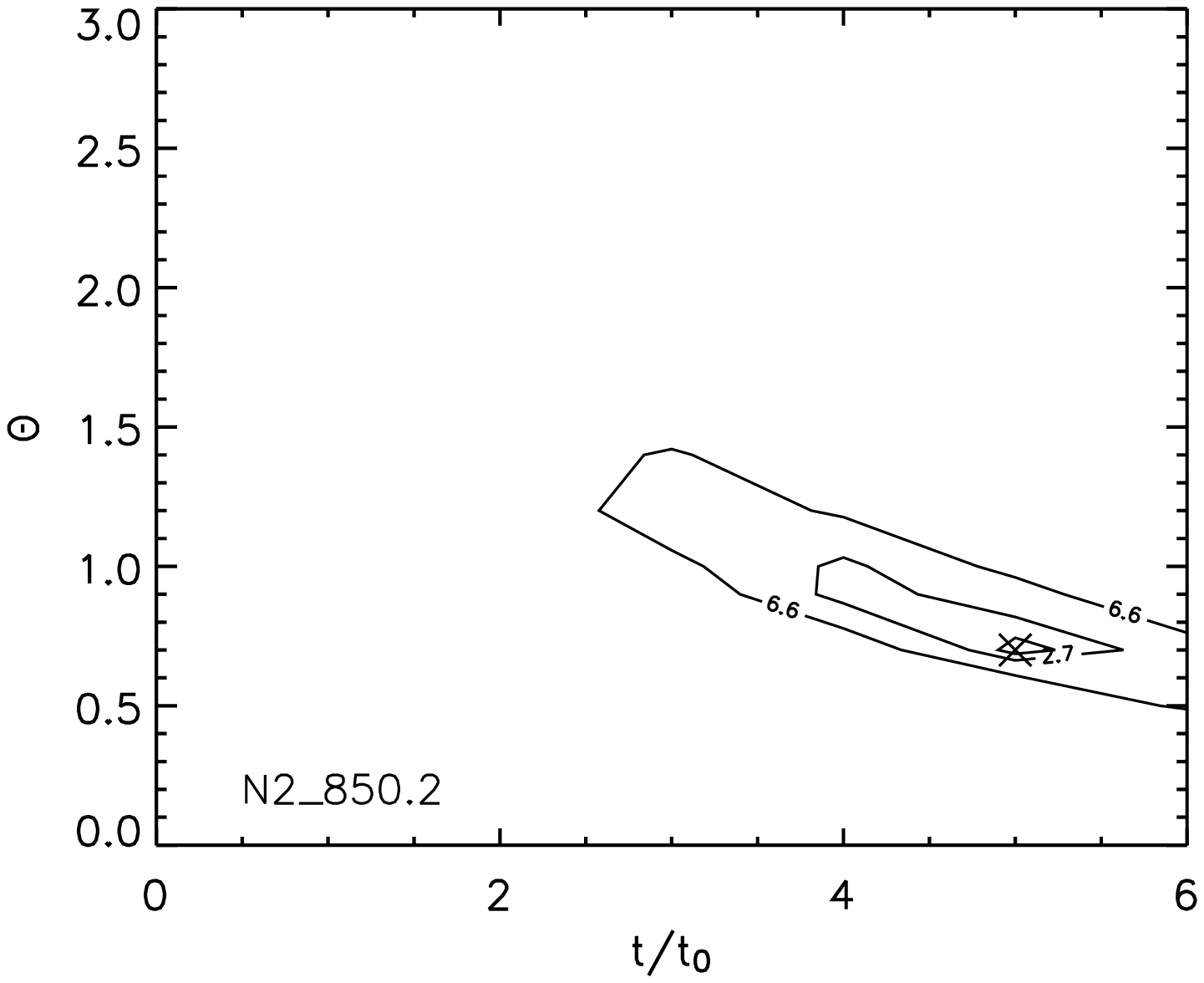}}
  \resizebox{6cm}{!}{\includegraphics{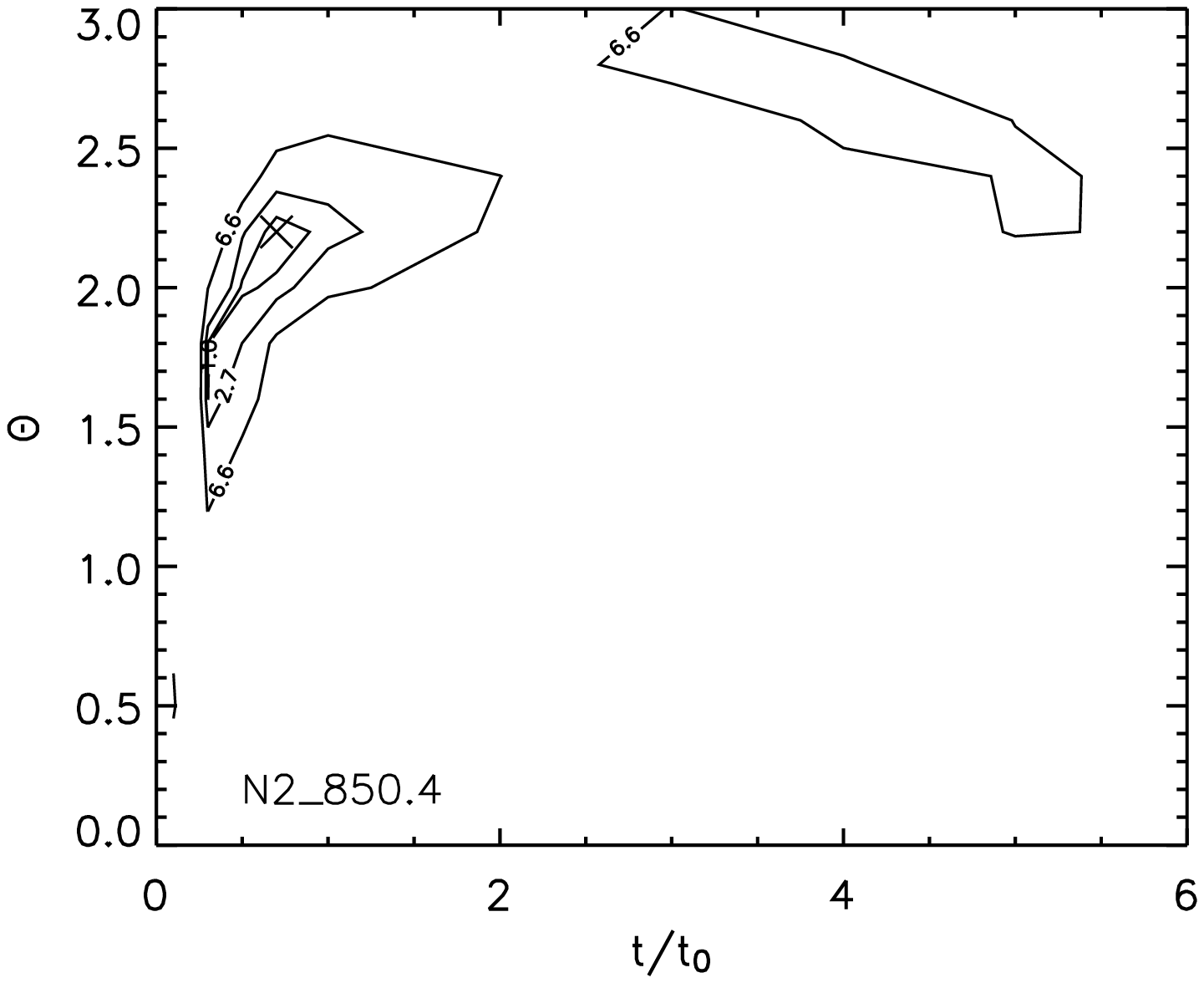}}
 \caption{The contour maps of 
$\Delta \chi^2$ for the sample used in Section 4. 
The contours are depicted at $\Delta \chi^2 =1.0,$ 2.71, and 6.63, 
which correspond to the 
probability of 68.3 \%, 90 \%, and 99 \%, respectively (when 
projected to each axis). 
}
 \label{chi2a}
\end{figure*}

\addtocounter{figure}{-1}
\begin{figure*}
  \resizebox{6cm}{!}{\includegraphics{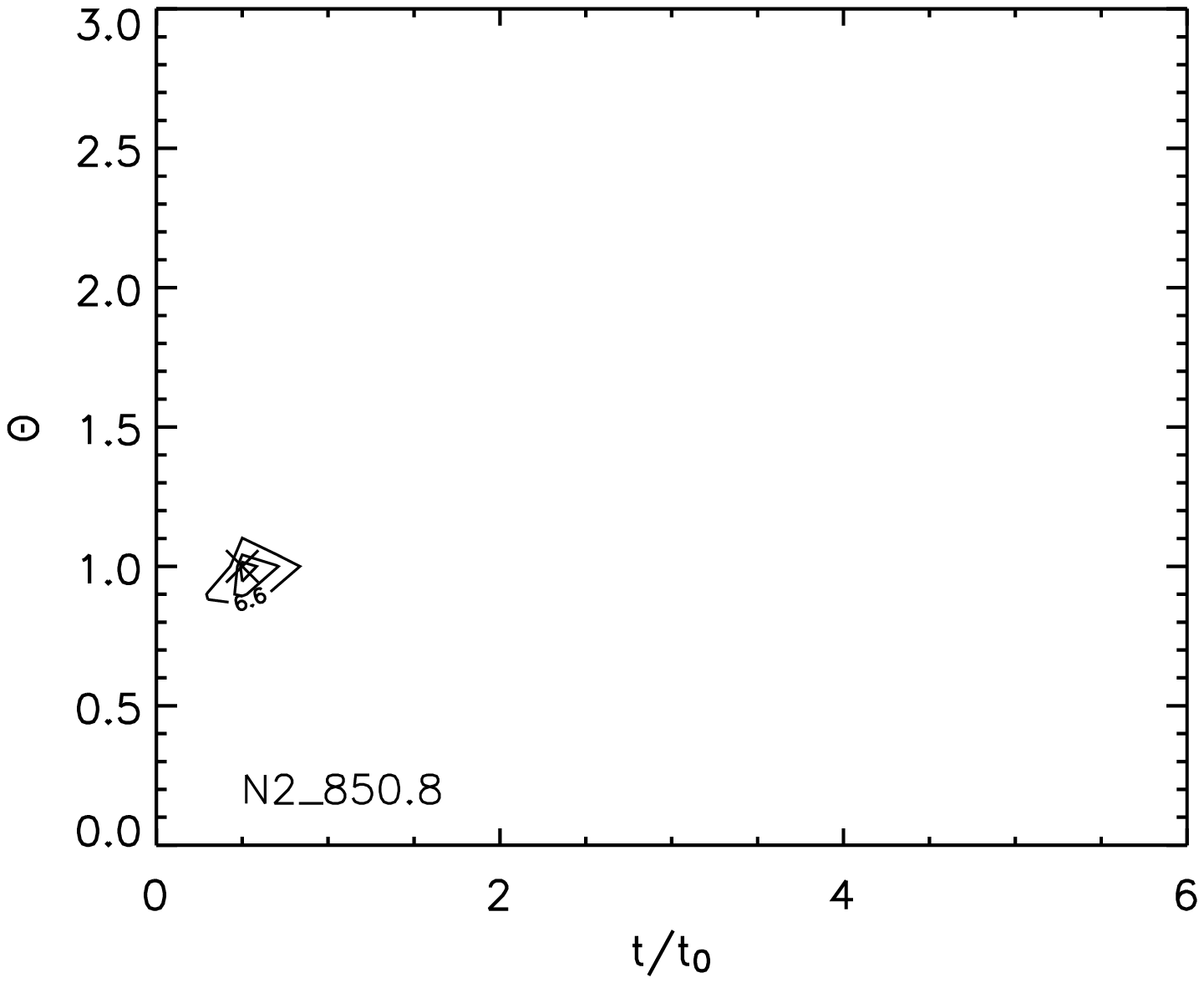}}
  \resizebox{6cm}{!}{\includegraphics{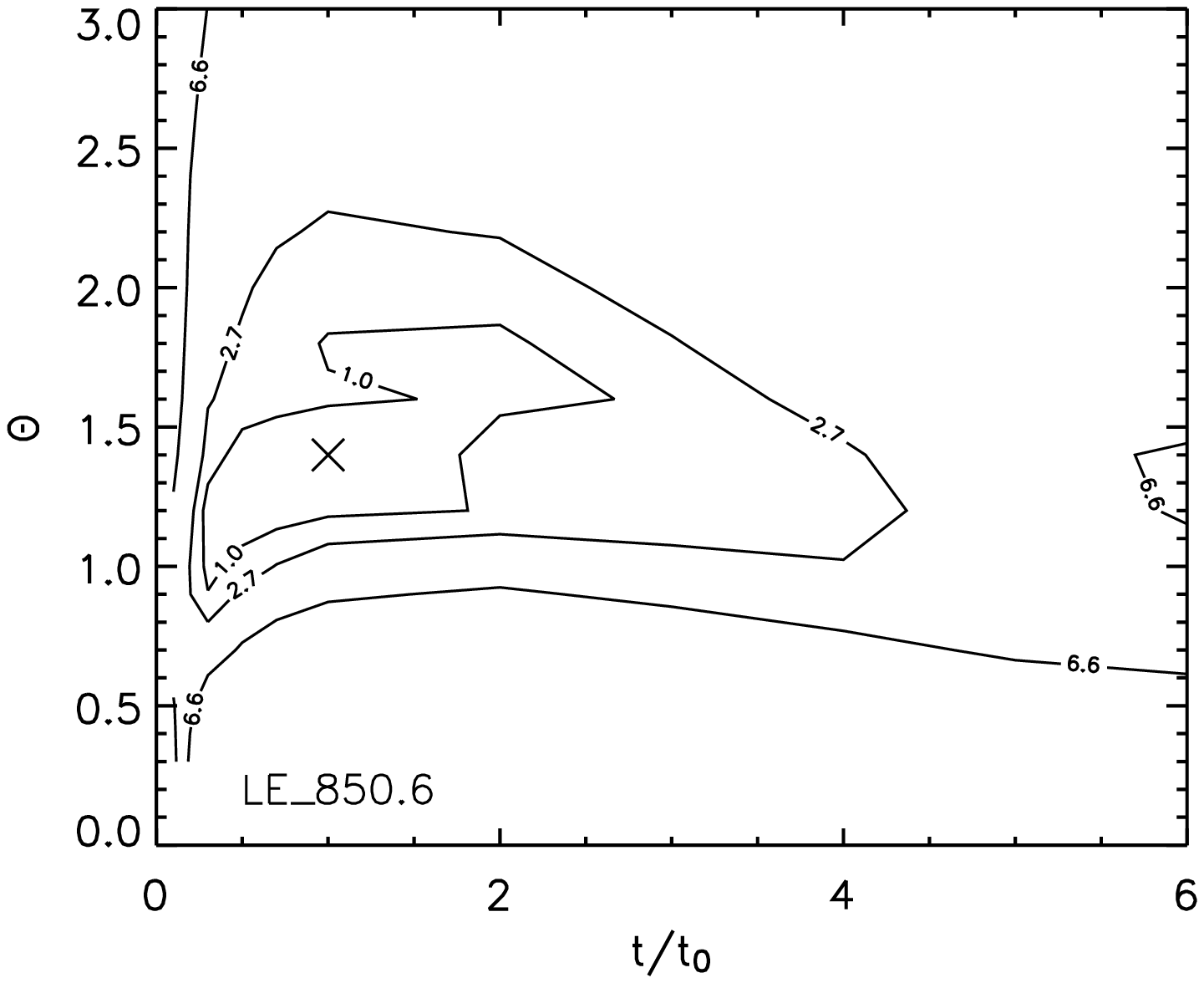}}
  \resizebox{6cm}{!}{\includegraphics{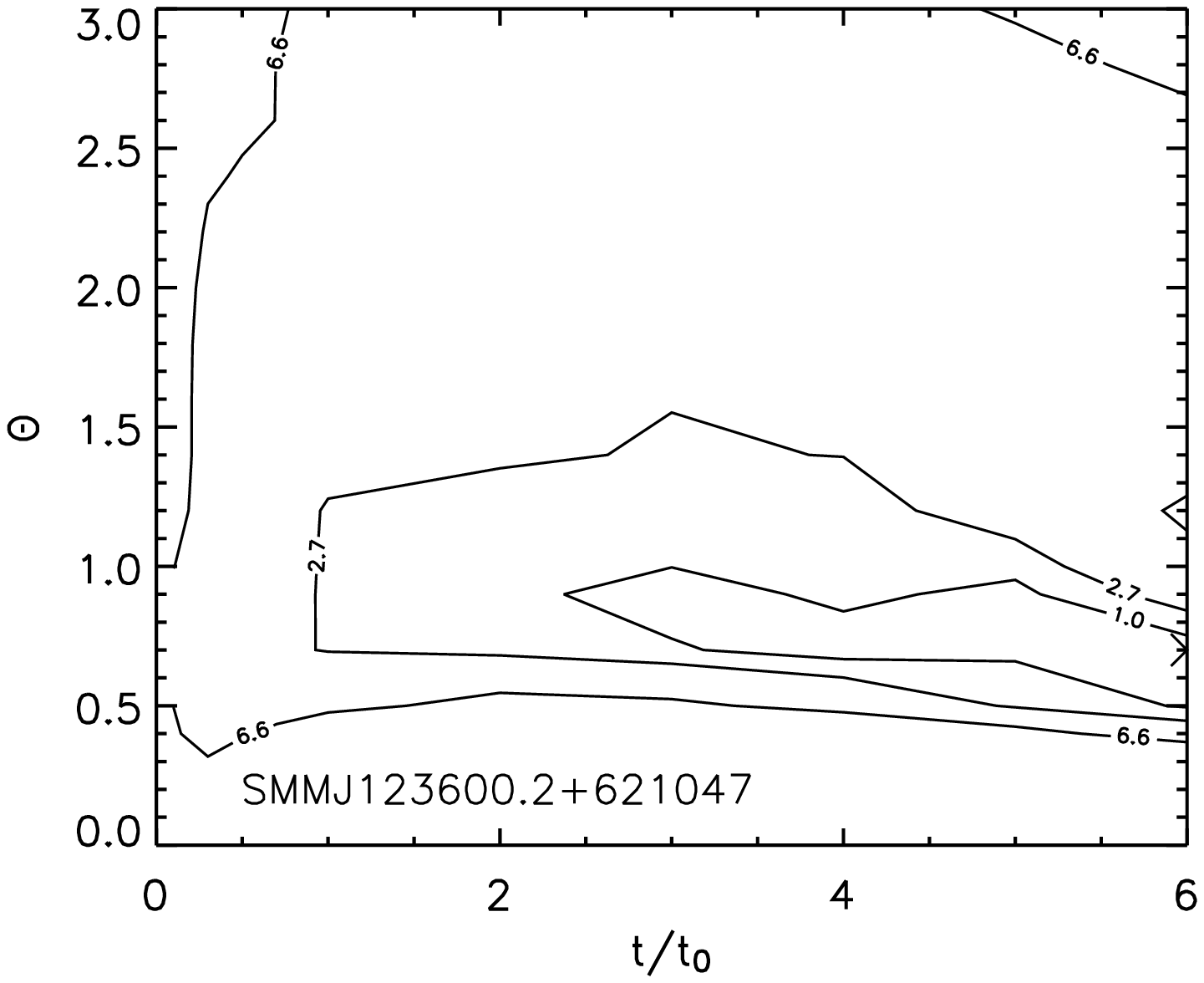}}
  \resizebox{6cm}{!}{\includegraphics{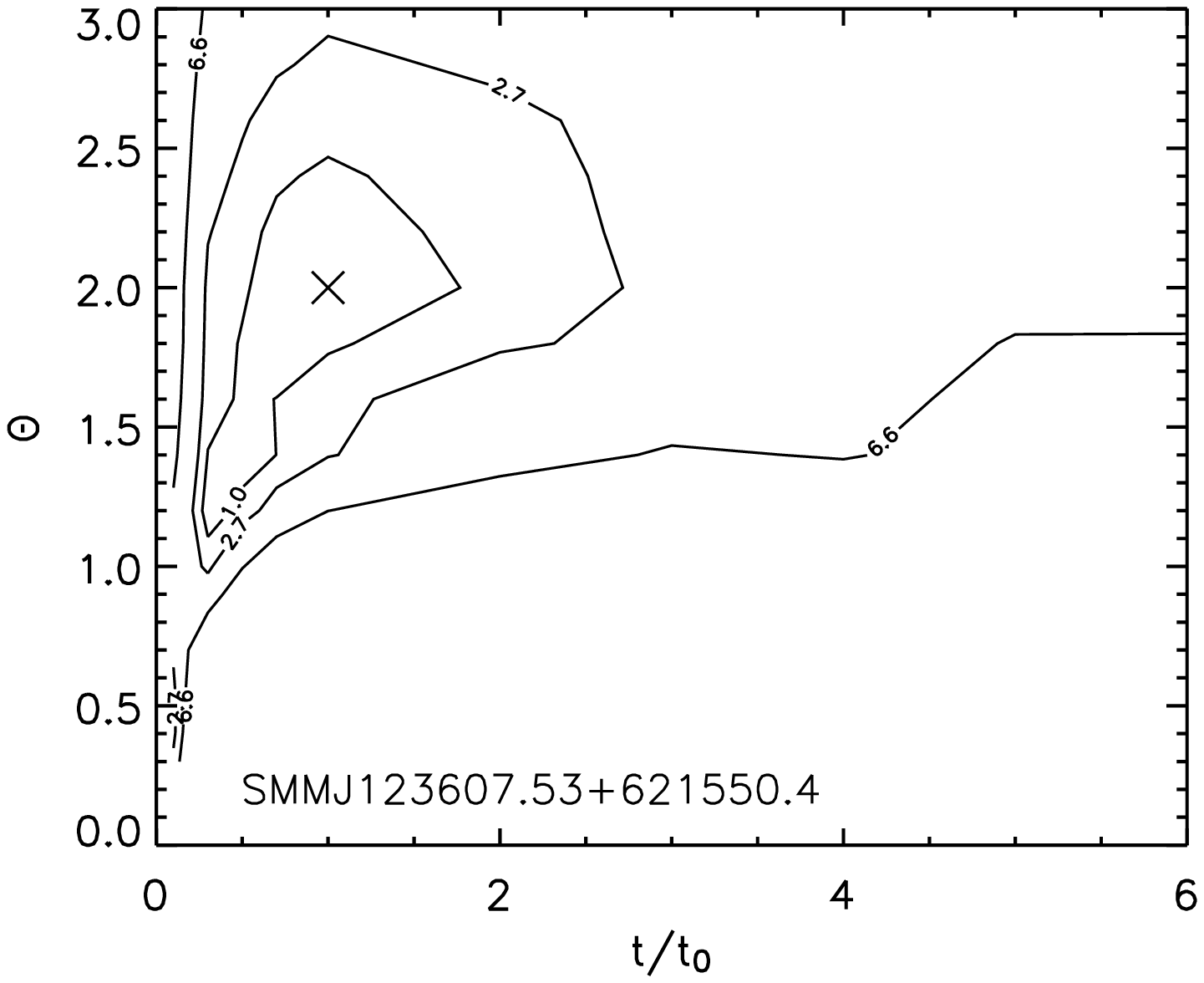}}
  \resizebox{6cm}{!}{\includegraphics{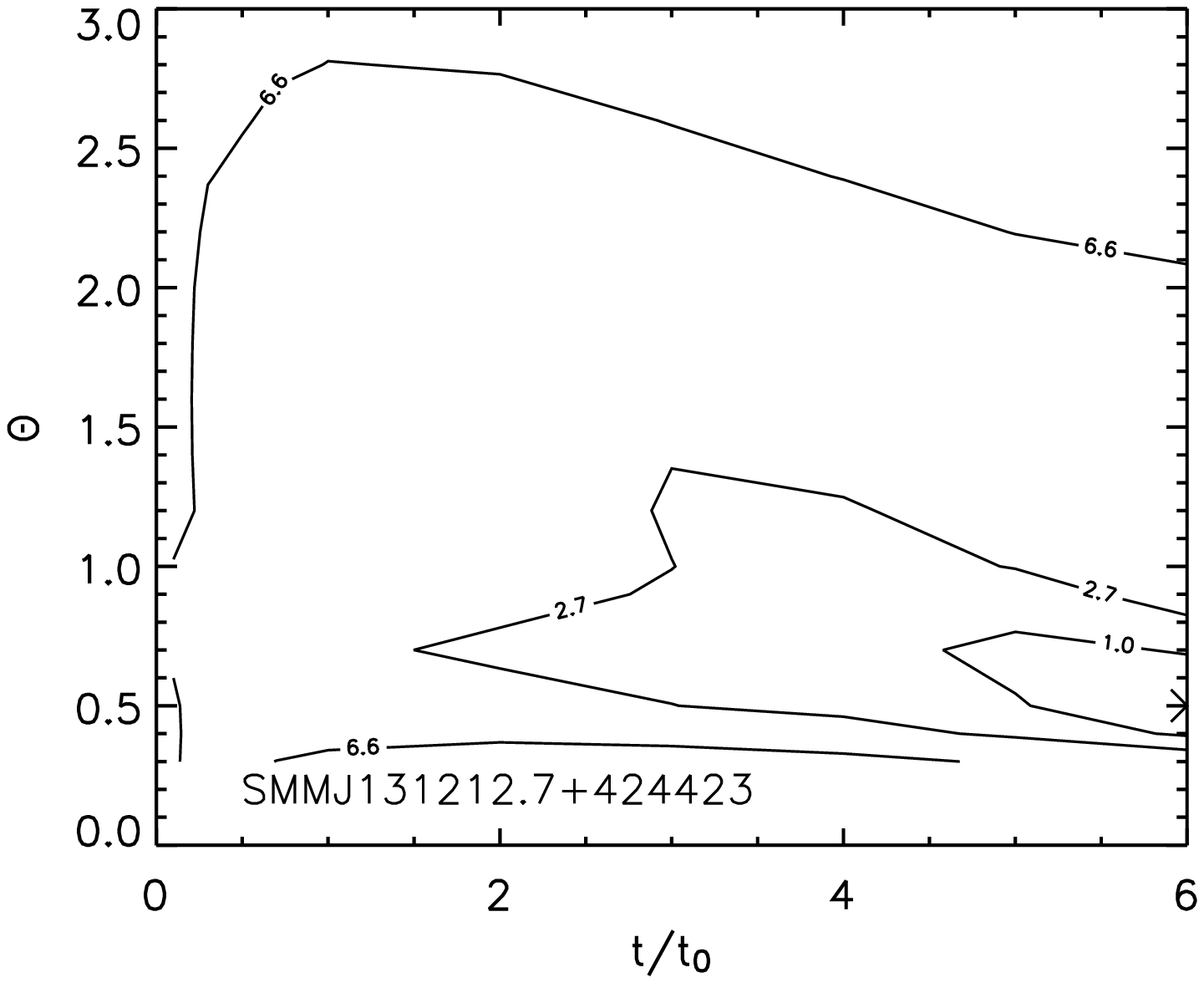}}
  \resizebox{6cm}{!}{\includegraphics{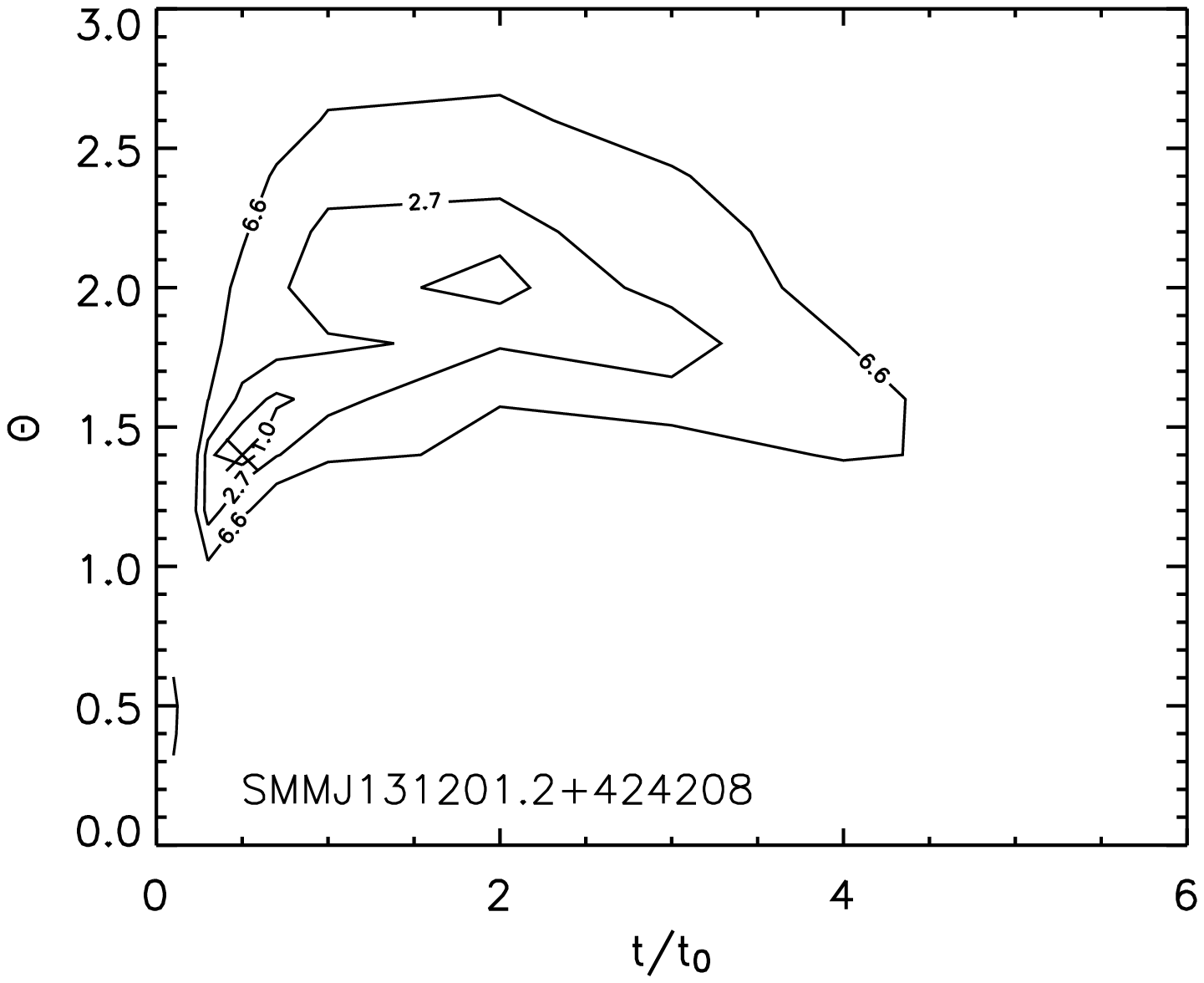}}
  \resizebox{6cm}{!}{\includegraphics{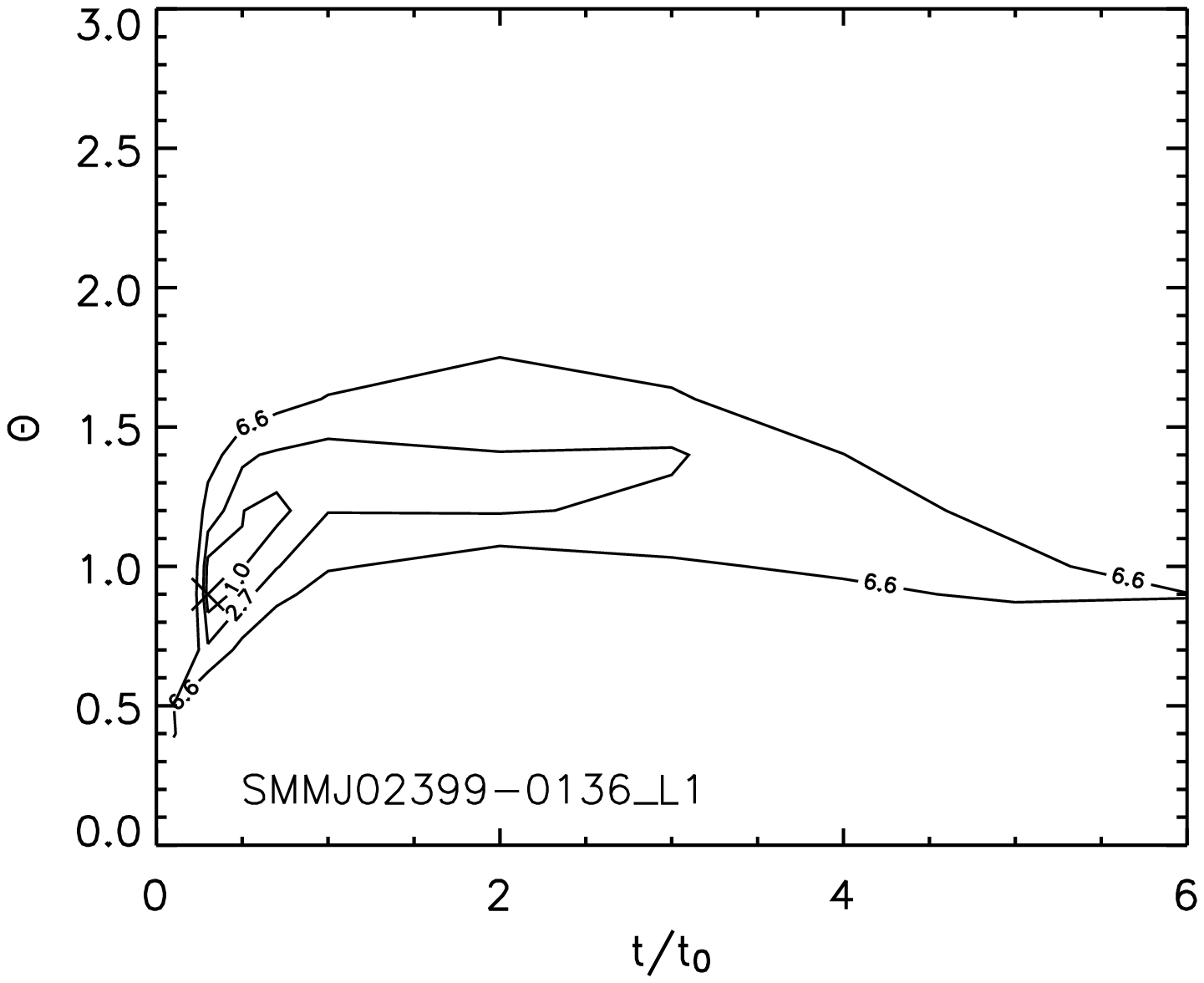}}
  \resizebox{6cm}{!}{\includegraphics{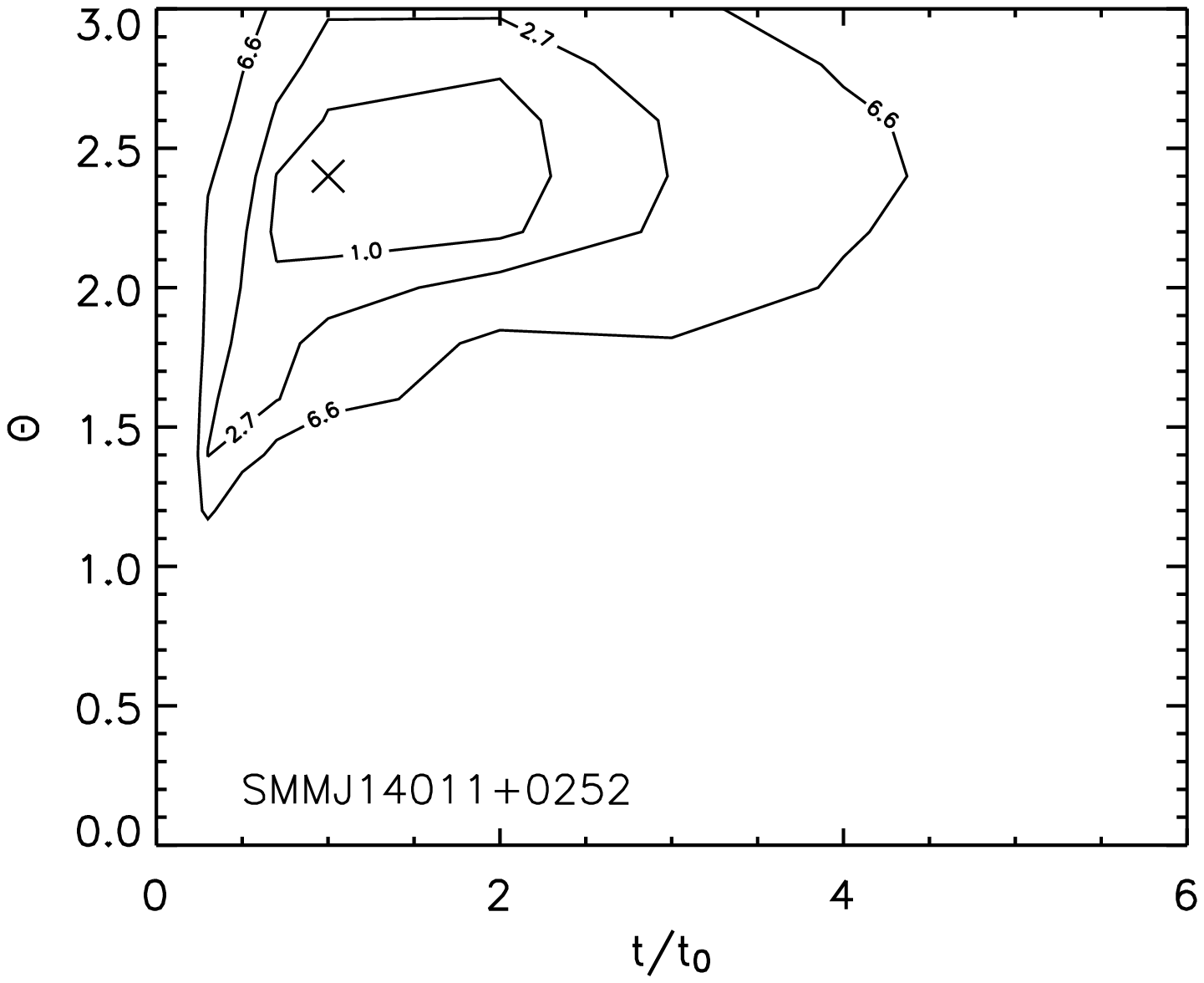}}
 \caption{{\it - continued.}
}
 \label{chi2b}
\end{figure*}

\addtocounter{figure}{-1}
\begin{figure*}
  \resizebox{6cm}{!}{\includegraphics{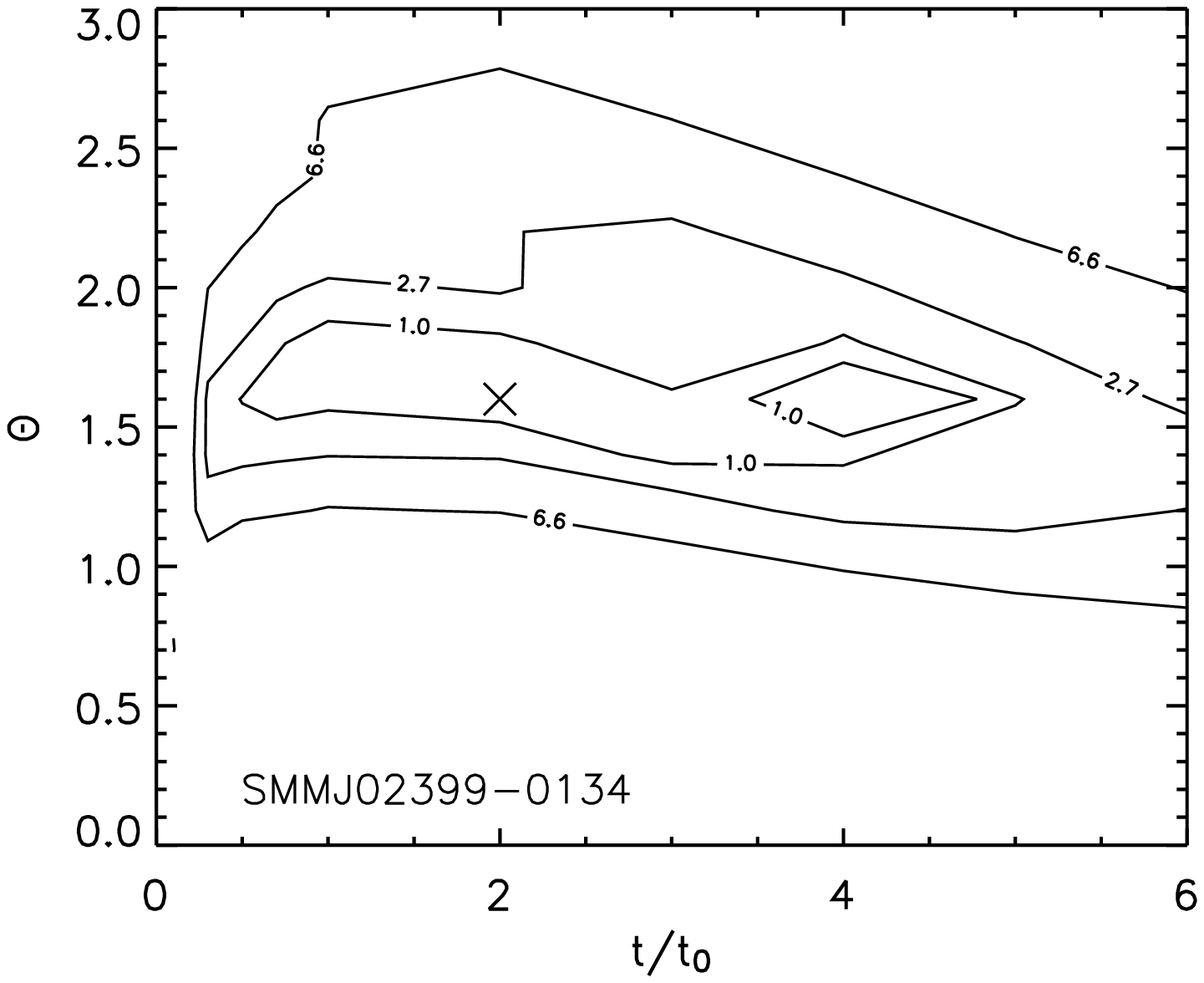}}
  \resizebox{6cm}{!}{\includegraphics{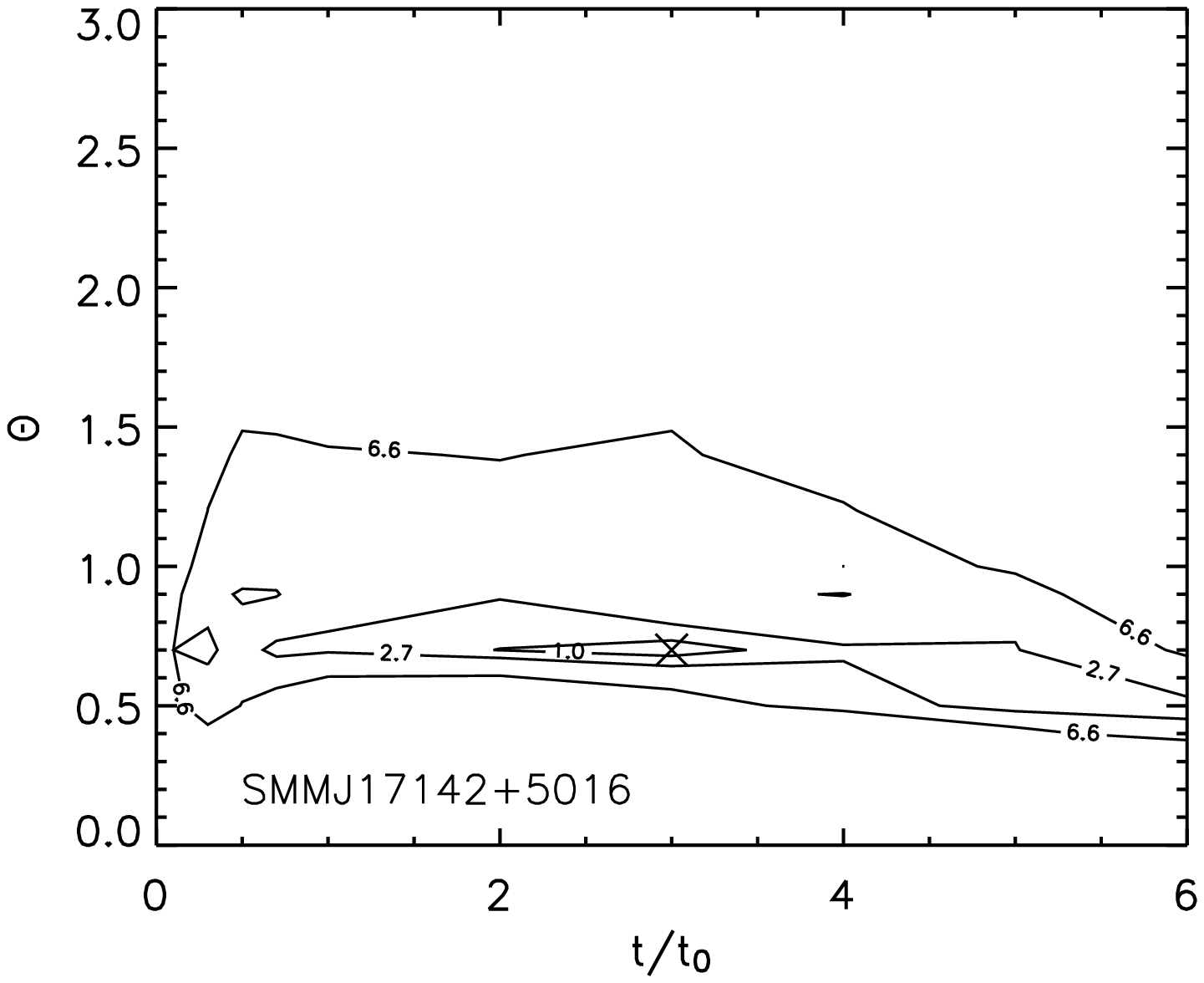}}
  \resizebox{6cm}{!}{\includegraphics{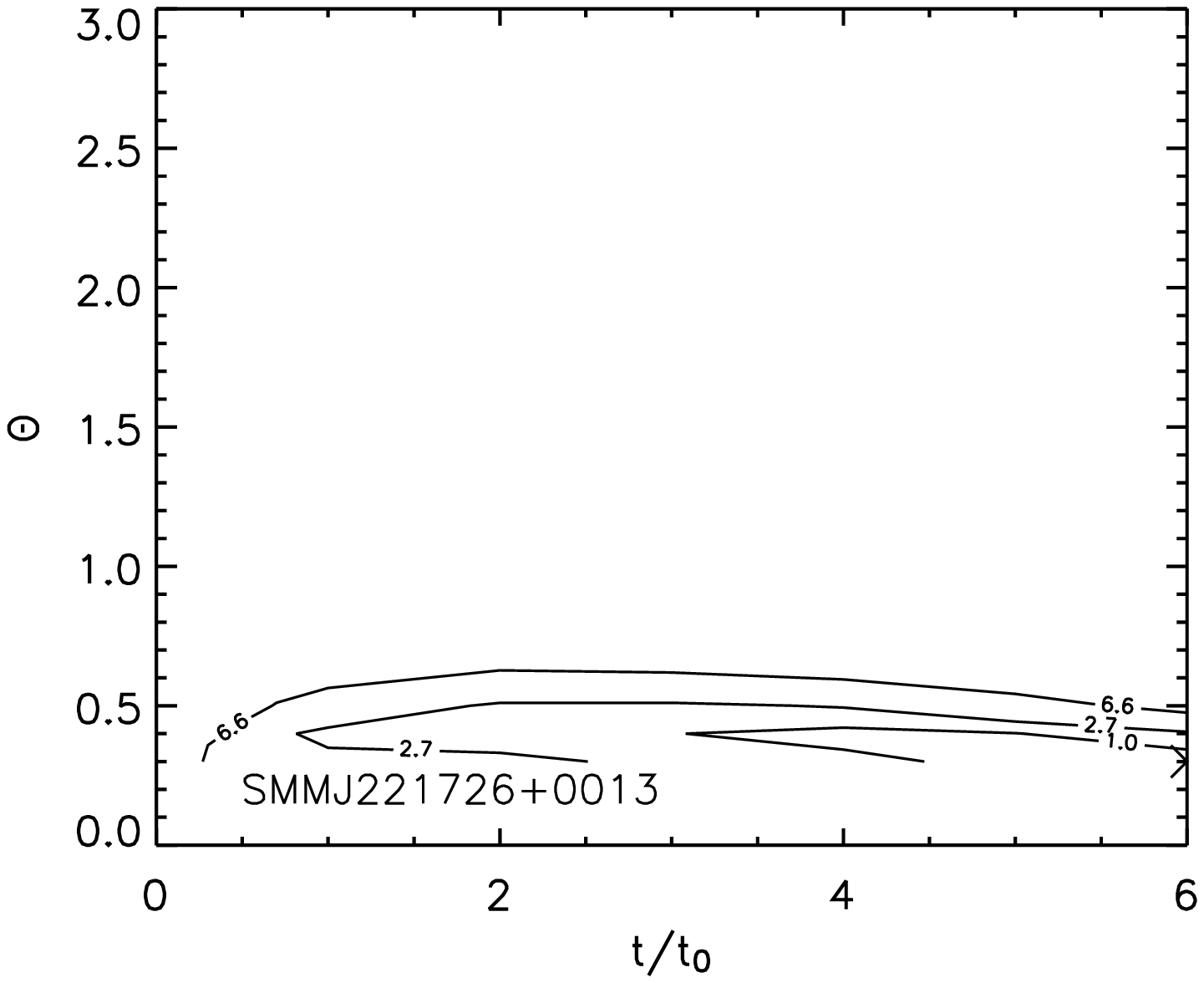}}
  \resizebox{6cm}{!}{\includegraphics{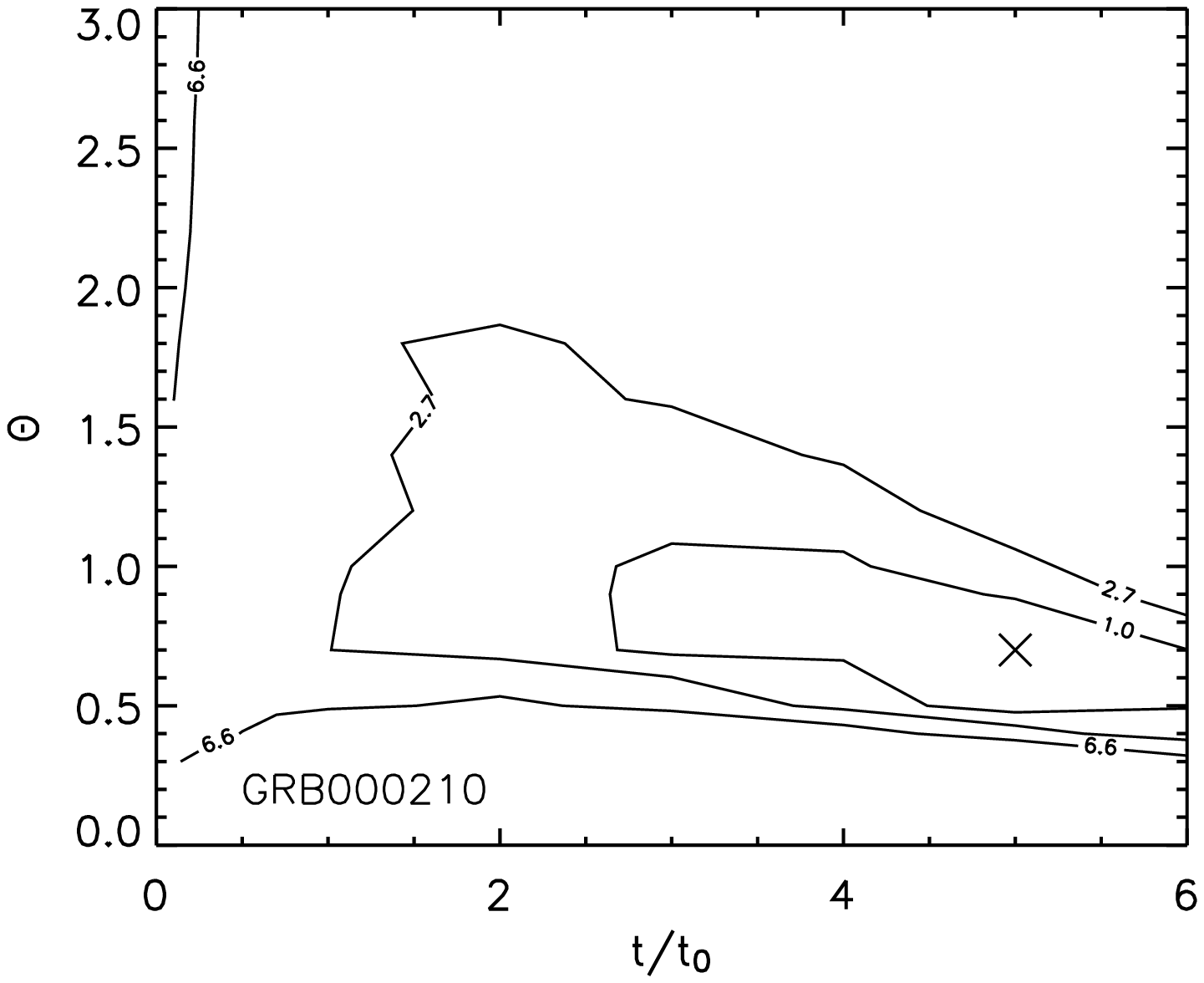}}
  \resizebox{6cm}{!}{\includegraphics{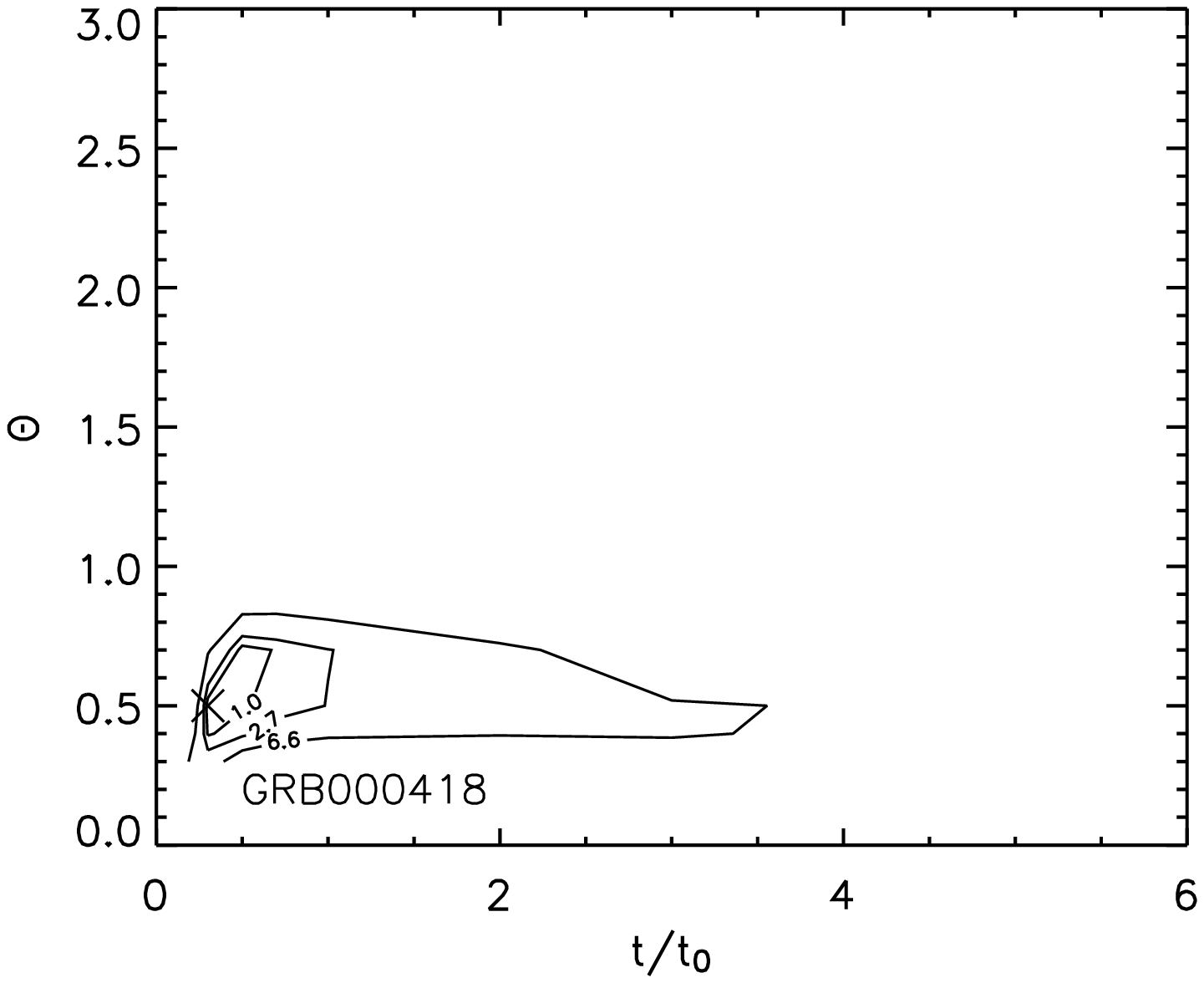}}
  \resizebox{6cm}{!}{\includegraphics{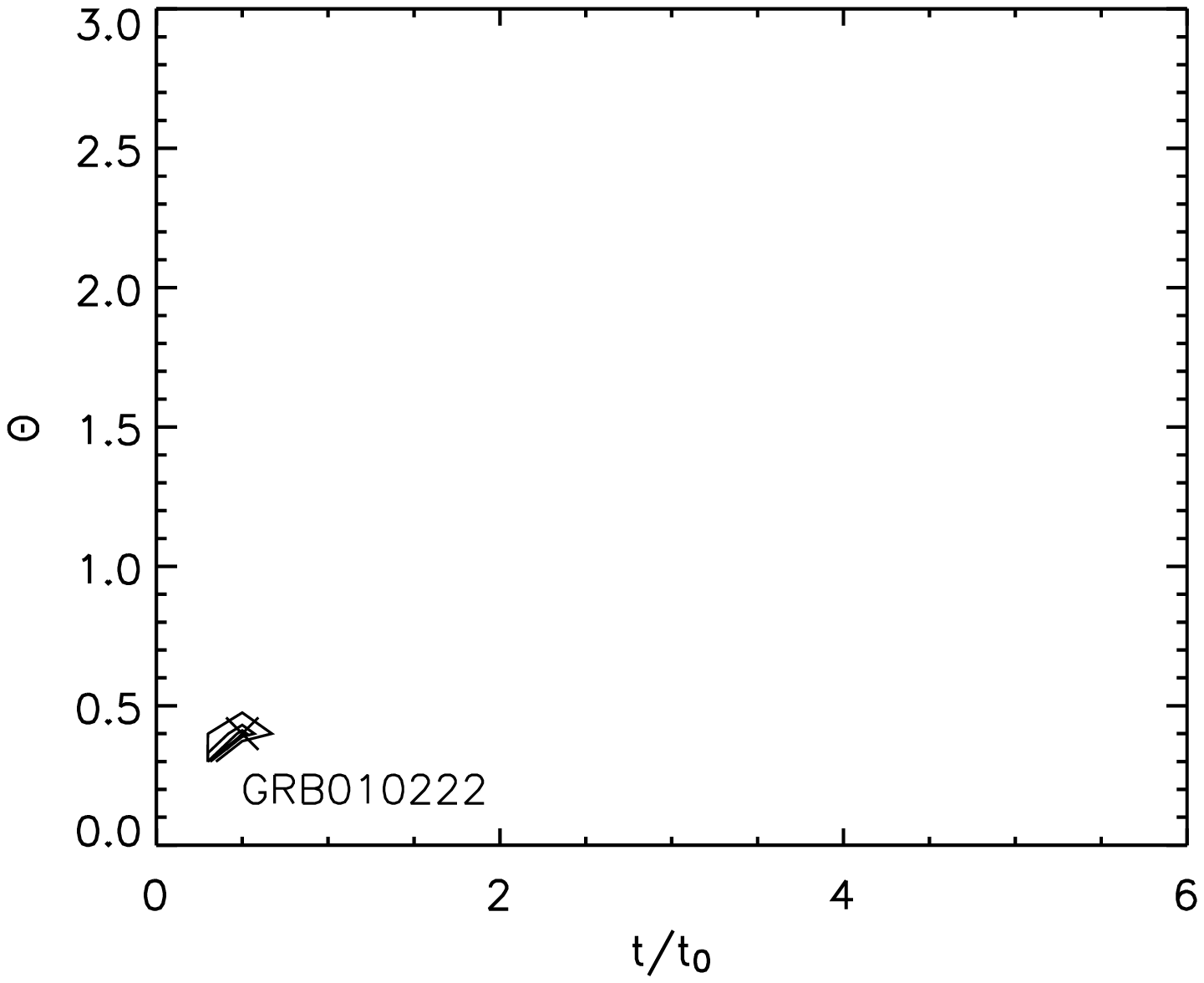}}
 \caption{{\it - continued.}
}
 \label{chi2b}
\end{figure*}

\end{document}